\documentclass[11pt,a4paper]{article}
\pdfoutput=1
\usepackage{cancel}
\usepackage{enumitem}
\usepackage{etex}
\usepackage{amsthm}
\usepackage{geometry}
\usepackage{dcolumn}
\usepackage{epsf}
\usepackage{mathrsfs}
\usepackage{multirow}
\usepackage{booktabs}
\usepackage{tabularx}
\usepackage{array}
\usepackage{slashed}
\usepackage{float}
\usepackage{leftidx}
\usepackage{setspace}
\usepackage{verbatim}
\usepackage{adjustbox}
\usepackage{tikz}
\usepackage[caption=false]{subfig}
\usepackage{arydshln}
\usepackage{jheppub}
\usepackage{shuffle}
\usepackage{amsmath}
\usepackage{cases}
\usepackage{lipsum}
\usepackage{tabularx}
\usepackage{graphicx}
\usepackage{lscape}

\numberwithin{equation}{section}
\usetikzlibrary{arrows,decorations.markings,shapes.arrows,patterns,positioning}

\newcommand{\bea}{\begin{eqnarray}}
\newcommand{\eea}{\end{eqnarray}}
\newcommand{\bean}{\begin{eqnarray*}}
\newcommand{\eean}{\end{eqnarray*}}
\newcommand{\nn}{\nonumber\\}
\newcommand{\Sl}{\sum\limits}
\newcommand{\blue}{\color{blue}}

\def\W #1{\widetilde{#1}}

\def\Label#1{\label{#1}%
  \smash{\hbox to0pt{\raise1ex\hbox{\tiny[#1]}\hss}}}
\def\Label#1{\label{#1}}
\renewcommand{\eqref}[1]{eq.~(\ref{#1})}
\newcommand{\figref}[1]{Fig.~\ref{#1}}
\newcommand{\tabref}[1]{table~\ref{#1}}
\newcommand{\secref}[1]{section~\ref{#1}}

\def\Sl{\sum\limits}

\newcommand{\ctobedelete}[1]{}

\allowdisplaybreaks

\title{Local vertices, quadratic propagators and double-copy structure of one-loop integrands from forward limits}

\author[a]{Chongsi Xie} \author[a,b]{Yi-Jian Du\footnote{Corresponding author}}

\affiliation[a]{Department of Physics, School of Physics and Technology,
Wuhan University, \\
No.299 Bayi Road, Wuhan 430072, China}
\affiliation[b]{Hubei Key Laboratory of Nuclear Solid Physics, School of Physics and Technology, Wuhan University,\\
No.299 Bayi Road, Wuhan 430072, China}
\emailAdd{chongsi.xie@whu.edu.cn} \emailAdd{yijian.du@whu.edu.cn}

\date{\today}
\abstract{By worldsheet approach, $n$-point one-loop integrand can be expressed as a combination of  $(n+2)$-point tree-level bi-adjoint scalar (BS) amplitudes under forward limit. The integrands constructed by this approach have two closely related features which differ from conventional Feynman diagrams. First, the denominators of loop propagators are linear functions of the loop momentum. Second, the local vertex expression is not manifest. In our previous work, a systematic approach was proposed to handle the nonlocal terms in the one-loop integrand of Yang-Mills-scalar (YMS) theory. Upon canceling the nonlocalities, quadratic propagator forms of both  YMS and Yang-Mills (YM) integrands  are naturally obtained. In this paper, we generalize the calculation to theories involving gravitons by introducing the one-loop double Yang-Mills-scalar (dYMS) integrands. The cancellation of the nonlocalities of the dYMS integrand in the forward limit coincides with the emergence of local multi-point vertices. We provide two equivalent methods for extracting the vertices and give the final expression of the dYMS integrand with quadratic propagators. In this formula,  tree-level effective subcurrents are attached to the loop propagator line via local vertices. Each of the effective subcurrents exhibits double-copy structure, in the sense that it is expressed as a combination of tree-level BS subcurrents associated with two copies of kinematic coefficients. The quadratic propagator formulas for Einstein-Yang-Mills (EYM) and gravity (GR) integrands are further derived, by the help of the formula for dYMS. The extraction of local vertices in one-loop dYMS integrand also applies at tree-level, thus we have the corresponding  expressions of tree-level dYMS, EYM, and GR amplitudes.
}


\begin{document}
\maketitle \flushbottom

\section{Introduction}



 Over the past few decades, many hidden structures of scattering amplitudes have been revealed. Among these new progresses, amplitude relations such as the Kleiss-Kuijf (KK) \cite{Kleiss:1988ne} and Bern-Carrasco-Johansson (BCJ) \cite{Bern:2008qj,Bern:2010ue} relations for tree-level color-ordered amplitudes, as well as the Kawai-Lewellen-Tye (KLT) \cite{Kawai:1985xq} relation between tree amplitudes of distinct theories have been discovered. Relations between scattering amplitudes play an important role in simplifying calculations and in uncovering hidden structures of theories. In \cite{Fu:2017uzt,Chiodaroli:2017ngp,Teng:2017tbo,Du:2017gnh} (which were inspired by earlier work \cite{Stieberger:2016lng,Nandan:2016pya,Schlotterer:2016cxa}), a recursive expansion relation for  tree-level Einstein-Yang-Mills (EYM) amplitudes was proposed, by which, a color-ordered EYM amplitude is expressed as a combination of those amplitudes with fewer external gravitons. This relation results in expressions of EYM and gravity (GR) amplitudes in terms of color-ordered Yang-Mills (YM) amplitudes, which  are closely related to the construction of polynomial BCJ \cite{Bern:2008qj,Bern:2010ue} numerators \cite{Du:2017kpo}. Furthermore, the EYM recursion relation is also satisfied by Yang-Mills-scalar (YMS) amplitudes \cite{Fu:2017uzt}\footnote{An alternative approach to the expansion of YMS amplitudes (with massive scalars of the same mass), based on Hopf algebra, was supposed in \cite{Chen:2022nei}.} and amplitudes in other related theories \cite{Dong:2021qai}. More studies on the EYM recursive expansion including the refined graphic rule  \cite{Hou:2018bwm,Du:2019vzf}, the off-shell approach \cite{Wu:2021exa,Tao:2024vcz}, the gauge invariance identities \cite{Hou:2018bwm,Du:2019vzf,Du:2022vsw} as well as the soft-limit construction \cite{Zhou:2022orv,Zhou:2023iae,Hu:2023koy,Du:2024xva} have been provided.

The recursive expansion relation for tree-level EYM amplitudes was derived/proved in different approaches, such as Britto-Cachazo-Feng-Witten (BCFW) \cite{Britto:2004ap,Britto:2005fq} recursion, Feynman diagrams, and the Cachazo-He-Yuan (CHY) \cite{Cachazo:2013gna,Cachazo:2013hca,Cachazo:2013iea,Cachazo:2014nsa,Cachazo:2014xea} formula. { The latter is essentially based on a worldsheet interpretation, which can also be derived from the ambitwistor string whose correlators reproduce the CHY integrand \cite{Mason:2013sva}}. In particular, in the framework of the CHY formula, the full amplitude is expressed by an integral over scattering variables that satisfy scattering equations. The integrand depends on theories and is factorized into two half integrands. From the aspect of the CHY formula, the recursive expansion formula of EYM amplitudes is reflected as recursive relations between half integrands \cite{Fu:2017uzt,Teng:2017tbo,Du:2017kpo,Du:2017gnh}. This finally reduces any half integrand as a combination of the Parke-Taylor (PT) \cite{Parke:1986gb} factors which are the half integrands of bi-adjoint scalar (BS) theory.

A significant development of the world-sheet approach is to generalize the formulas satisfied by tree-level amplitudes to one-loop level.  As shown in \cite{Adamo:2013tsa,Mason:2013sva,He:2015yua,Cachazo:2015aol,Wen:2020qrj}, both CHY formula and ambitwistor string approach are extended to one-loop, through so-called {\it forward limit} approach, which expresses a one-loop integrand with $n$ external particles by $(n+2)$-point tree amplitudes with two particles carrying the plus/minus loop momentum. Along this line, a one-loop integrand is given as a combination of  $(n+2)$-point tree-level BS amplitudes, while the expansion coefficients follow from the tree-level expansion formula. The KLT relation, double-copy expressions, and more about the recursive expansion formula at one-loop level were proposed from this forward limit approach \cite{He:2016mzd,He:2017spx}.

Although the forward limit approach has been shown to be an effective method for calculating the one-loop integrands, the integrand constructed by this method has two prominent features that are quite different from conventional Feynman diagram approach: (i) The denominators of loop propagators are linear functions of loop momentum \cite{He:2015yua,Cachazo:2015aol,Geyer:2015jch,Geyer:2017ela,Edison:2020uzf} but those in a conventional Feynman diagram are quadratic functions; (ii) The locality of the expansion formula from forward limit is not manifest, but a Feynman diagram is constructed by local vertices. These features prevent one from making use of the conventional treatments that are applied in a Feynman diagram approach. What's more, a general construction of BCJ numerators, with respect to quadratic propagators, remains as an open problem.

In order to overcome the problem of linear propagators, earlier efforts have been made, see  \cite{He:2015yua,Cachazo:2015aol,Geyer:2015jch,Geyer:2017ela,Edison:2020uzf,Feng:2022wee,Baadsgaard:2015hia,Cardona:2016bpi,Cardona:2016wcr,Gomez:2016cqb,Gomez:2017lhy,Ahmadiniaz:2018nvr,Agerskov:2019ryp,Farrow:2020voh,Porkert:2022efy}.  As pointed out in \cite{Feng:2022wee}, once a half integrand from the worldsheet approach is expanded in terms of tensorial PT factors, a full integrand with quadratic propagators are generally obtained. A further study \cite{Dong:2023stt} showed that this tensorial PT factor approach was essentially related to a general construction of BCJ numerators with respect to quadratic propagators. Nevertheless, there is still a lack of a general approach to the tensorial PT factor expansion. In our previous work  \cite{Xie:2024pro}, another treatment of the linear propagator problem was provided. We showed that the nonlocal expressions of YMS integrands can be localized schematically, yielding quadratic propagator formulas as a direct consequence. Through the relations between YM and YMS, a general formula of quadratic propagator form of the YM integrand was further provided \cite{Xie:2024pro}. Another interesting approach to such quadratic-propagator expressed integrand \cite{Cao:2024olg} is to consider the single cuts of integrands by the help of surfaceology \cite{Jashi:2024bay,Arkani-Hamed:2024nhp}. It seems that the formula presented in \cite{Xie:2024pro} can be regarded as an explicit solution of the approach in \cite{Cao:2024olg}.

In this work, we extend our earlier approach in \cite{Xie:2024pro} to the one-loop integrands of theories with gravitons (coupled to dilatons and B-fields). We point out that the one-loop integrands of EYM and GR can be expanded in terms of the so-called double Yang-Mills-scalar (dYMS) integrands which consist of two copies of the left half integrands of YMS. Therefore, a quadratic propagator form of the dYMS integrands directly means a quadratic propagator form of the integrands of theories with gravitons. By focusing on $(n+2)$-point dYMS tree amplitude under forward limit (which are called partial integrands in this paper), we show a systematic way to localize the expressions that benefit from the off-shell BCJ relations and the refined graphic rule. After this step, {\it local multi-point vertices} for dYMS are induced, and the full local expression of the partial integrands is derived. These local partial integrands produce a quadratic propagator formula of the full dYMS integrand, according to the partial fraction identity. Since the off-shell BCJ relations can be regarded as results of gauge invariance, the localization of this work is in fact a technical approach to achieving locality by gauge invariance, which was earlier discussed in \cite{Arkani-Hamed:2016rak}. Consequently, the final expression of one-loop integrand is expressed as a combination of such terms that each is given by attaching tree-level double-copy objects to the loop (with quadratic propagators) via local vertices.

Now let us  provide an overview of the main features of the local formula of the dYMS integrand with quadratic propagators. {In this formula, the external particles are partitioned into disjoint sets, each of which is involved in a tree-level effective current that is} planted at the quadratic propagator line via local vertices. All possible partitions are summed over. The explicit expressions of the local vertices can be produced systematically from the  cancellation of nonlocalities. The tree-level effective currents, {which can be either obtained directly from the graphic rule or generated in a recursive way inherited from the EYM recursive expansion}, have an evident {\it double-copy structure}.  To be specific, this double-copy structure means that a tree-level effective current is a combination of tree-level Berends-Giele BS currents with two copies of kinematic coefficients. Under on-shell limit, this double-copy expression turns into the familiar one, which expresses the full tree amplitude in terms of tree-level BS amplitudes and the coefficients can be used to generate tree-level BCJ numerators. Therefore, the final result of one-loop dYMS (hence EYM and GR) integrand is given by a sum of terms, each involving such tree-level double-copy objects attached to the quadratic propagator loop, via local vertices.
The expressions of one-loop integrands proposed in this work can be regarded as an intermediate one between (i) the one directly derived from the forward limit of tree-level double-copy expression which involves linear propagators \cite{He:2016mzd,He:2017spx,Geyer:2017ela}, and (ii) the expected quadratic propagator expressed BCJ double-copy which will result in an expansion formula in terms of one-loop BS integrands (with quadratic propagators). We expect that this new formula may help to study the general construction of BCJ numerators with respect to quadratic propagators.

The structure of this paper is given as follows. In  \secref{sec:review}, a helpful review of one-loop integrands, graphic rule, and relations between Berends-Giele currents is provided. We provide the main idea of this work in  \secref{sec:MainIdea}. The dYMS partial integrands are classified according to the number of $\mathsf{W}$ elements (denoted by  $|\mathsf{W}|$) which have two copies of polarization vectors. In \secref{sec:W0}, we show the localization of the partial integrand with linear propagators and the full integrand with quadratic propagators when  $|\mathsf{W}|=0$. The double-copy structures and convention for contractions of Lorentz indices are established in \secref{sec:5}. The more complicated cases, with $|\mathsf{W}|=1$ and $|\mathsf{W}|=2$  are discussed in \secref{sec:W1} and \secref{sec:W2}. In \secref{sec:W3}, we provide an alternative approach to extracting the local vertices with  $|\mathsf{W}|=3$. In \secref{sec:MoreDiscussions}, we supply discussions on tree amplitudes. A summary  of this work  and further discussions are provided in \secref{sec:Conclusions}. More cancellations of the  $|\mathsf{W}|=2$ example is involved in the appendix. 
\section{Review of one-loop integrands, graphic rule, graph-based relations and the local form of the YMS integrand}\label{sec:review}

In this section, we provide a review of useful definitions and results, including the one-loop integrands of various theories, the graphic rule, graph-based relations of BS currents, and the local expression of the YMS integrand. All these results are helpful in this work.

\subsection{One-loop integrands with forward limit}

In the framework of the worldsheet approach, a one-loop integrand is obtained by taking the forward limit of the tree-level amplitude. Generally speaking,
 one-loop $n$-point amplitude $M_n^{\text{1-loop}}$ is given by the following integral
\bea
M_n^{\text{1-loop}}=\int {\text{d}^Dl}\,\mathcal{I}(l)=\int {\text{d}^Dl\over l^2}\lim\limits_{k_{\pm}\to \pm l}\int \text{d}\mu_{\,n+2}^{\,\text{tree}}\,I_{\,\text{L}}^{\,\text{1-loop}}\,I_{\,\text{R}}^{\,\text{1-loop}},
\Label{Eq:LoopCHY}
\eea
where $l^{\mu}$ denotes the loop momentum in $D$ dimensions. The loop integrand $\mathcal{I}(l)$ of this amplitude, upto ${1\over l^2}$, is expressed by an integral over scattering variables ${z_i}$ ($i=1,...,n$), where each of the half integrands $I_{\,\text{L}}^{\,\text{1-loop}}$ and $I_{\,\text{R}}^{\,\text{1-loop}}$ is the one for $(n+2)$-point tree-level amplitudes from worldsheet approach. The notation $\lim\limits_{k_{\pm}\to \pm l}$ in (\ref{Eq:LoopCHY}) implies the {\it forward limit}, i.e., two of the $(n+2)$ particles, $+$ and $-$ carry the loop momentum $\pm l^{\mu}$, while other $n$ particles are those external particles for the one-loop amplitude. The integral measure $\text{d}\mu_{\,n+2}^{\,\text{tree}}$ implies that the integration over scattering variables ${z_i}$  should be carried out under the constraint of scattering equations
\bea
{l\cdot k_i\over z_i}+\Sl_{\substack{j=1\\ j\neq i}} {k_i\cdot k_j\over z_{ij}}=0,
\eea
in which $k^{\mu}_i$ ($i=1,...,n$) refers to the external momentum of the massless external particle and $z_{ij}\equiv z_i-z_j$.

In the remainder of this section, we first review the integrand of BS theory and the tree-level Berends-Giele currents of BS, which together form the backbone of the  subsequent discussion. We then provide the expansion relations for half integrands of different theories, by which one can finally express any integrand of YMS, YM, EYM, or GR as a combination of the BS ones. The local expression of the YMS integrand given in \cite{Xie:2024pro} is also reviewed.

\subsubsection{The BS integrand and tree-level Berends-Giele currents of BS}
In BS theory, we express the color stripped half integrands $I_{\,\text{L}}^{\,\text{1-loop}}$ and $I_{\,\text{R}}^{\,\text{1-loop}}$ with respect to permutations of external particles $\pmb\gamma=(\gamma_1...\gamma_n)$ and $\pmb\sigma=(\sigma_1...\sigma_n)$ as  $I_{\,\text{L}}^{\,\text{1-loop}}(+,\pmb\gamma,-)$ and $I_{\,\text{R}}^{\,\text{1-loop}}(+,\pmb\sigma,-)$, while the corresponding one-loop integrand $\mathcal{I}^{\text{BS}}$ in (\ref{Eq:LoopCHY}) is labeled as $\mathcal{I}^{\text{BS}}\big(\pmb{\gamma}\big|\pmb{\sigma}\big)$. The explicit expression of the half integrands are given by
\bea
I_{\,\text{L}}^{\,\text{1-loop}}(+,\pmb\gamma,-)=\text{PT}(+,\pmb\gamma,-)+\text{cyc}(\pmb{\gamma}),~~~I_{\,\text{R}}^{\,\text{1-loop}}(+,\pmb\sigma,-)=\text{PT}(+,\pmb\sigma,-)+\text{cyc}(\pmb{\sigma}),~~~\Label{Eq:BSHalfIntegrands}
\eea
where the $+$ and $-$ refer to the two particles under the forward limit. In general, a PT factor $\text{PT}(\pmb\rho)$  for a given permutation $\pmb\rho=(\rho_1...\rho_m)$ is defined by
\bea
\text{PT}(\pmb{\rho})\equiv{1\over {z_{\rho_1\rho_2}z_{\rho_2\rho_3}...z_{\rho_{m-1}\rho_m}z_{\rho_m\rho_1}}}.~~~\Label{Eq:PTtree}
\eea
When we substitute the half integrands (\ref{Eq:BSHalfIntegrands}) into (\ref{Eq:LoopCHY}) and take the integration over scattering variables, the one-loop BS integrand turns into 
\bea
\mathcal{I}^{\text{BS}}\big(\pmb{\gamma}\big|\pmb{\sigma}\big)&\!\!=\!\!&\left[\,\mathcal{A}^{\text{BS}}\big(+,\pmb{\gamma},-\big|+,\pmb{\sigma},-\big)+\text{cyc}(\pmb{\gamma})\right]+\text{cyc}(\pmb{\sigma})\nn
&\!\!\equiv\!\!&\Bigg[\Sl_{\small\substack{(A_1A_2...A_i)={\pmb{\gamma}}\\\small (\W A_1\W A_2...\W A_i)={\pmb{\sigma}}\\ \small A_j=\W A_j}}{1\over l^2}{1\over s_{A_1,l}}{1\over s_{A_1A_2,l}}\cdots{1\over s_{A_1A_2\cdots A_{i-1},l}}\phi_{A_1|\W A_1}\phi_{A_2|\W A_2}\cdots \phi_{A_i|\W A_i}+\text{cyc}(\pmb{\gamma})\Bigg]+\text{cyc}(\pmb{\sigma})\nn
&\!\!=\!\!&\left[  \Sl_{\small\substack{(A_1A_2...A_i)={\pmb{\gamma}}\\\small (\W A_1\W A_2...\W A_i)={\pmb{\sigma}}\\ \small A_j=\W A_j}}\begin{minipage}{2.9cm}\includegraphics[width=2.9cm]{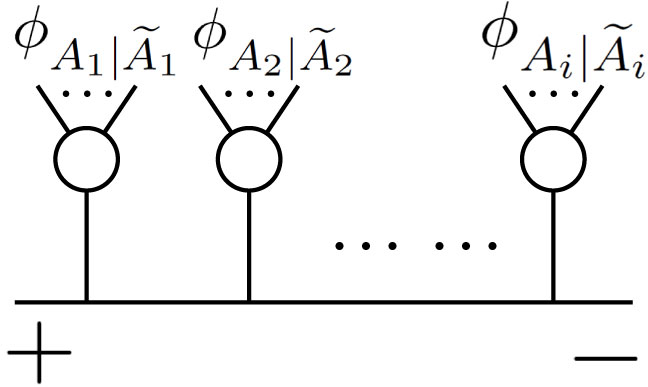}\end{minipage}+\text{cyc}(\pmb{\gamma})\right]+\text{cyc}(\pmb{\sigma}).\Label{Eq:One-loopBasis-1}
\eea
The $\mathcal{A}^{\text{BS}}\big(+,\pmb{\gamma},-\big|+,\pmb{\sigma},-\big)$ on the first line of the above expression, called the {\it partial integrand} of BS,  is defined by
\bea
 \mathcal{A}^{\text{BS}}\big(+,\pmb{\gamma},-\big|+,\pmb{\sigma},-\big)\equiv{1\over l^2}\lim\limits_{k^{\mu}_{\pm}\to \pm l^{\mu}}{A}_{\text{tree}}^{\text{BS}}\big(+,\pmb{\gamma},-\big|+,\pmb{\sigma},-\big).\Label{Eq:BSPartialIntegrand}
\eea 
This definition is further generalized to the partial integrand for an arbitrary theory
\bea
\mathcal{A}\equiv{1\over l^2}\lim\limits_{k^{\mu}_{\pm}\to \pm l^{\mu}}{A}_{\text{tree}}.\Label{Eq:PartialIntegrand}
\eea
On the second line of (\ref{Eq:One-loopBasis-1}), we summed over all possible {\it divisions}  $(A_1A_2...A_i)={\pmb{\gamma}}$ and $(\W A_1\W A_2...\W A_i)={\pmb{\sigma}}$ (including the case $i=1$) of the permutations  $\pmb{\gamma}$ and $\pmb{\sigma}$, respectively. For fixed divisions of $\pmb{\gamma}$ and $\pmb{\sigma}$, $A_j=\W A_j$ requires that subsets $A_j$ and $\W A_j$ contain the same elements. For example, if $\pmb\gamma=(a_1a_2a_3a_4a_5)$ and $\pmb\sigma=(a_2a_1a_3a_5a_4)$, the  four possible divisions of $\pmb\gamma$ and $\pmb\sigma$ such that $A_j=\W A_j$ are explicitly given by 
\bea
&&(a_1a_2|a_3|a_4a_5),~~~(a_2a_1|a_3|a_5a_4);~~~~~(a_1a_2a_3|a_4a_5),~~~(a_2a_1a_3|a_5a_4);\nn
&&(a_1a_2|a_3a_4a_5),~~~~(a_2a_1|a_3a_5a_4);~~~~~~(a_1a_2a_3a_4a_5),~~~~(a_2a_1a_3a_5a_4).
\eea
Once the divisions are fixed, the partial integrand is explicitly written as the product of propagators ${1\over s_{A_1\cdots A_j,l}}$ between $+$, $-$ and Berends-Giele subcurrents $\phi_{A_l|\W A_l}$ which will be introduced soon. The propagator ${1\over s_{A_1\cdots A_j,l}}$ is defined by
\bea
{1\over s_{A_1\cdots A_j,l}}\equiv{1\over {2 l\cdot (k_{A_1}+\cdots +k_{A_j})+(k_{A_1}+\cdots +k_{A_j})^2}},\Label{Eq:LinearPropagator}
\eea
where $k^{\mu}_{A_j}$ is the total momentum of elements in $A_j$. Such a propagator is  mentioned as {\it linear propagator}, for the denominator is a linear function of the loop momentum $l^{\mu}$.
Each term on the second line of (\ref{Eq:One-loopBasis-1}) is represented by {\it linear propagator expressed Feynman diagram} (LPFD) on the third line, with the additional factor ${1\over l^2}$ implicitly absorbed into the LPFD.  

Now we define the Berends-Giele currents of BS on the rhs. of (\ref{Eq:One-loopBasis-1}). For a given pair of ordered subsets $A$ and $\widetilde{A}$  (which can be any pair of $A_j$ and $\widetilde{A}_j$ for $j=1,...,i$ in (\ref{Eq:One-loopBasis-1})), the Berends-Giele subcurrent $\phi_{A|\widetilde{A}}$ of BS is recursively defined by \cite{Mafra:2016ltu}
\bea
\phi_{A|\widetilde{A}}={1\over s_A}\Sl_{\substack{A=A_LA_R\\ \widetilde{A}=\widetilde{A}_L\widetilde{A}_R}}\Bigl[\,\phi_{A_L|\widetilde{A}_L}\phi_{A_R|\widetilde{A}_R}-\phi_{A_R|\widetilde{A}_L}\phi_{A_L|\widetilde{A}_R}\,\Bigr],~\Label{Eq:BScurrent}
\eea
where $s_A\equiv k_A^2$.  We have summed over divisions $A=A_LA_R$, $\widetilde{A}=\widetilde{A}_L\widetilde{A}_R$ in (\ref{Eq:BScurrent}) so that in the first term $|A_L|=|\W A_L|$, $|A_R|=|\W A_R|$,  or in the second term  $|A_R|=|\W A_L|$, $|A_L|=|\W A_R|$.  Note that throughout this paper, the subsets $A_j$ ($\widetilde{A}_j$), $A_L$ ($\W A_L$) and  $A_R$ ($\W A_R$) resulting from any division are always assumed to be non-empty. The starting point of this recursive definition is $\phi_{a|a}=1$, $\phi_{a|b}=0$ $(a\neq b)$. Two consequent results are the following: (i) The BS current $\phi_{A|\widetilde{A}}$ is symmetric under $A\leftrightarrow \W A$. (ii) If the elements in $A$ and $\widetilde{A}$ in $\phi_{A|\widetilde{A}}$ are not identical to each other, the current must vanish. Another helpful property of the BS currrents is the generalized $U(1)$-decoupling identity \cite{Du:2011js}
\bea
\phi_{A\shuffle B|\widetilde{C}}=0, \Label{Eq:GenU1}
\eea
 where $A$ and $B$ are two ordered non-empty sets, $\W C$ is an arbitrary permutation of elements in $A\cup B$. The shuffle permutations $A\shuffle B$ of $A$ and $B$ are defined by all possible permutations obtained by merging $A$ and $B$, such that the relative ordering in each set is preserved.
{\small\begin{table}[]
    \centering
  
   \scalebox{0.95}{\begin{tabular}{|c|c|c|c|c|}
        \hline
        \small Theories   & \small YM &\small YMS & \small EYM &\small GR        \\[1pt] \hline
      \small Amplitudes  &\small $A^{\text{YM}}(\pmb{\sigma})$ &\small $A^{\text{YMS}}\big(\pmb{\sigma}||\mathsf{G}\big|\pmb{\rho}_{\mathsf{S}\cup \mathsf{G}}\big)$  & \small $A^{\text{EYM}}\big(\pmb{\sigma}||\mathsf{H}\big)$ & \small $M^{\text{GR}}(\mathsf{H})$  \\[1pt] \hline
    \small$I^{\text{1-loop}}_{\text{L}}$ & \small $\text{PT}(+,\pmb{\sigma},-)+\text{cyc}(\pmb{\sigma})$ & \small $\text{PT}(+,\pmb{\sigma},-)\text{Pf}\,[\,\Psi_{\mathsf{G}}]+\text{cyc}(\pmb{\sigma})$ & \small $\text{PT}(+,\pmb{\sigma},-)\text{Pf}\,[\,\Psi_{\mathsf{H}}]+\text{cyc}(\pmb{\sigma})$&\small $\text{Pf}\,'[\Psi_{n+2}]$\\[1pt] \hline
      \small$I^{\text{1-loop}}_{\text{R}}$ & \small $\text{Pf}\,'[\Psi_{n+2}]$ & \small$\text{PT}\left(+,\pmb{\rho}_{\mathsf{S}\cup \mathsf{G}},-\right)+\text{cyc}(\pmb{\rho}_{\mathsf{S}\cup \mathsf{G}})$ & \small
    $\text{Pf}\,'[\Psi_{n+2}]$ &\small $\text{Pf}\,'[\Psi_{n+2}]$\\[1pt] \hline
      \small$I^{\text{tree}}$ & \small $\text{PT}(\pmb{\sigma})\,\text{Pf}\,'[\Psi_{n}]$ & \small$\text{PT}(\pmb{\sigma})\text{Pf}\,[\,\Psi_{\mathsf{G}}]\,\text{PT}\left(\pmb{\rho}_{\mathsf{S}\cup \mathsf{G}}\right)$ & \small
    $\text{PT}(\pmb{\sigma})\text{Pf}\,[\,\Psi_{\mathsf{H}}]\,\text{Pf}\,'[\Psi_{n}]$ &\small $\left(\text{Pf}\,'[\Psi_{n}]\right)^2$\\[1pt] \hline
    \end{tabular} }
    \caption{One-loop half integrands  and tree-level full integrands for YM, YMS, EYM and GR. Here, we only provide the YMS and EYM half integrands that result in pure scalar loop and pure gluon loop, respectively.} \label{table:LoopIntegrands}
\end{table}}
As an example, for $A=(a_1a_2)$ and $B=(b_1b_2)$, all possible elements of  $A\shuffle B$ are 
\bea
(a_1a_2b_1b_2),~~~(a_1b_1a_2b_2),~~~(a_1b_1b_2a_2),~~~(b_1a_1a_2b_2),~~~(b_1a_1b_2a_2),~~~(b_1b_2a_1a_2).
\eea
More complicated identities for BS currents, which are based on graphs, will be introduced later.

We will see that the one-loop GR, EYM, YM, and YMS integrands can be expanded in terms of $\mathcal{A}^{\text{BS}}\big(+,\pmb{\gamma},-\big|+,\pmb{\sigma},-\big)$ in (\ref{Eq:One-loopBasis-1}). The expression of BS integrand therefore plays a key role in understanding integrands of other theories.

\subsubsection{Expansion relations for half integrands}

The one-loop half integrands $I_{\,\text{L}}^{\,\text{1-loop}}$ and $I_{\,\text{R}}^{\,\text{1-loop}}$ in YMS, YM, GR, and EYM are displayed in \tabref{table:LoopIntegrands}. Particularly,  the reduced Pfaffian $\text{Pf}\,'[\Psi_{n+2}]$ (which includes information of both on-shell external particles and the particles $\pm$ carrying momenta $\pm l^{\mu}$) is formally introduced as 
\bea
\text{Pf}\,'[\Psi_{n+2}]\equiv{(-1)^{i+j}\over z_{ij}}\text{Pf}\left[\,\Psi_{ij}^{ij}\,\right].~~~~~\Label{Eq:ReducedPf}
\eea
The $\Psi_{n+2}$ in the above expression is a $2(n+2)\times 2(n+2)$ skew matrix 
\[\Psi=\begin{pmatrix}
A&-C^T\\
C&B\\
\end{pmatrix},~~~\Label{Eq:Psi}
\]
where $\Psi_{ij}^{ij}$ means the $i$-, $j$-th rows and columns are deleted. The blocks $A$, $B$ and $C$ denote $(n+2)\times (n+2)$ matrices containing the external kinematic information which includes the Lorentz contractions of polarizations $\epsilon_j^{\mu}$ and/or momenta $k_j^{\mu}$ (for $j=\pm$, $k_{j}^{\mu}=\pm l^{\mu}$)
\bea
A_{ab}=\Bigg\{
            \begin{array}{cc}
              \frac{k_a\cdot k_b}{z_{ab}} &, a\ne b \\
               0 &,a=b \\
            \end{array}\,,~~~~
B_{ab}=\Bigg\{
            \begin{array}{cc}
               \frac{\epsilon_a\cdot \epsilon_b}{z_{ab}} &, a\ne b \\
               0 &,a=b \\
            \end{array}\,,~~~~
C_{ab}=\Bigg\{
            \begin{array}{cc}
              \frac{\epsilon_a\cdot k_b}{z_{ab}} &, a\ne b \\
              -\Sl_{c\ne a}\frac{\epsilon_a\cdot k_c}{z_{ac}} &,a=b \\
            \end{array}\,.
\Label{Eq:ABC}
\eea
In the forward limit approach (\ref{Eq:LoopCHY}) to integrands of YM, EYM and GR, a summation of independent polarizations $\Sl_{+,-}\epsilon^{\mu}_{-}\epsilon^{\nu}_{+}$ in the reduced Pfaffian is also implied. This summation is further replaced according to the following rule (see e.g., \cite{Porkert:2022efy})
\bea
\Sl_{+,-}\epsilon^{\mu}_{-}\epsilon^{\nu}_{+}\to\Delta^{\mu\nu},~~\eta_{\mu\nu}\Delta^{\mu\nu}\to D-2,~~V_{\mu}W_{\nu}\Delta^{\mu\nu}\to V\cdot W, \Label{Eq:Polarizations}
\eea
where $V^{\mu}$ and $W^{\nu}$ are two arbitrary vectors. The $\text{Pf}\,[\,\Psi_{\mathsf{G}}]$ and $\text{Pf}\,[\,\Psi_{\mathsf{H}}]$ in the $I_L^{\text{1-loop}}$ for YMS and EYM amplitudes (i.e., in the third row of \tabref{table:LoopIntegrands}) stand for the Pfaffian of the matrices $\Psi_{\mathsf{G}}$ and $\Psi_{\mathsf{H}}$, which are obtained by keeping the rows and columns with respect to elements in $\mathsf{G}$ and $\mathsf{H}$, in $\Psi_{n+2}$. In the current work, we only consider those YMS and EYM integrands in which the loop propagators are only scalars and gluons, respectively. Thus, the corresponding $\pm$ are only scalars and gluons. In the YMS case, the $+$ and $-$ are involved in the PT factors in both left and right half integrands. In EYM and YM, $+$, $-$ are contained in the PT factor of $I_L$, and the reduced Pfaffian in the $I_R$. For GR, $+$ and $-$ live in the reduced Pfaffians in both half integrands.

There are two important properties of the half integrands in \tabref{table:LoopIntegrands}, which inherit from the tree-level amplitudes \cite{Hou:2018bwm,Du:2019vzf,Fu:2017uzt,Du:2017kpo}. {\it The first property} is that the reduced Pfaffian can be expanded as a combination of PT factors, each is multiplied by a lower-point Pfaffian
\bea
\text{Pf}\,'[\,\Psi_{n+2}\,]&=&(D-2)\,\text{PT}(+,-)\,\text{Pf}\,\left[\,\Psi_{\{1,...,n\}}\,\right]\Label{Eq:ExpandReducedPfaffian2}\\
%
%
&&+\Sl_{(i_1i_2)\in\text{S}_{2}\setminus\text{Z}_{2}}\text{Tr}[F_{i_1}\cdot F_{i_2}]\Big[\,\text{PT}(+,i_1,i_2,-)+\text{cyc}(i_1i_2)\Big]\,\text{Pf}\,\left[\,\Psi_{\{1,...,n\}\setminus \{i_1,i_2\}}\,\right]\nn
&&-\cdots\nn
&&+(-1)^n\Sl_{(i_1i_2...i_n)\in \text{S}_{n}\setminus\text{Z}_{n}}\text{Tr}[ F_{i_1}\cdot F_{i_2}\cdot...\cdot F_{i_n}]\,\Big[\,\text{PT}(+,i_1,i_2,...,i_n,-)+\text{cyc}(i_1i_2...i_n)\,\Big],\nonumber
\eea
where (\ref{Eq:Polarizations}) is applied. In the above expression, the $\pm$  serve as the two ends of the PT factors, while each $\text{Tr}[F_{i_1}\cdot F_{i_2}\cdot...\cdot F_{i_l}]$ is defined by 
\bea
\text{Tr}[F_{i_1}\cdot F_{i_2}\cdot...\cdot F_{i_l}]\equiv {{(F_{i_1})}^{\mu_1}_{~\,\mu_2} {(F_{i_2})}^{\mu_2}_{~\,\mu_3}... {(F_{i_l})}^{\mu_{l}}_{~\,\mu_1}},
\eea
which has cyclic symmetry with respect to $i_1,i_2,...,i_l$, and $\text{Tr}[F_{i}]=0$ due to the antisymmetry of the  field strength tensor $F_{i}^{\mu\nu}=k_i^{\mu}\epsilon_i^{\nu}-\epsilon_i^{\mu}k_i^{\nu}$. 
{\it The second property} is that each PT factor $\text{PT}(+,\pmb{\sigma},-)$ multiplied by a Pfaffian can be expanded in terms of pure PT factors according to the {\it graphic rule }
\bea
\text{PT}(+,\pmb{\sigma},-)\,\text{Pf}\,\left[\,\Psi_{\mathsf{G}}\,\right]=\Sl_{\mathcal{F}}\,\mathcal{C}[\,\mathcal{F}]\,\Bigg[\,\Sl_{\pmb{\rho}[\,\mathcal{F}]}\text{PT}\left(+,\pmb{\rho}[\,\mathcal{F}],-\right)\,\Bigg], \Label{Eq:Expansion1}
\eea
 which holds only on
the support of the scattering equations. In the above expression, all possible graphs $\mathcal{F}$ constructed by graphic rule are summed over. Each graph $\mathcal{F}$ is associated with a kinematic coefficient $\mathcal{C}[\mathcal{F}]$ (as a product of Lorentz contractions of polarizations and/or momenta) and a set of permutations $\left(+\pmb{\rho}[\,\mathcal{F}]-\right)$.  We will review the details about the graphic rule in the coming subsection.

\subsection{Graphic rule and graph-based relations }
Now we present the graphic rule \cite{Hou:2018bwm,Du:2019vzf,Du:2017kpo} for (\ref{Eq:Expansion1}) and introduce two graph-based relations \cite{Wu:2021exa} for BS currents.
\subsubsection{Graphic rule}\label{sec:GraphicRule}
%
\begin{figure}
\centering
\includegraphics[width=0.55\textwidth]{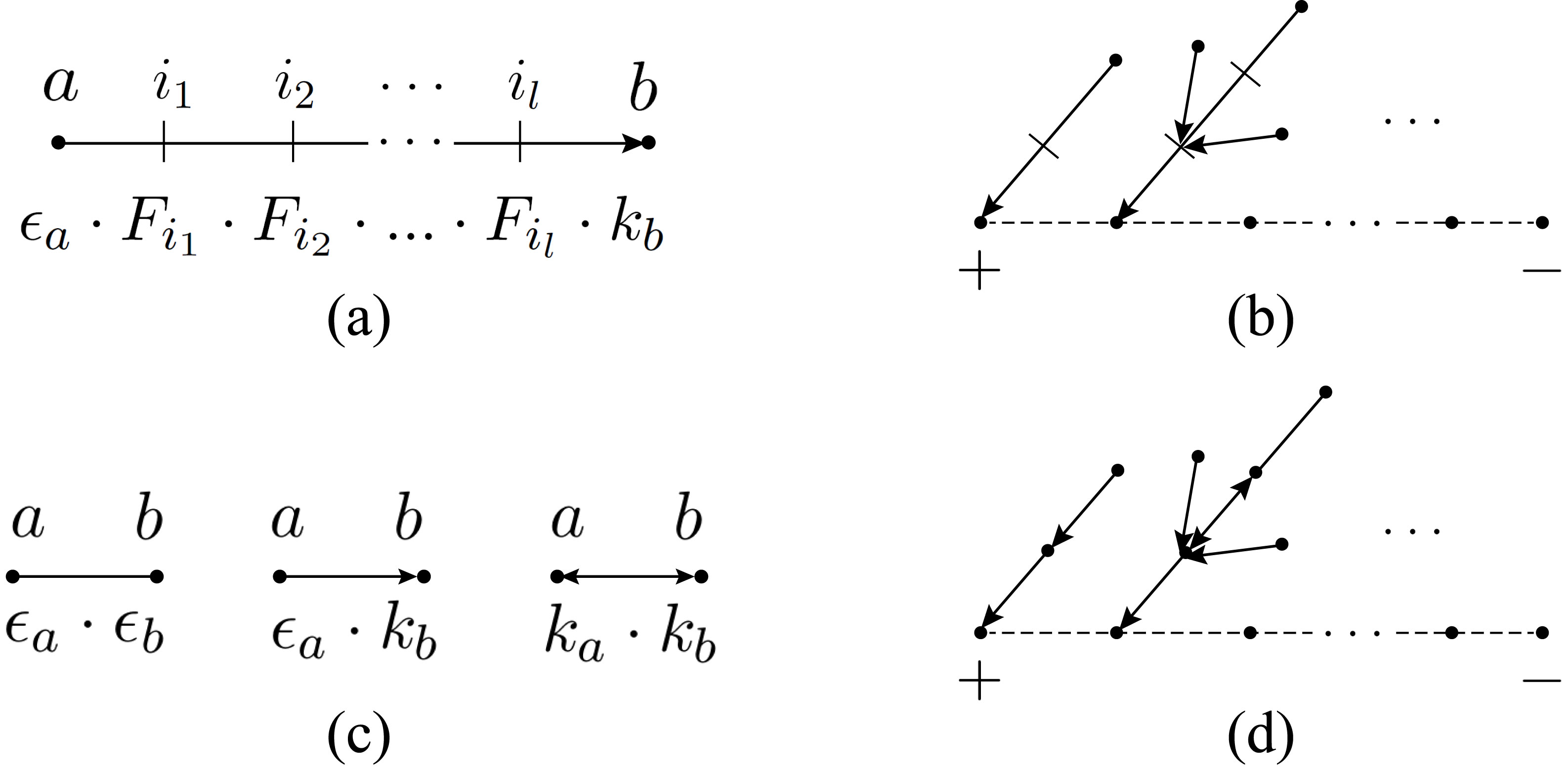}
\caption{In the graphic rule, (a) is a typical chain that constructs a connected graph (b). Correspondingly, when the  field strength tensors are expanded explicitly, the lines in (c) form refined graphs such as (d) in the refined graphic rule.}
\label{Fig:graphicrule1}
\end{figure}

All possible graphs summed over in (\ref{Eq:Expansion1}) for $\text{PT}(+,\sigma_1,\sigma_2,...\sigma_r,-)\,\text{Pf}\,\left[\,\Psi_{\mathsf{G}}\,\right]$ with $s$ gluons in $\mathsf{G}$ can be obtained by the following graphic rule:
\begin{itemize}
\item[]{\bf Step-1} Connect all scalars and $+$, $-$ particles in order $(+\sigma_1\sigma_2...\sigma_r-)$ via dashed lines. 
\item[]{\bf Step-2} Define the reference set $\mathsf{R}$ and the root set $\mathcal{R}$,  consisting respectively of all gluons and of all scalars (excluding the $-$ particle), by
\bea
\mathsf{R}\equiv(\gamma_1\gamma_2~...~\gamma_s),~~~~~\mathcal{R}\equiv\{+,\sigma_1,\sigma_2,...,\sigma_r\}.
\eea
Here, the position of each element in $\mathsf{R}$ is called its {\it weight}, and the rightmost element in $\mathsf{R}$ is the highest-weight gluon.  Thus, the reference set $\mathsf{R}$ specifies the {\it reference order} of the gluons as $\gamma_1\prec\gamma_2\prec...\prec\gamma_s$. 
\item[]{\bf Step-3} Construct a straight chain, as illustrated in \figref{Fig:graphicrule1} (a), that starts from the highest-weight particle $a$ in $\mathsf{R}$ and ends at a certain particle $b$ in $\mathcal{R}$. The associated kinematic factor is 
\bea
\epsilon_a\cdot F_{i_1} \cdot  F_{i_2} \cdot...\cdot  F_{i_l} \cdot k_b,
\eea
where $i_1,i_2,...,i_l$ are arbitrarily chosen gluons from the set $\mathsf{G}\setminus a$, and $F_{i}^{\mu\nu}=k_i^{\mu}\epsilon_i^{\nu}-\epsilon_i^{\mu}k_i^{\nu}$ .
\item[]{\bf Step-4} Redefine the reference set and the root set as
\bea
\mathsf{R}\equiv(\gamma_1\gamma_2~...~\gamma_s)\setminus\{a,i_1,i_2,...,i_l\},~~~~~\mathcal{R}\equiv\{+,\sigma_1,\sigma_2,...,\sigma_r\}\cup\{a,i_1,i_2,...,i_l\}.
\eea
Repeat the previous step until the reference set $\mathsf{R}$ is empty. We then get a fully connected graph $\mathcal{F}$, shown in \figref{Fig:graphicrule1} (b).
\item[]{\bf Step-5} Consider all possible ways of connecting the particles to obtain the complete set of graphs.
\end{itemize}
For a given graph $\mathcal{F}$, the corresponding coefficient $\mathcal{C}[\mathcal{F}]$ and permutations $(+\pmb{\rho}[\mathcal{F}]-)$ are given by:
\begin{itemize}
\item The coefficient $\mathcal{C}[\mathcal{F}]$ is the product of the kinematic factors for all chains.
\item The permutations $(+\pmb{\rho}[\mathcal{F}]-)$ are defined in the following way: (i) The first and last elements are fixed as $+$ and $-$, respectively.
(ii) For any two particles $x$ and $y$ living on  the same path towards $+$, $x$ appears to the left of $y$ in the $\pmb{\rho}[\mathcal{F}]$ whenever $x$ is closer to $+$ than $y$ along the path.
(iii) If there are two or more branches attaching to the same particle, the orders corresponding to these branches must be shuffled together.
\end{itemize}
For example, a possible term for $\text{PT}(+,x_1,x_2,-)\,\text{Pf}\,\left[\,\Psi_{\{p,q,r\}}\,\right]$ with the reference order $p\prec q\prec r$ is given as follows
\bea
\begin{minipage}{3.6cm}\includegraphics[width=3.6cm]{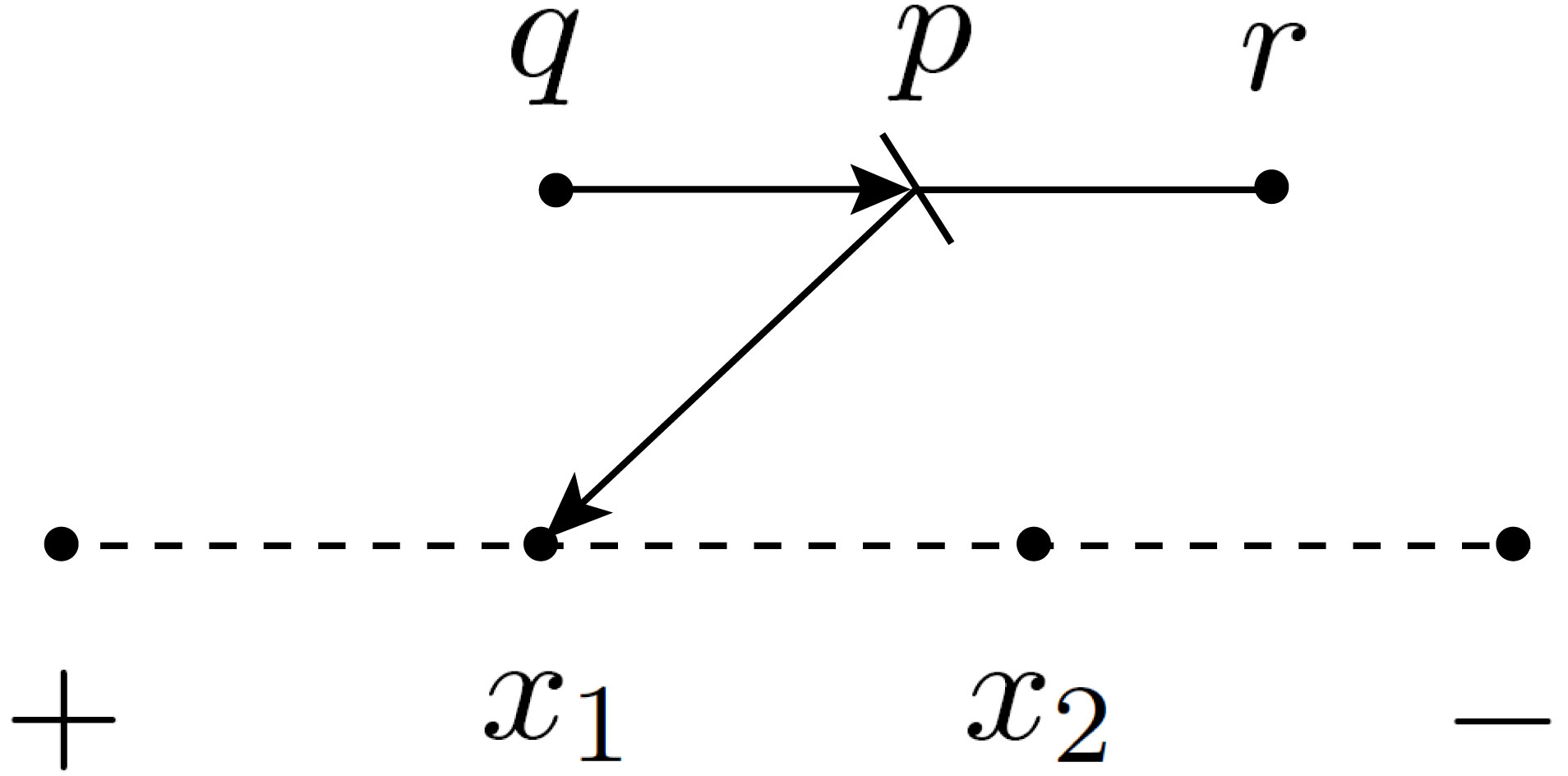}\end{minipage}~\to~ 
(\epsilon_r\cdot F_p \cdot k_{x_1})(\epsilon_q\cdot k_p)\text{PT}(+,x_1,\{x_2\}\shuffle\{p,\{q\}\shuffle\{r\}\},-).
\Label{Eq:graphiceg}
\eea

There exist equivalent rules to generate all possible graphs summed over in (\ref{Eq:Expansion1}). For instance, expanding every $F_{i}^{\mu\nu}$ into $k_i^{\mu}\epsilon_i^{\nu}-\epsilon_i^{\mu}k_i^{\nu}$ and introducing the three types of lines in \figref{Fig:graphicrule1} (c) to represent Lorentz contractions of momenta and/or polarization vectors, we recover the {\it refined graphic rule} in \cite{Hou:2018bwm,Du:2019vzf}. A typical refined graph is shown by \figref{Fig:graphicrule1} (d). By further expressing a connected graph as a composition of subgraphs, we arrive at the {\it refined graphic rule of subgraphs} given in \cite{Xie:2024pro}. Moreover, by adding and subtracting identical graphs (termed spurious graphs), we can adapt the refined graphic rule for computational convenience, which is the {\it second version} of the refined graphic rule of subgraphs presented in \cite{Xie:2024pro}. In the refined graphic rule, the permutations for a connected graph is generated exactly as in the original graphic rule, while the coefficient is the product of the factors for all lines, with an extra sign $(-1)^{N}$ where $N$ denotes the number of arrows pointing away from the node $+$. A concrete example is that the term in (\ref{Eq:graphiceg}) under the graphic rule can be expressed as the sum of the following two terms under the refined graphic rule
\bea
\begin{minipage}{3.6cm}\includegraphics[width=3.6cm]{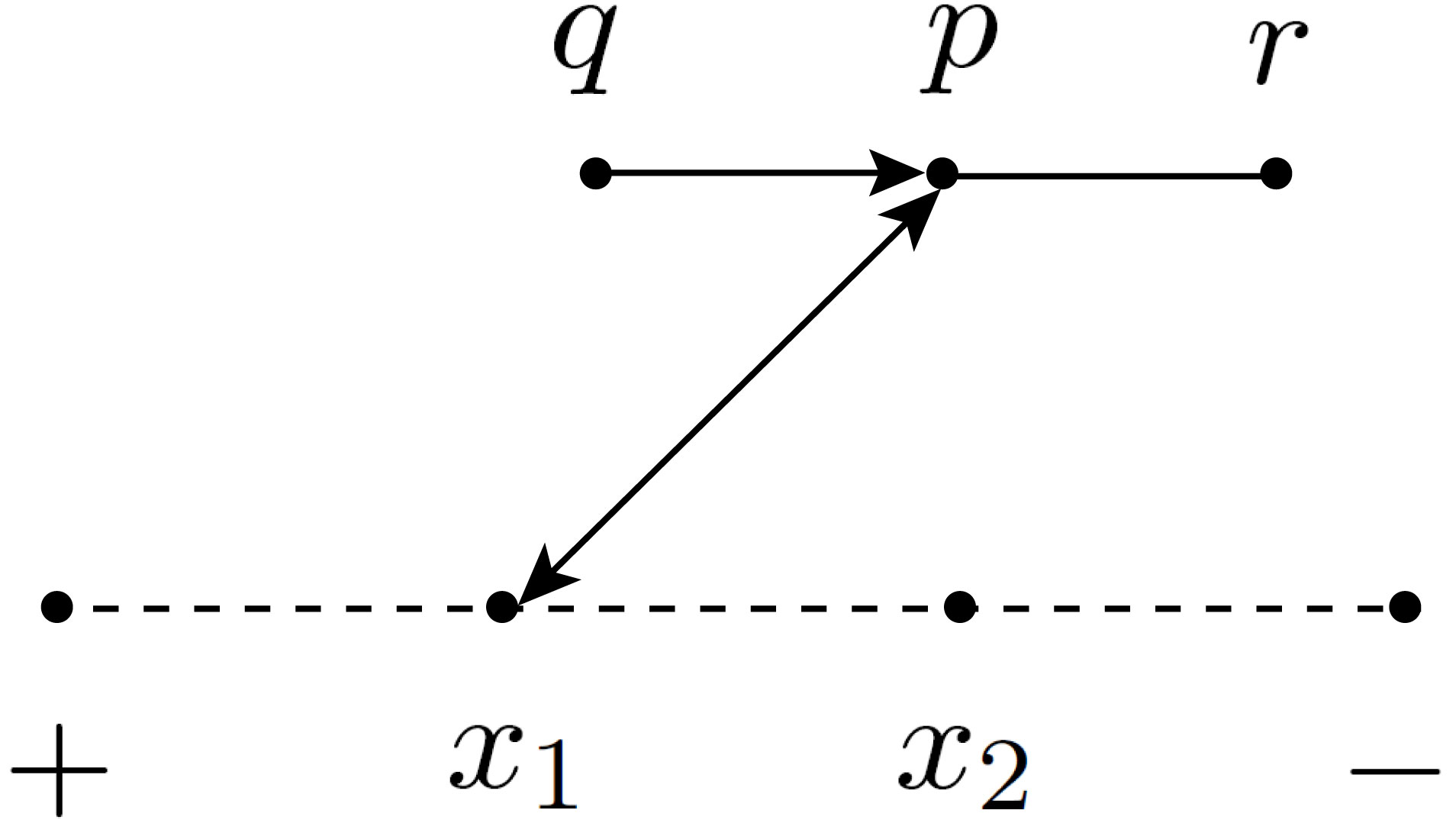}\end{minipage}&\to& (-k_p\cdot k_{x_1})(\epsilon_q\cdot k_p)(\epsilon_p\cdot\epsilon_r)\text{PT}(+,x_1,\{x_2\}\shuffle\{p,\{q\}\shuffle\{r\}\},-),\nn
\begin{minipage}{3.6cm}\includegraphics[width=3.6cm]{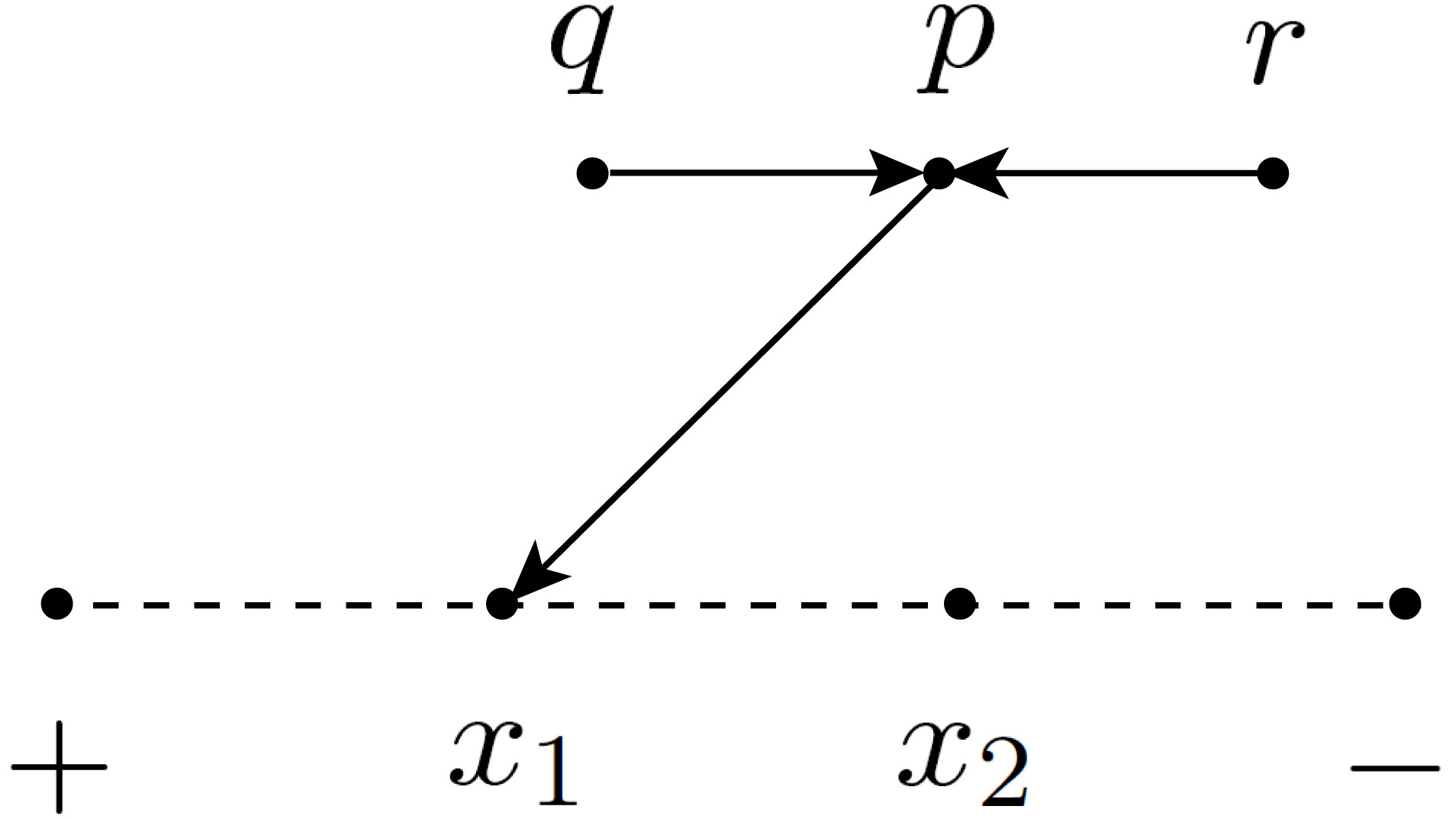}\end{minipage}&\to& (\epsilon_p\cdot k_{x_1})(\epsilon_q\cdot k_p)(\epsilon_r\cdot k_p)\text{PT}(+,x_1,\{x_2\}\shuffle\{p,\{q\}\shuffle\{r\}\},-).
\eea

In this paper, all graphs are presented within the refined graphic rules, which means Lorentz contractions are explicitly given by the three types of lines shown in \figref{Fig:graphicrule1} (c). We omit the detailed construction rules for the refined graphs, since all refined graphs (except for the spurious graph that is added and then subtracted) can be obtained by expanding the  field strength tensors $F_{i}$ under the original graphic rule.

\subsubsection{Graph-based relations satisfied by the BS currents}

Having introduced the graphic rule, we know that permutations can be established by graphs. Based on this, there are two identities \cite{Wu:2021exa} satisfied by the BS currents (\ref{Eq:BScurrent}).

\begin{figure}
\centering
\includegraphics[width=0.55\textwidth]{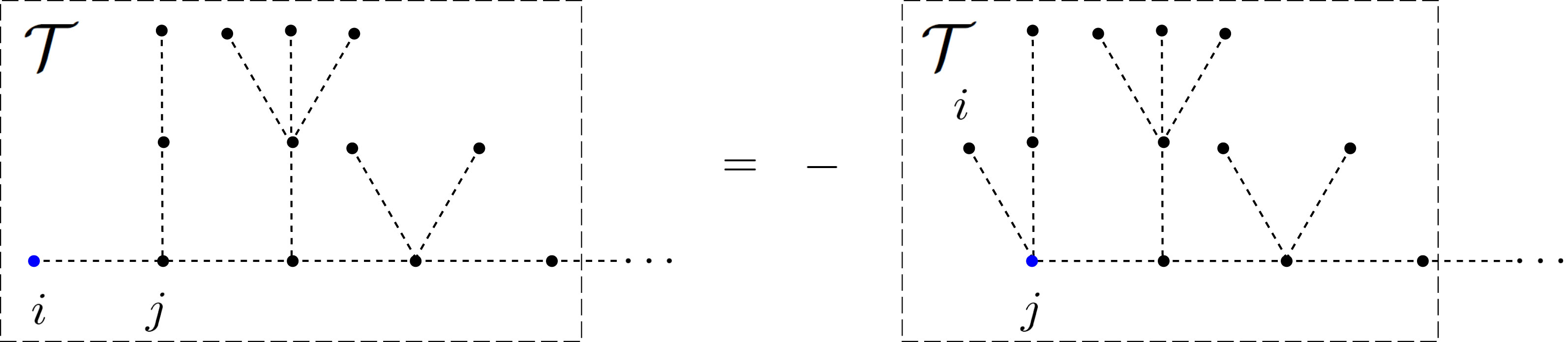}
\caption{The first graph-based relation }
\label{Fig:GraphBasedProperty1}
\end{figure}
\begin{figure}
\centering
\includegraphics[width=0.95\textwidth]{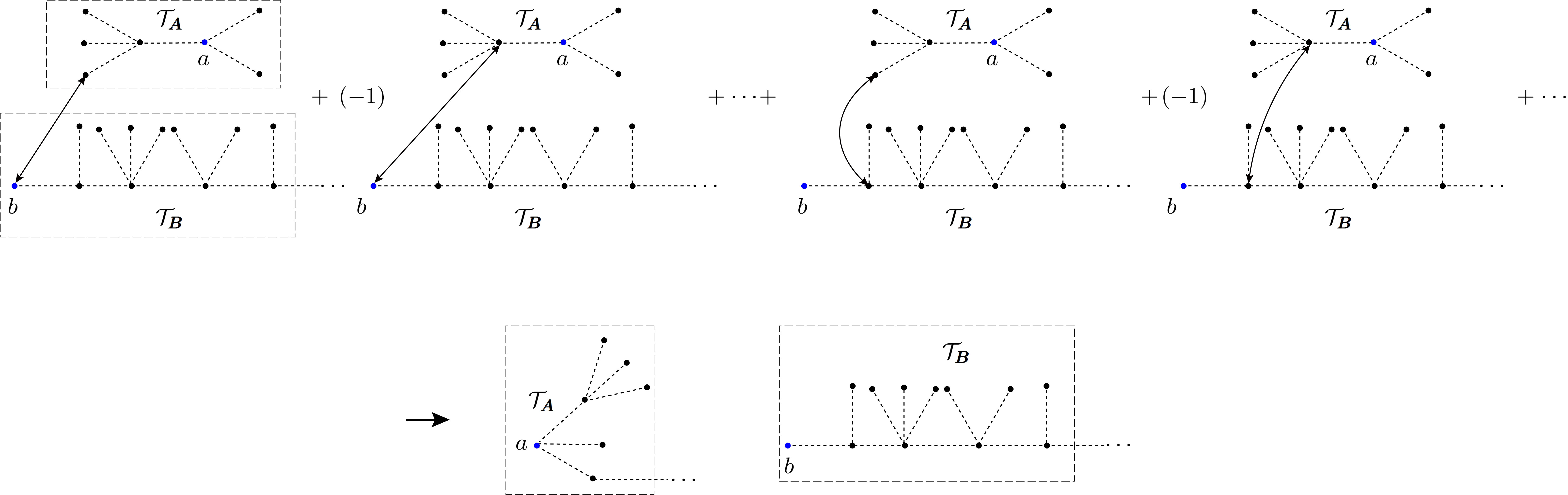}
\caption{The second graph-based relation }
\label{Fig:GraphBasedProperty2}
\end{figure}

The first identity, as shown by \figref{Fig:GraphBasedProperty1}, is 
\bea
\Sl_{\pmb{\sigma}\in\pmb{\rho}\left. [\mathcal{T}]\right|_{i}}\phi_{\pmb{\sigma}\,|\,\W A}=-\Sl_{\pmb{\gamma}\in\pmb{\rho}\left.[\mathcal{T}]\right|_{j}}\phi_{\pmb{\gamma}\,|\,\W A},~\Label{Eq:GraphBased1}
\eea
where $\pmb{\rho}[{\mathcal{T}}]|_{i}$ denotes permutations established by the connected tree graph $\mathcal{T}$ {(without considering the line styles)}, with the leftmost element  chosen to be $i$. The node $j$ on the rhs. is supposed to be a node adjacent to $i$. For example, in (\ref{Eq:graphiceg}), if $i$ is taken as $p$, the node $j$ adjacent to $p$ can be $q$, $r$ or $x_1$. 

The second identity, illustrated in \figref{Fig:GraphBasedProperty2}, is called  {\it off-shell BCJ relation} \cite{Du:2011js,Frost:2020eoa,Wu:2021exa,Du:2022vsw}
\bea
\Sl_{x\in \mathcal{T}_A, y\in \mathcal{T}_{{B}}}(-1)^{|ax|}\Sl_{\pmb{\gamma}\in\pmb{\rho}\left.[\left(\mathcal{T}_{{A}}\oplus\mathcal{T}_{{B}}\right)]\right|_b}s_{xy}\phi_{\pmb{\gamma}|\pmb{\sigma}}&=&\Sl_{\pmb{\alpha}\in \pmb{\rho}\left.[\mathcal{T}_{A}]\right|_a}\Sl_{\pmb{\beta}\in \pmb{\rho}\left.[\mathcal{T}_{B}]\right|_b} \Big[\phi_{\pmb{\beta}|\pmb{\sigma}_{1,i}}\phi_{\pmb{\alpha}|\pmb{\sigma}_{i+1,l}} -\phi_{\pmb{\alpha}|\pmb{\sigma}_{1,{l-i}}}\phi_{\pmb{\beta}|\pmb{\sigma}_{l-i+1,l}}\Big], \Label{Eq:OffBCJ1}
\eea
in which, we have two connected tree graphs $\mathcal{T}_{{A}}$ and $\mathcal{T}_{{B}}$. When connecting two nodes $x\in \mathcal{T}_A$ and $y\in \mathcal{T}_{{B}}$ by a $k_x\cdot k_y$ line, we get a graph $\mathcal{T}_{{A}}\oplus\mathcal{T}_{{B}}$. The first summation on the LHS implies that all graphs $\mathcal{T}_{{A}}\oplus\mathcal{T}_{{B}}$ corresponding to all choices of $(x,y)$ pairs are summed over. We further chose two nodes  $a\in \mathcal{T}_{{A}}$ and $b\in \mathcal{T}_{{B}}$, arbitrarily. The sign $(-1)^{|ax|}$ on the LHS depends on the distance $|ax|$, i.e., the number of lines between the nodes $a$ and $x$. The second summation is taken over permutations established by the graph $\mathcal{T}_{{A}}\oplus\mathcal{T}_{{B}}$, while $b$ is regarded as the leftmost element. On the rhs., $\pmb{\alpha}\in \pmb{\rho}\left.[\mathcal{T}_{A}]\right|_a$ and $\pmb{\beta}\in \pmb{\rho}\left[\mathcal{T}_{B}\right]|_b$ are the permutations established by $\mathcal{T}_{A}$ and $\mathcal{T}_{B}$ when $a$ and $b$ are considered as the leftmost elements, respectively. 
Permutation $\pmb{\sigma}=(\sigma_1\dots\sigma_l)$ in $\phi_{\,\pmb{\gamma}\,|\,\pmb{\sigma}\,}$ is divided into two parts $\pmb{\sigma}_{1,i}=({\sigma}_{1}\dots{\sigma}_{i})$ and $\pmb{\sigma}_{i+1,l}=({\sigma}_{i+1}\dots{\sigma}_{l})$, where the number of nodes in $\mathcal{T}_{B}$ is assumed to be $i$ and the number of nodes in $\mathcal{T}_{{A}}\oplus\mathcal{T}_{{B}}$ is $l$. In the second term on the rhs, $\pmb{\sigma}$ is divided into $\pmb{\sigma}_{1,l-i}=({\sigma}_{1}\dots{\sigma}_{l-i})$ and $\pmb{\sigma}_{l-i+1,l}=({\sigma}_{1-i+1}\dots{\sigma}_{l})$. {Since the two tree graphs $\mathcal{T}_{{A}}$ and $\mathcal{T}_{{B}}$ only set up permutations, the relation (\ref{Eq:OffBCJ1}) is independent of the line styles in $\mathcal{T}_{{A}}$ and $\mathcal{T}_{{B}}$.}

\subsection{The local expressions of YMS integrand}\label{sec:2.3YMSreview}

In this subsection, we review the local expressions of one-loop YMS partial integrands and full integrands, which were derived in \cite{Xie:2024pro}. 

According to (\ref{Eq:LoopCHY}), (\ref{Eq:PartialIntegrand}) and the half integrands displayed in \tabref{table:LoopIntegrands}, the one-loop YMS integrand $\mathcal{I}^{\text{YMS}}_{\text{1-loop}}(\pmb\gamma||\mathsf{G}|\pmb{\rho})$ can be expressed via partial integrands as follows
\bea
\mathcal{I}^{\text{YMS}}_{\text{1-loop}}(\pmb\gamma||\mathsf{G}|\pmb{\rho})=\left[\mathcal{A}^{\text{YMS}}(+,\pmb\gamma,-||\mathsf{G}|+,\pmb{\rho},-)+\text{cyc}(\pmb\gamma)\right]+\text{cyc}(\pmb{\rho}).\Label{Eq:GenResultNew0}
\eea
As  pointed out in \cite{Xie:2024pro}, the partial integrand (\ref{Eq:PartialIntegrand}) of YMS can be reduced into 
\bea
\mathcal{A}^{\text{YMS}}(+,\pmb\gamma,-||\mathsf{G}|+,\pmb{\rho},-)&=&{1\over l^2}\lim\limits_{k^{\mu}_{\pm}\to \pm l^{\mu}}{A}_{\text{tree}}^{\text{YMS}}(+,\pmb\gamma,-||\mathsf{G}|+,\pmb{\rho},-)\nn
&=&\Sl_{\small\substack{\{A_1,A_2,...,A_I\}\\\small (\W A_1\W A_2...\W A_I)={\pmb{\rho}}\\ \small A_j=\W A_j}}{1\over l^2}\,J[A_1]\, {1\over s_{A_1,l}}\,J[A_2]\cdots\,{1\over s_{A_1...A_{I-1},l}}\,J[A_I].\Label{Eq:YMSPartial0}
\eea
In the second line of the above expression, all possible {\it partitions} $\{A_1,A_2,...,A_I\}$ of external particles and possible divisions $(\W A_1\W A_2...\W A_I)$ of permutation $\pmb{\rho}$ have been summed over, where $A_j=\W A_j$ requires  $A_j$ and $\W A_j$ contain the same elements (in other words this constraint is an equality in the sense of sets). Partition $\{A_1,A_2,...,A_I\}$ assigns each external particle to one of the $I$ distinct sets, such that the relative ordering of elements in $\pmb\gamma$ is preserved. For instance, if $\pmb\gamma=(x_1x_2)$ and $\mathsf{G}=\{p,q\}$, the possible partitions $\{A_1,A_2,...,A_I\}$ are
\bea
I=1:&\!\!&\{\{x_1,x_2,p,q\}\};\nn
I=2:&\!\!&\{\{x_1,x_2\}, \{p,q\}\};~~~~~~~~~\{\{p,q\},\{x_1,x_2\}\};~~~~~~~~~\{\{x_1,p\}, \{x_2,q\}\};~~~~~~~~~\{\{x_1,q\}, \{x_2,p\}\};\nn
&\!\!&\{\{x_1\}, \{x_2,p,q\}\};~~~~~~~~~\{\{x_1,p,q\}, \{x_2\}\};~~~~~~~~~\{\{x_1,x_2,p\}, \{q\}\};~~~~~~~~~\{\{q\},\{x_1,x_2,p\}\};\nn
&\!\!&\{\{x_1,x_2,q\}, \{p\}\};~~~~~~~~~\{\{p\},\{x_1,x_2,q\}\};\nn
I=3:&\!\!&\{\{x_1,x_2\}, \{p\}, \{q\}\};~~~~~~\{\{p\}, \{x_1,x_2\}, \{q\}\};~~~~~~\{\{p\}, \{q\},\{x_1,x_2\}\}; ~~~~~~\{\{p,q\}, \{x_1\}, \{x_2\}\};\nn
&\!\!&\{\{q\}, \{x_1,p\}, \{x_2\}\};~~~~~~\{\{x_1,p\}, \{q\}, \{x_2\}\};~~~~~~\{\{x_1,p\},\{x_2\},\{q\}\};~~~~~~\{\{x_1\}, \{p,q\}, \{x_2\}\};\nn
&\!\!&\{\{q\}, \{x_1\}, \{x_2,p\}\};~~~~~~\{\{x_1\}, \{q\}, \{x_2,p\}\};~~~~~~\{\{x_1\},\{x_2,p\},\{q\}\};~~~~~~\{\{x_1\},\{x_2\},\{p,q\}\};\nn
&\!\!&(p\leftrightarrow q~\text{for the first 9 terms});\nn
I=4:&\!\!&\{\{x_1\}, \{x_2\}, \{p\}, \{q\}\};~~~\{\{x_1\}, \{p\}, \{x_2\}, \{q\}\};~~~\{\{p\},\{x_1\}, \{x_2\},  \{q\}\};~~~\{\{p\}, \{q\}, \{x_1\}, \{x_2\}\}; \nn
&\!\!&\{\{x_1\}, \{p\}, \{q\}, \{x_2\}\};~~~\{\{p\}, \{x_1\},  \{q\}, \{x_2\}\};~~~(p\leftrightarrow q).
\eea
Supposing the permutation $\pmb{\rho}$  in (\ref{Eq:YMSPartial0}) is $(x_1x_2pq)$, for the specific division $(\W A_1\W A_2)=(x_1x_2\,|\,pq)$ where $I=2$, only the first partition for $I=2$, i.e.,  $\{\{x_1,x_2\}, \{p,q\}\}$ satisfies the constraint $A_j=\W A_j$ ($j=1,2$) since $A_j$ has the same elements as $\W A_j$. Thus the partition $\{\{x_1,x_2\}, \{p,q\}\}$ and the division $(x_1x_2pq)$ provide a nonvanishing term in the summation of (\ref{Eq:YMSPartial0}).

Each $J[A_i]$ in (\ref{Eq:YMSPartial0}) is defined by
\bea
J[A_i]&=&T (A_i),~~~~~~~~~~~~~~~~~~~~~~~~~~~~~~~~~~~~~~~~~~~~~\,~~~~~~~~~~~~\text{(if $A_i$ contains  scalars)}\Label{Eq:GenResultNewJ1}\\
J[A_i]&=&T (A_i)\cdot X_{A_i}+\left(-{1\over 2}\right)\,\Sl_{\small\substack{\{A_{i1},A_{i2}\}\\\small (\W A_{i1}\W A_{i2})=\W A_{i}\\ \small A_{ij}=\W A_{ij}}}T (A_{i1})\cdot T (A_{i2}),~~~~\text{(if $A_i$ contains only gluons)}\Label{Eq:GenResultNewJ2}
\eea
where $X^{\mu}_{A_i}$ denotes the momentum $(l+k_{A_1}+...+k_{A_{i-1}})^{\mu}$ of the linear propagator ${1\over s_{A_1...A_{i-1},l}}$ attaching to the  subcurrent $T^{\mu}(A_i)$ from left.  In the second term of (\ref{Eq:GenResultNewJ2}), the weights of $A_{i1}$ and $A_{i2}$ (i.e., the weights of the highest weight gluons in the sets $A_{i1}$ and $A_{i2}$) should satisfy $A_{i1}\prec A_{i2}$.
 The {\it {YMS} effective current} $T (A_i)$ in (\ref{Eq:GenResultNewJ1}) and {\it YM effective current} $T^{\mu} (A_i)$ in (\ref{Eq:GenResultNewJ2}) are respectively given by
\bea
T (A_i)=\Sl_{\mathcal{F}}\mathcal{C}[\mathcal{F}\,]\Sl_{\pmb{\rho}[\mathcal{F}\,]}\phi_{\pmb{\rho}[\mathcal{F}\,]\left|\W A_i\right.},~~~~~~~T^{\mu} (A_i)=\Sl_{\mathcal{F}}\mathcal{C}^{\mu}[\mathcal{F}\,]\Sl_{\pmb{\rho}[\mathcal{F}\,]}\phi_{\pmb{\rho}[\mathcal{F}\,]\left|\W A_i\right.}.\Label{Eq:EffectiveCurrent}
\eea
In the first equality, $\mathcal{F}$ are the graphs constructed by the graphic rule, where all scalars (connected via dashed lines following their order in permutation $\pmb\gamma$) are considered as roots. In the second equality, $\mathcal{F}$ are graphs with the off-shell leg considered as root, where the off-shell leg only gives the Lorentz index and does not appear in the resultant permutation $\pmb{\rho}[\mathcal{F}\,]$. The $ \mathcal{C}[\mathcal{F}\,]$ or  $\mathcal{C}^{\mu}[\mathcal{F}\,]$ refers to 
the kinematic coefficients for graph $\mathcal{F}$, depending on whether $\mathcal{F}$ contains scalars or not. For example, if $A_i$ contains scalar $x_1$ and gluon $g_1$, the corresponding coefficient $\mathcal{C}[\mathcal{F}\,]$ and permutation $\pmb{\rho}[\mathcal{F}\,]$ are $\epsilon_{g_1}\cdot k_{x_1}$ and $(x_1g_1)$. If $A_i$ contains solely one gluon $g_1$, the corresponding coefficient $\mathcal{C}^{\mu}[\mathcal{F}\,]$ and permutation $\pmb{\rho}[\mathcal{F}\,]$ are $\epsilon_{g_1}^{\mu}$ and $(g_1)$.  From the definition of the local structure $J[A_i]$ and effective currents $T (A_i),T^{\mu} (A_i)$, it is not hard to see (\ref{Eq:YMSPartial0}) is just the sum of LPFDs, in which the effective currents are attached to the linear propagator line between $+$ and $-$, through the following local vertices:
\bea
\begin{minipage}{1.5cm}  \includegraphics[width=1.5cm]{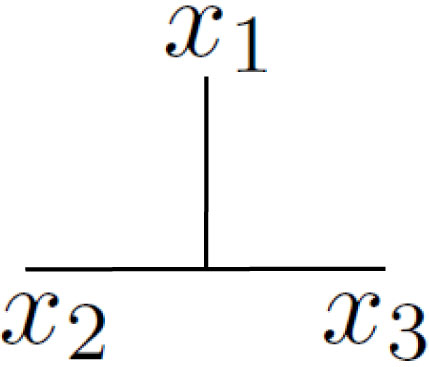} \end{minipage}=1,~~~~~~~~~~\begin{minipage}{1.5cm}  \includegraphics[width=1.5cm]{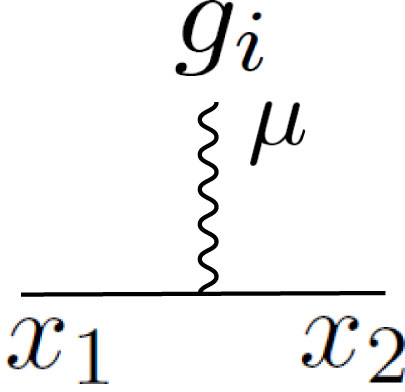} \end{minipage}= k_{x_1}^{\mu},~~~~~~~~~~
\begin{minipage}{1.5cm}  \includegraphics[width=1.5cm]{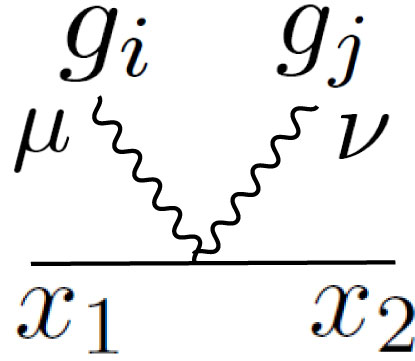} \end{minipage}= \left(-{1\over 2}\right)\eta^{\mu\nu}~~~(g_i\prec g_j).\Label{Eq:reviewvertex}
\eea
Hence, the expressions (\ref{Eq:GenResultNewJ1}) and (\ref{Eq:GenResultNewJ2}) are naturally {\it local along the linear loop propagator line}. In other words, there do not exist contractions between different effective currents separated by loop propagators.

Based on the partial fraction identity, we have the following useful identity
\bea
&&{N(l)\over l^2s_{A_1,l}s_{A_{12},l}\dots s_{A_{12\dots m-1},l}}+{N(l+k_{A_m})\over l^2s_{A_m,l}s_{A_{m1},l}\dots s_{A_{m1\dots m-2},l}}+\dots+{N(l+k_{A_{23\dots m}})\over l^2s_{A_2,l}s_{A_{23},l}\dots s_{A_{23\dots m},l}}\nn
&\cong& {N(l)\over l^2l_{A_1}^2l_{A_{12}}^2\dots l_{A_{12\dots m-1}}^2},
\Label{Eq:partial}
\eea
where $l_{A_{12\dots i}}^{\mu}\equiv l^{\mu}+k_{A_{12\dots i}}^{\mu}=l^{\mu}+k^{\mu}_{A_{1}}+k^{\mu}_{A_{2}}+...+k^{\mu}_{A_{i}}$, $N(l)$ is an $l$-dependent numerator, and $\cong$ means that the left- and right-hand sides of the above expression are equal after integration over the loop momentum. With the help of this identity, the local expression of the partial integrand (\ref{Eq:YMSPartial0}), after cyclic sums  corresponding to $\text{cyc}(\pmb{\gamma})$ and $\text{cyc}(\pmb{\rho})$ in (\ref{Eq:GenResultNew0}), gives the full integrand
\bea
\mathcal{I}^{\text{YMS}}_{\text{1-loop}}(\pmb\gamma||\mathsf{G}|\pmb{\rho})\cong\,\Sl_{\substack{\{A_1,A_2,...,A_I\}\\ (\W A_1\,\W A_2\,...\,\W A_I)\\ A_i=\W A_i,\gamma_1\in A_1}}{1\over l^2}\,J[A_1]\, {1\over l^2_{A_1}}\,J[A_2]\cdots\,{1\over l^2_{A_1...A_{I-1}}}\,J[A_I],\Label{Eq:GenResultNew}
\eea
where we sum over all possible partitions and divisions with respect to each permutation in $\text{cyc}(\pmb\gamma)$ and $\text{cyc}(\pmb\rho)$ (the partitions and divisions in (\ref{Eq:YMSPartial0}) are considered for fixed permutations $\pmb\gamma$ and $\pmb\rho$), and further require that the leftmost element $\gamma_1$ in $\pmb\gamma$ belongs to $A_1$ to avoid redundancy. In (\ref{Eq:GenResultNew}), {\it the one-loop integrand has a local expression along the propagator line which already consists of quadratic propagators.}

\section{Double YMS integrands, nonlocality and linear propagators in EYM and GR}\label{sec:MainIdea}

Applying the expansion relations (\ref{Eq:ExpandReducedPfaffian2}), (\ref{Eq:Expansion1}) to the left and right half integrands in \tabref{table:LoopIntegrands}, the one-loop EYM and GR integrands from the forward limit approach (\ref{Eq:LoopCHY}) are directly expressed in terms of BS partial integrands (\ref{Eq:BSPartialIntegrand}) multiplied by kinematic coefficients from each half. According to  (\ref{Eq:One-loopBasis-1}), each BS partial integrand can further be written as the sum of LPFDs where tree-level BS subcurrents are attached to the loop propagator line from $+$ to $-$.  Such expressions of EYM and GR integrands possess two significant features:
\begin{itemize} 
 \item [(i)] The denominator of a loop propagator is a linear function of loop momentum.
\item [(ii)] In both {\it left} and {\it right} coefficients (the kinematic coefficients obtained from the left and right half integrands), there exist Lorentz contractions between particles that are separated by loop propagators in LPFD. Such contractions are not achieved by local vertices, and thus the integrand has {\it nonlocalities} along the loop propagator line.
 \end{itemize} 
 Contrarily, in a conventional Feynman diagram, the loop propagators are quadratic ones, while there does not exist nonlocality since Lorentz contractions are realized via local vertices. 
 
In this section, we provide a general discussion for treating nonlocalities and extracting quadratic propagators in the forward limit form (\ref{Eq:LoopCHY}) of EYM and GR integrands. The expansion of EYM and GR integrands to BS partial integrands can be viewed as a three-step process
\bea
\text{EYM, GR integrands}&\pmb{\xrightarrow{~~~}}&\text{BS partial integrands}\nn
\blue \pmb{\downarrow}~~~~~~~~~~&&~~~~~~~~~~\blue \pmb{\uparrow}\nn
\text{dYMS integrands}&\blue \pmb{\xrightarrow{~~~}}&\text{dYMS partial integrands}
\eea
 in which the EYM and GR integrands are first expanded into so-called {\it double YMS (dYMS) integrands}, then into dYMS partial integrands, and finally into BS partial integrands. We point out that the coefficients accompanying to each dYMS integrand are independent of the loop momentum, thus {\it once the linear propagators in dYMS integrands have been transformed into the quadratic ones, we naturally get the quadratic propagators in  EYM and GR integrands}. A further investigation shows that the cancellation of nonlocalities in dYMS partial integrands (hence full integrands) is the most crucial step for extracting quadratic propagators. This localization follows from two steps: (i) expressing dYMS partial integrands by the local expression (\ref{Eq:YMSPartial0}) of YMS ones and canceling the nonlocality caused by the left coefficients, (ii) further canceling the nonlocalities caused by the right coefficients.

In the remainder of this section, we introduce the dYMS partial integrands and full integrands, and then express the forward limit form of EYM and GR integrands by combinations of dYMS ones. We further show that the nonlocalities of dYMS partial integrands associated with the left coefficients are canceled out when the local expression (\ref{Eq:YMSPartial0}) of YMS is considered. In order to cancel the nonlocalities caused by the right coefficients, we provide a further rearrangement of the dYMS partial integrand.

\subsection{Expanding EYM and GR integrands in terms of double YMS integrands}

In this work, we introduce one-loop {\it double Yang-Mills-scalar (dYMS) integrand} $\mathcal{I}^{\,\text{dYMS}}$ as 
\bea
\mathcal{I}^{\,\text{dYMS}}\big(\big.\pmb{\sigma}||\mathsf{G}\,\big|\,\pmb{\rho}||\W{\mathsf{G}}\,\big)=\left[\mathcal{A}^{\,\text{dYMS}}\big(\big.+,\pmb{\sigma},-||\mathsf{G}\,\big|\,+,\pmb{\rho},-||\W{\mathsf{G}}\,\big)+\text{cyc}(\pmb{\sigma})\right]+\text{cyc}(\pmb{\rho}),\Label{Eq:doubleYMSIntegrand}
\eea
where the {\it partial integrand} of dYMS, $\mathcal{A}^{\,\text{dYMS}}$ is defined by the following integral over scattering variables
\bea
\mathcal{A}^{\,\text{dYMS}}\big(\big.+,\pmb{\sigma},-||\mathsf{G}\,\big|\,+,\pmb{\rho},-||\W{\mathsf{G}}\,\big)\equiv{1\over l^2}\,\lim\limits_{k_{\pm}\,\to \pm l}\int \text{d}\mu\,I^{\,\text{dYMS}}_L\,I^{\,\text{dYMS}}_R. \Label{Eq:doubleYMS}
\eea
In the above, both half integrands $I_L$ and $I_R$ are defined by {\it the left half integrands of YMS} in \tabref{table:LoopIntegrands} 
\bea
I^{\,\text{dYMS}}_L=\text{PT}(+,\pmb{\sigma},-)\,\text{Pf}\,[\Psi_{\mathsf{G}}],~~~~I^{\,\text{dYMS}}_R=\text{PT}(+,\pmb{\rho},-)\,\text{Pf}\,[\Psi_{\W{\mathsf{G}}}]. \Label{Eq:ILR}
\eea
In \eqref{Eq:doubleYMS} and \eqref{Eq:ILR}, the two permutations $\pmb{\sigma}$, $\pmb{\rho}$ and the two sets $\mathsf{G}$, $\W{\mathsf{G}}$ are introduced based on the following four sets of particles
\bea
\mathsf{X}^{\text{S}\otimes\text{S}}\equiv\{x_1,x_2,\dots,x_i\},\mathsf{Y}^{\text{G}\otimes\text{S}}\equiv\{y_1,y_2,\dots,y_j\},\mathsf{Z}^{\text{S}\otimes\text{G}}\equiv\{z_1,z_2,\dots,z_k\},\mathsf{W}^{\text{G}\otimes\text{G}}\equiv\{w_1,w_2,\dots,w_l\},~~\Label{Eq:ParticleTypes}
\eea
in which the superscripts $\text{A}\otimes\text{B}$ where $\text{A},\text{B}=\text{S}~\text{(scalar)}~\text{or}~\text{G}~\text{(gluon)}$ record the roles played by the particles in the left and right half integrands. Each left or right superscript G of a set implies an element therein has a polarization vector $\epsilon^{\mu}$ or ${\W\epsilon}^{\,\W\mu}$.
Hence, {\it elements in $\mathsf{X}\cup\mathsf{Z}$ and  $\mathsf{X}\cup\mathsf{Y}$ respectively play as the scalars in the half integrands $I^{\,\text{dYMS}}_L$ and $I^{\,\text{dYMS}}_R$, while elements in $\mathsf{Y}\cup \mathsf{W}$ and $\mathsf{Z}\cup \mathsf{W}$ play as the gluons in $I^{\,\text{dYMS}}_L$ and $I^{\,\text{dYMS}}_R$}. Apparently, from this definition, all elements in $I^{\,\text{dYMS}}_L$ must be the same  as that in  $I^{\,\text{dYMS}}_R$, with $\pmb{\sigma}$, $\pmb{\rho}$, $\mathsf{G}$ and $\W{\mathsf{G}}$ concretely specified as
\bea
\pmb{\sigma}\in\text{perms~}(\mathsf{X}\cup \mathsf{Z}),~~\pmb{\rho}\in\text{perms~}(\mathsf{X}\cup \mathsf{Y}),~~\mathsf{G}=\mathsf{Y}\cup \mathsf{W},~~\W{\mathsf{G}}=\mathsf{Z}\cup \mathsf{W},
\eea
where $\text{perms~}(A)$ denotes the collection of permutations of elements in the set $A$. For example, in the dYMS partial integrand $\mathcal{A}^{\,\text{dYMS}}\big(\big.+,x_1,z_1,-||\{y_1,w_1\}\,\big|\,+,x_1,y_1,-||\{z_1,w_1\}\,\big)$, the particle sets are given by $\mathsf{X}^{\text{S}\otimes\text{S}}=\{x_1\},\mathsf{Y}^{\text{G}\otimes\text{S}}=\{y_1\},\mathsf{Z}^{\text{S}\otimes\text{G}}=\{z_1\},\mathsf{W}^{\text{G}\otimes\text{G}}=\{w_1\}$,
where particles $y_1$, $z_1$ and $w_1$ carry polarizations $\epsilon_{y_1}^{\mu}$, ${\W\epsilon}_{z_1}^{\,\W\mu}$ and $\epsilon_{w_1}^{\mu}{\W\epsilon}_{w_1}^{\,\W\mu}$, respectively. The $\pmb{\sigma}$, $\pmb{\rho}$, $\mathsf{G}$ and $\W{\mathsf{G}}$ in (\ref{Eq:doubleYMS}) are explicitly given by
\bea
\pmb{\sigma}=(x_1z_1),~~\pmb{\rho}=(x_1y_1),~~\mathsf{G}=\{y_1\}\cup \{w_1\},~~\W{\mathsf{G}}=\{z_1\}\cup \{w_1\}.
\eea

For EYM and GR, when we expand the corresponding half integrands in  \tabref{table:LoopIntegrands} according to (\ref{Eq:ExpandReducedPfaffian2}), (\ref{Eq:Expansion1}) and further consider the definition of dYMS integrand (\ref{Eq:doubleYMSIntegrand}), the full EYM and GR integrands can always be expanded as combinations of dYMS ones. 
Particularly, the EYM integrand $\mathcal{I}^{\,\text{EYM}}(\pmb{\rho}||\mathsf{H})$ reads
\bea
\mathcal{I}^{\,\text{EYM}}(\pmb{\rho}||\mathsf{H})&=&(D-2)\,\mathcal{I}^{\,\text{dYMS}}\big(\emptyset||\mathsf{G}\cup\mathsf{H}\,\big|\,\pmb{\rho}||\mathsf{H}\big)\nn
&&+\Sl_{l=2}^{n}(-1)^l\Sl_{{\small{(j_1...j_l)}\in\text{S}_l\setminus\text{Z}_l}}\text{Tr}\big[{F}_{j_1}\cdot...\cdot {F}_{j_l}\big]\mathcal{I}^{\,\text{dYMS}}\big(j_1,...,j_l||(\mathsf{G}\cup\mathsf{H})\setminus\{j_1,...,j_l\}\,\big|\,\pmb{\rho}||\mathsf{H}\big),\Label{Eq:EYMDoubleYMS1}
\eea
where $\mathsf{G}$ and $\mathsf{H}$ respectively denote the gluon set and the graviton set.
One can find the four sets (\ref{Eq:ParticleTypes}) in each dYMS integrand in the above expression according to their definitions.  Obviously, for the dYMS integrand $\mathcal{I}^{\,\text{dYMS}}\big(\emptyset||\mathsf{G}\cup\mathsf{H}\,\big|\,\pmb{\rho}||\mathsf{H}\big)$ on the first line, we have
\bea
 \mathsf{X}^{\text{S}\otimes\text{S}}=\emptyset,~~\mathsf{Y}^{\text{G}\otimes\text{S}}=\mathsf{G},~~\mathsf{Z}^{\text{S}\otimes\text{G}}=\emptyset,~~\mathsf{W}^{\text{G}\otimes\text{G}}=\mathsf{H}.
 \eea
For each $\mathcal{I}^{\,\text{dYMS}}\big(j_1,...,j_l||(\mathsf{G}\cup\mathsf{H})\setminus\{j_1,...,j_l\}\,\big|\,\pmb{\rho}||\mathsf{H}\big)$ on the second line, the four sets  (\ref{Eq:ParticleTypes}) are respectively    presented by 
\bea
\mathsf{X}^{\text{S}\otimes\text{S}}=\{j_1,...,j_l\}\cap\mathsf{G},\,\mathsf{Y}^{\text{G}\otimes\text{S}}=\mathsf{G}\setminus\{j_1,...,j_l\},\,\mathsf{Z}^{\text{S}\otimes\text{G}}=\{j_1,...,j_l\}\setminus\mathsf{G},\,\mathsf{W}^{\text{G}\otimes\text{G}}=\mathsf{H}\setminus\{j_1,...,j_l\}.
\eea
Analogously to the EYM case, the gravity integrand $\mathcal{I}^{\,\text{GR}}$ is written in terms of double YMS ones
\bea
\mathcal{I}^{\,\text{GR}}(\mathsf{H})&=&(D-2)^2\,\mathcal{I}^{\,\text{dYMS}}\big(\emptyset||\mathsf{H}\,\big|\,\emptyset||\mathsf{H}\big)\Label{Eq:GRDoubleYMS1}\\
&&+(D-2)\Sl_{l=2}^{n}(-1)^l\Sl_{{\small{(i_1...i_l)}\in\text{S}_l\setminus\text{Z}_l}}\text{Tr}\left[\,\W{F}_{i_1}\cdot...\cdot \W{F}_{i_l}\,\right]\mathcal{I}^{\,\text{dYMS}}\big(\emptyset||\mathsf{H}\,\big|\,i_1,...,i_l||\mathsf{H}\setminus\{i_1,...,i_l\}\big)\nn
&&+(D-2)\Sl_{l=2}^{n}(-1)^l\Sl_{{\small{(j_1...j_l)}\in\text{S}_l\setminus\text{Z}_l}}\text{Tr}\big[{F}_{j_1}\cdot...\cdot {F}_{j_l}\big]\mathcal{I}^{\,\text{dYMS}}\big(j_1,...,j_l||\mathsf{H}\setminus\{j_1,...,j_l\}\,\big|\,\emptyset||\mathsf{H}\big)\nn
&&+\Sl_{l,m=2}^{n}(-1)^{l+m}\Sl_{\substack{\small{(j_1...j_l)}\in\text{S}_l\setminus\text{Z}_l\\\small{(i_1...i_m)}\,\in\,\text{S}_m\setminus\text{Z}_m}}\text{Tr}\big[{F}_{j_1}\cdot...\cdot {F}_{j_l}\big]\nn
&&~~~~~\times\mathcal{I}^{\,\text{dYMS}}\big(j_1,...,j_l||\mathsf{H}\setminus\{j_1,...,j_l\}\,\big|\,i_1,...,i_m||\mathsf{H}\setminus\{i_1,...,i_m\}\big)\times\text{Tr}\left[\,\W{F}_{i_1}\cdot...\cdot \W{F}_{i_m}\,\right],\nonumber
\eea
where each $\W{F}^{\W\mu\W\nu}_{a}$ is defined by $k_a^{\,\W\mu}{\W\epsilon}_a^{\,\W\nu}-{\W\epsilon}_a^{\,\W\mu}k_a^{\,\W\nu}$,
the four sets (\ref{Eq:ParticleTypes}) in each dYMS integrand can also be read off according to definition.  To sum up, in (\ref{Eq:EYMDoubleYMS1}) or (\ref{Eq:GRDoubleYMS1})  {\it (i) a graviton (gluon) does not always play as a $\mathsf{W}$ ($\,\mathsf{Y}$) element in the dYMS integrands on the rhs., since its polarization may be absorbed into the expansion coefficients; (ii) the expansion coefficients are independent of the loop momentum $l^{\mu}$.}

\subsection{Nonlocality and linear propagators in dYMS, EYM and GR integrands}

As demonstrated previously, both EYM and GR integrands are expressed by dYMS ones with {\it coefficients independent of loop momentum}, and a full dYMS integrand is further written in terms of partial integrands, as shown by (\ref{Eq:doubleYMSIntegrand}). Now let us show that {\it the expression of dYMS partial integrands (\ref{Eq:doubleYMS}) has nonlocality, and the loop propagators (except for a $1\over l^2$) in both dYMS partial and full integrands are linear ones.} To see this, by inserting the half integrands (\ref{Eq:ILR}) into the dYMS partial integrand (\ref{Eq:doubleYMS}), applying (\ref{Eq:Expansion1}) and using the definition of BS partial integrands (\ref{Eq:BSPartialIntegrand}), we express the dYMS partial integrand as a combination of BS ones
\bea
\mathcal{A}^{\,\text{dYMS}}\big(\big.+,\pmb{\sigma},-||\mathsf{G}\,\big|\,+,\pmb{\rho},-||\W{\mathsf{G}}\,\big)&=&\Sl_{{\mathcal{F}},\,\W{\mathcal{F}}}\,\Sl_{\pmb{\beta}[{\mathcal{F}}],\pmb{\gamma}[\W{\mathcal{F}}]}\,\mathcal{C}[{\mathcal{F}}]\,\mathcal{A}^{\text{BS}}\big(+,\pmb{\beta}[\mathcal{F}],-\big|+,\pmb{\gamma}[\W{\mathcal{F}}],-\big)\,\mathcal{C}\big[\W{\mathcal{F}}\big],\Label{Eq:doubleYMS1}
\eea
where $\mathcal{F}$ and $\W{\mathcal{F}}$ stand for the graphs corresponding to the left and the right half integrands in (\ref{Eq:ILR}). The $\pmb{\beta}[{\mathcal{F}}]$, $\pmb{\gamma}[\W{\mathcal{F}}]$ and $\mathcal{C}[{\mathcal{F}}]$, $\mathcal{C}\big[\W{\mathcal{F}}\big]$ refer to permutations and coefficients defined by $\mathcal{F}$, $\W{\mathcal{F}}$ according to graphic rule. In this paper, we mention $\mathcal{F}$ ($\W{\mathcal{F}}$), $\pmb{\beta}[{\mathcal{F}}]$ ($\pmb{\gamma}[\W{\mathcal{F}}]$) and $\mathcal{C}[{\mathcal{F}}]$ ($\mathcal{C}\big[\W{\mathcal{F}}\big]$) as left (right) graph, left (right) permutation and left (right) coefficient.
A further substitution of the linear propagator expressed Feynman diagram (LPFD) expression (\ref{Eq:One-loopBasis-1}) of $\mathcal{A}^{\text{BS}}$ shows that $\mathcal{A}^{\,\text{dYMS}}$ can be written as 
\bea
\mathcal{A}^{\,\text{dYMS}}\big(\big.+,\pmb{\sigma},-||\mathsf{G}\,\big|\,+,\pmb{\rho},-||\W{\mathsf{G}}\,\big)&=&\Sl_{{\mathcal{F}},\,\W{\mathcal{F}}}\,\Sl_{\pmb{\beta}[{\mathcal{F}}],\pmb{\gamma}[\W{\mathcal{F}}]}\,\Sl_{\small\substack{(A_1A_2...A_I)={\pmb{\beta}}\\\small (\W A_1\W A_2...\W A_I)={\pmb{\gamma}}}}\mathcal{C}[{\mathcal{F}}]\,{\phi_{A_1|\W A_1}\phi_{A_2|\W A_2}\cdots \phi_{A_I|\W A_I}\over l^2s_{A_1,l}s_{A_1A_2,l}\cdots s_{A_1A_2\cdots A_{I-1},l}}\,\mathcal{C}\big[\W{\mathcal{F}}\big]\nn
&=&\Sl_{{\mathcal{F}},\,\W{\mathcal{F}}}\,\Sl_{\pmb{\beta}[{\mathcal{F}}],\pmb{\gamma}[\W{\mathcal{F}}]}\,\Sl_{\small\substack{(A_1A_2...A_I)={\pmb{\beta}}\\\small (\W A_1\W A_2...\W A_I)={\pmb{\gamma}}}}\mathcal{C}[{\mathcal{F}}]\,\begin{minipage}{2.5cm}  \includegraphics[width=2.5cm]{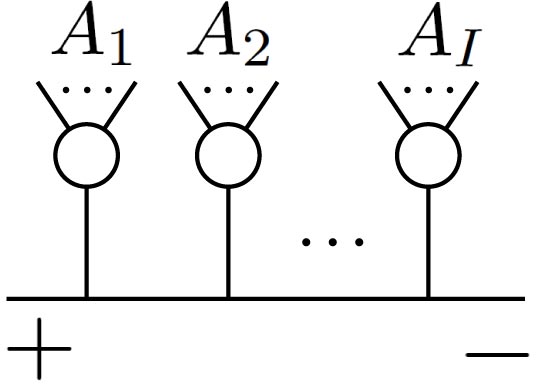} \end{minipage}\,\mathcal{C}\big[\W{\mathcal{F}}\big].\Label{Eq:doubleYMS20}
\eea
 Generally speaking, in each term of the summation, both left and right coefficients in  (\ref{Eq:doubleYMS20}) may contain contractions between the polarizations and/or momenta of particles belonging to distinct subsets, e.g., $\epsilon_{w_1}\cdot \epsilon_{w_2}$ where $w_1$ and $w_2$ are divided into subsets $A_1$ and $A_3$ of the division $(A_1A_2A_3A_4)$, respectively. These contractions are separated by loop propagators in the LPFD and cannot be achieved by local vertices on loop propagator line, thus  (\ref{Eq:doubleYMS20}) has {\it nonlocality} crossing the propagator line between $+$ and $-$. These nonlocal coefficients in general lack cyclicity, which means that, after expanding all partial integrands contained in the full dYMS integrand (\ref{Eq:doubleYMSIntegrand}) into the LPFDs of BS, we can not obtain sufficient linear propagator terms sharing the proper coefficient to apply (\ref{Eq:partial}) and yield a quadratic propagator contribution. For instance, returning to the previous case with the nonlocal coefficient $\epsilon_{w_1}\cdot \epsilon_{w_2}$, if $w_2$ is assumed to be the highest-weight particle, only the LPFDs for the divisions $(A_1A_2A_3A_4)$ and $(A_4A_1A_2A_3)$ exist; those for $(A_3A_4A_1A_2)$ and $(A_2A_3A_4A_1)$ are precisely excluded from the full dYMS integrand by the graphic rule. Hence, the nonlocal coefficient $\epsilon_{w_1}\cdot \epsilon_{w_2}$ does not exhibit the desired cyclicity, preventing the conversion of the associated linear propagator terms into the quadratic propagator form. The nonlocality and linear propagators in dYMS are further transferred into the EYM and GR integrands via (\ref{Eq:EYMDoubleYMS1}) and (\ref{Eq:GRDoubleYMS1}). 

In this work, we provide a systematic approach to canceling the nonlocalities in the dYMS partial integrands and extracting quadratic propagators of the full dYMS integrands. We show that the local expression of the dYMS partial integrand is the key point for obtaining the quadratic propagator form of the full dYMS integrands, since it implies cyclic structures and the relation (\ref{Eq:partial}) between linear and quadratic propagators can be applied. Once a quadratic propagator form of dYMS integrand has been constructed, those of EYM and GR integrands arise naturally from  (\ref{Eq:GenResultNewJ1}) and (\ref{Eq:GenResultNewJ2}).

\subsection{Expressing dYMS  by YMS, canceling the nonlocalities associated with the left coefficients}
As discussed above, the nonlocalities involved in the partial integrand (\ref{Eq:doubleYMS20}) come from both left and right coefficients $\mathcal{C}\big[{\mathcal{F}}\big]$ and $\mathcal{C}\big[\W{\mathcal{F}}\big]$. For a given right graph $\W{\mathcal{F}}$ and a given right permutation  $\pmb{\gamma}\big[\W{\mathcal{F}}\big]$, the summations over the left graphs, left permutations, left and right divisions give a YMS integrand which already has local form (\ref{Eq:YMSPartial0}). Therefore, the nonlocalities associated with the left coefficients are canceled out and the dYMS partial integrand (\ref{Eq:doubleYMS20}) turns into 
\bea
\mathcal{A}^{\,\text{dYMS}}\big(\big.+,\pmb{\sigma},-||\mathsf{G}\,\big|\,+,\pmb{\rho},-||\W{\mathsf{G}}\,\big)&\!=\!&\Sl_{\W{\mathcal{F}}}\,\mathcal{C}[\W{\mathcal{F}}]\,\Sl_{\pmb{\gamma}[\W{\mathcal{F}}]}\,\mathcal{A}^{\text{YMS}}\big(+,\pmb{\sigma},-||\mathsf{G}\,\big|+,\pmb{\gamma}[\W{\mathcal{F}}],-\big)\Label{Eq:doubleYMS2}\\
&\!=\!&\Sl_{\W{\mathcal{F}}}\,\mathcal{C}[\W{\mathcal{F}}]\,\Sl_{\pmb{\gamma}[\W{\mathcal{F}}]}\Sl_{\small\substack{\{A_1,A_2,...,A_I\}\\\small (\W A_1\W A_2...\W A_I)=\pmb{\gamma}[\W{\mathcal{F}}]\\ \small A_j=\W A_j}}{1\over l^2}J[A_1] {1\over s_{A_1,l}}J[A_2]\cdots{1\over s_{A_1...A_{I-1},l}}J[A_I]\,, \nonumber
\eea
 where the ordered set $\W A_j$ ($j=1,...,I$) is absorbed into the definition of $J[A_j]$, as shown by (\ref{Eq:GenResultNewJ1}), (\ref{Eq:GenResultNewJ2}) and (\ref{Eq:EffectiveCurrent}). Each term of (\ref{Eq:doubleYMS2}) can be understood by LPFD of YMS as follows: 
 \begin{itemize}
\item [(i)] Each YMS effective current {(defined by the first equality of (\ref{Eq:EffectiveCurrent}))} is attached to the linear propagator line via a three-point vertex defined by (\ref{Eq:reviewvertex}).

\item  [(ii)] The YM effective currents {(defined by the second equality of (\ref{Eq:EffectiveCurrent}))} are attached to the linear propagator line via either a three-point vertex or a four-point vertex defined by (\ref{Eq:GenResultNewJ2}) and  (\ref{Eq:reviewvertex}). 
\end{itemize}

Although there is no nonlocality caused by the left coefficient in (\ref{Eq:doubleYMS2}), there still exist nonlocalities associated with the right coefficients which require a further cancellation. In the next subsection, we perform a rearrangement of the summations in (\ref{Eq:doubleYMS2}) so that it is in an appropriate  form to cancel the right nonlocalities. We further show the cancellation of the nonlocalities for $|\mathsf{W}|=0,1,2$ in the coming sections, and present the result for the $|\mathsf{W}|=3$ case.
 
\subsection{Further rearrangement of the dYMS partial integrand (\ref{Eq:doubleYMS2})}
In this subsection, we perform a rearrangement of the orders of the summations on the second line of \eqref{Eq:doubleYMS2}. Specifically, the order between the summations over graphs $\W{\mathcal{F}}$, permutations $\pmb{\gamma}[\W{\mathcal{F}}]$, partitions $\{A_1,A_2,...,A_I\}$ and divisions $(\W A_1\W A_2...\W A_I)$ can be rearranged as follows 
\bea
&&\Sl_{\W{\mathcal{F}}}\,\mathcal{C}[\W{\mathcal{F}}]\,\Sl_{\pmb{\gamma}[\W{\mathcal{F}}]}\,\Sl_{\small\substack{\{A_1,A_2,...,A_I\}\\\small (\W A_1\W A_2...\W A_I)=\pmb{\gamma}[\W{\mathcal{F}}]\\ \small A_j=\W A_j}}{1\over l^2}J[A_1] {1\over s_{A_1,l}}J[A_2]\cdots{1\over s_{A_1...A_{I-1},l}}J[A_I]\,\nn
&\to& \Sl_{\{A_1,A_2,...,A_I\}}{1\over l^2}{1\over s_{A_1,l}}\cdots{1\over s_{A_1...A_{I-1},l}}\Bigg[\,\Sl_{\W{\mathcal{F}}}\,\mathcal{C}[\W{\mathcal{F}}]\,\Sl_{\pmb{\gamma}[\W{\mathcal{F}}]}\,\Sl_{\small\substack{\small (\W A_1\W A_2...\W A_I)=\pmb{\gamma}[\W{\mathcal{F}}]\\ \small A_j=\W A_j}}J[A_1]J[A_2]\cdots J[A_I]\,\Bigg].\Label{Eq:RearrangeSummation}
\eea
In the rewritten expression, we first sum over all possible left partitions $\{A_1,A_2,...,A_I\}$, and then sum over all right graphs $\W{\mathcal{F}}$, right permutations $\pmb{\gamma}[\W{\mathcal{F}}]$ and right divisions $(\W A_1\W A_2...\W A_I)$ for a fixed  left partition. If $J[A_i]$ in a given partition  $\{A_1,A_2,...,A_I\}$ contributes a four-point vertex term, i.e., the second term of (\ref{Eq:GenResultNewJ2}),  $A_i$ is further understood as the two subsets $A_{i1}$ and $A_{i2}$ in (\ref{Eq:GenResultNewJ2}). {\it In this case, the partition can be written more precisely as $\{A_1,A_2,...,A_{i1}\text{-}A_{i2},...,A_I\}$, which implies the four-point vertex structure  arising from the reduction of the left nonlocalities.}
\begin{figure}
\centering
\includegraphics[width=0.75\textwidth]{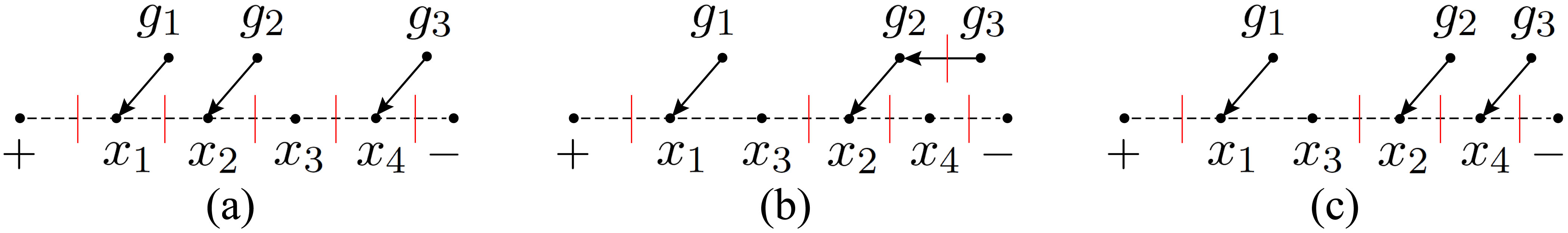}
\caption{For the given partition $\{A_1=\{x_1,x_3,g_1\},A_2=\{x_2,g_2\},A_3=\{x_4,g_3\}\}$, graphs (a) and (c) are decomposed into connected and disconnected subgraphs, whereas graph (b) is decomposed into connected subgraphs only. }
\label{Fig:connectgraph1}
\end{figure}
\begin{figure}
\centering
\includegraphics[width=0.55\textwidth]{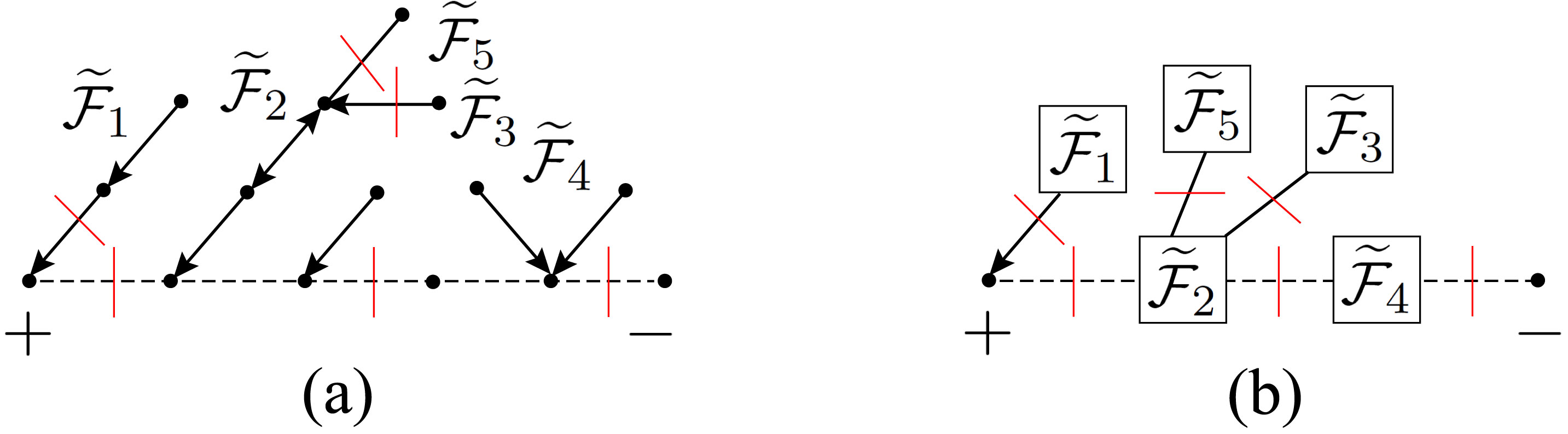}
\caption{A typical right graph (a) and its topology (b)}
\label{Fig:connectgraph2}
\end{figure}

Subject to the requirement $A_j=\W A_j$, the surviving right graphs that give nonvanishing contributions to a certain partition $\{A_1,A_2,...,A_{i-1},A_{i1}\text{-}A_{i2},A_{i+1},...,A_I\}$ on the second line of (\ref{Eq:RearrangeSummation}) satisfy the following conditions:
\begin{itemize}
\item [(i)] Decompose the right graph into $I+1$ subgraphs $\W{\mathcal{F}}_1,\W{\mathcal{F}}_2,...,\W{\mathcal{F}}_{i-1},\W{\mathcal{F}}_{i1},\W{\mathcal{F}}_{i2},\W{\mathcal{F}}_{i+1},...,\W{\mathcal{F}}_I$ such that each subgraph $\W{\mathcal{F}}_j$ contains the elements of $A_j$; then every resulting subgraph is {\it connected}. This connectivity is guaranteed by the permutations imposed on each graph by the graphic rule. As an example, consider the partition $\{A_1=\{x_1,x_3,g_1\},A_2=\{x_2,g_2\},A_3=\{x_4,g_3\}\}$. The permutation associated with \figref{Fig:connectgraph1} (a) is $(x_1\{g_1\}\shuffle\{x_2\{g_2\}\shuffle\{x_3x_4g_3\}\})$. Since the disconnected subgraph of elements $x_1$, $x_3$ and $g_1$ prevents these elements from being divided into the same subset $\W A_1$, this right graph yields no contribution to the given partition. In \figref{Fig:connectgraph1} (b), the disconnected subgraph of elements $x_4$, $g_3$ allows the full graph to generate both divisions $(\W A_1\W A_2\W A_3=(g_3x_4))$ and $(\W A_1\W A_2\W A_3=(x_4g_3))$ that satisfy $A_j=\W A_j$, namely $(\W A_1\W A_2\W A_3=(\{g_3\}\shuffle\{x_4\}))$. Consequently, $J[A_3]$ and the entire contribution vanish due to the generalized $U(1)$-decoupling identity $\phi_{x_4g_3|\{g_3\}\shuffle\{x_4\}}=0$. Only graphs whose subgraphs are all connected, as exemplified in \figref{Fig:connectgraph1} (c), survive. 
\item [(ii)] The relative positions of the connected subgraphs define a {\it topology}, and only graphs possessing the appropriate topology contribute to the given partition. For instance, the topology of the right graph in \figref{Fig:connectgraph2} (a), which consists of five connected subgraphs, is shown in \figref{Fig:connectgraph2} (b).  Due to the permutation prescribed by the graphic rule, the graph with this topology contributes only to the partitions $\{\{A_1\}\shuffle\{A_2,\{A_3\}\shuffle \{A_4\}\shuffle \{A_5\}\}\}$ while yielding no contribution to, for example, $\{A_1,A_3,A_2,A_4,A_5\}$ or $\{A_4,A_3,A_1,A_2,A_5\}$.
\end{itemize}
Since $J[A_j]$ on the second line of (\ref{Eq:RearrangeSummation}) can be expanded into BS currents via (\ref{Eq:GenResultNewJ1}), (\ref{Eq:GenResultNewJ2}) and (\ref{Eq:EffectiveCurrent}), and the left coefficients and permutations are likewise encoded in the connected graph $\mathcal{F}_j$ consisting of the elements in $A_j$, linking all left connected graphs rewrites the dYMS partial integrand as 
\bea
&&\mathcal{A}^{\,\text{dYMS}}\big(\big.+,\pmb{\sigma},-||\mathsf{G}\,\big|\,+,\pmb{\rho},-||\W{\mathsf{G}}\,\big)\Label{Eq:doubleYMS4}\\
&=&\Sl_{\small\{A_1, A_2, ..., A_I\}}\Sl_{\small\substack{{\mathcal{F}}_1,{\mathcal{F}}_2,...,{\mathcal{F}}_I,\W{\mathcal{F}}\\ \small {\mathcal{F}}_i=\W{\mathcal{F}}_i}}\,\begin{minipage}{4cm}  \includegraphics[width=4cm]{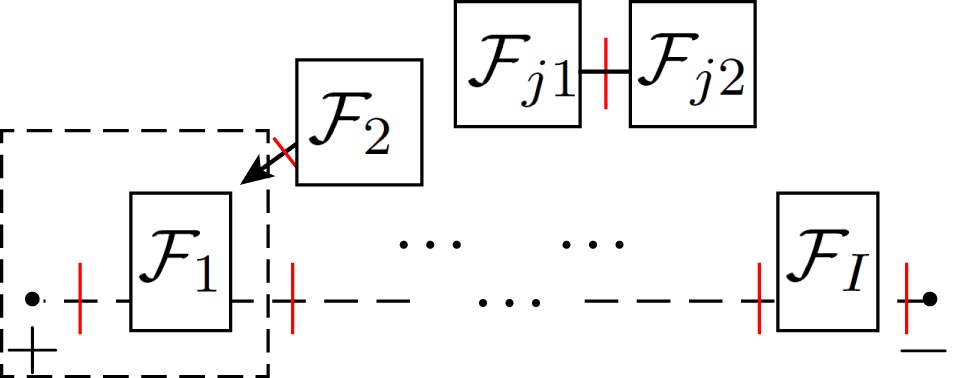} \end{minipage}\times\begin{minipage}{3cm}  \includegraphics[width=3cm]{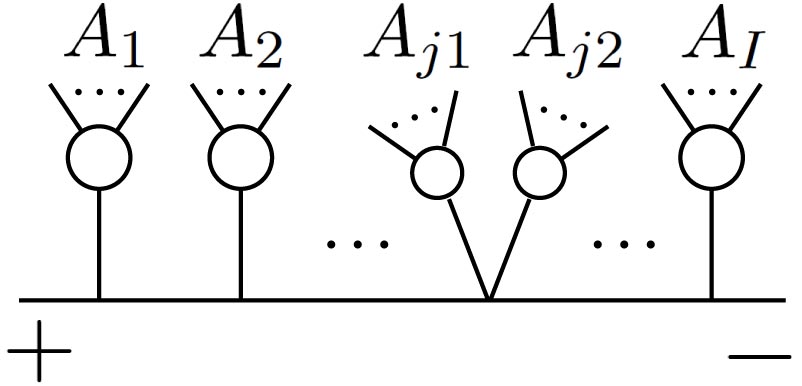} \end{minipage}\times\begin{minipage}{4cm} \includegraphics[width=4cm]{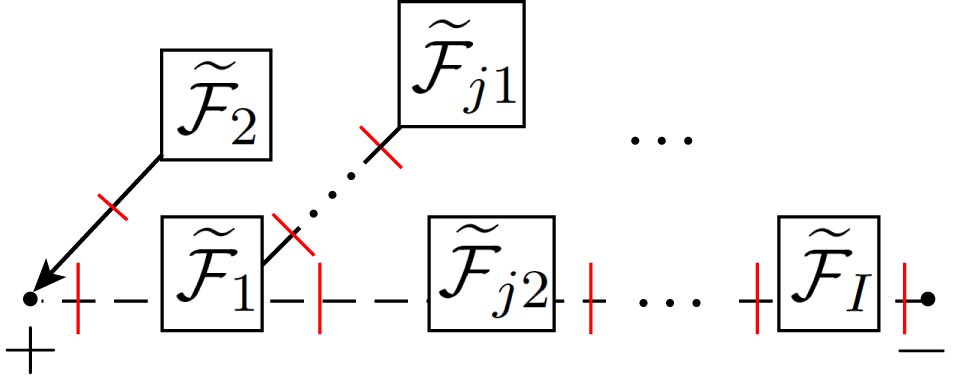} \end{minipage},\nonumber
\eea
In the above expression, we sum over all possible left partitions and, for each partition, sum over all nonvanishing left subgraphs and right graphs. Every right graph is decomposed into connected subgraphs, where the equality $\mathcal{F}_i=\W{\mathcal{F}}_i$ requires that $\mathcal{F}_i$ and $\W{\mathcal{F}}_i$ contain the same elements. The left and right coefficients, together with the left and right permutations of the BS currents in the LPFD, have already been encoded in the corresponding subgraphs. More concretely, in the typical left graph on the second line, $\mathcal{F}_1$ corresponds to the YMS three-point vertex structure of (\ref{Eq:GenResultNewJ1}), and $\mathcal{F}_2$ supplies the YM three-point vertex structure of (\ref{Eq:GenResultNewJ2}). The arrow line of $\mathcal{F}_2$ can attach to any node inside the dashed box, thereby providing momentum $X_{A_2}^{\mu}=l^{\mu}+k_{A_1}^{\mu}$ that contracts with  effective current $T^{\mu} (A_2)$. The independent connection of $\mathcal{F}_{j1}$ and $\mathcal{F}_{j2}$ finally corresponds to the four-point vertex structure in (\ref{Eq:GenResultNewJ2}).

As demonstrated in \cite{Xie:2024pro}, the summation over all possible right graphs in  (\ref{Eq:doubleYMS4}) can be reexpressed by two steps:  {\it Step-1}. Construct all possible right topologies $\W{\mathcal{T}}$, where each topology corresponds to the partition $A_1, ...,A_{j1}\text{-}A_{j2},...,A_I$. {\it Step-2.} For a given topology $\W{\mathcal{T}}$, sum over all possible subgraphs. Therefore, the double YMS partial integrand (\ref{Eq:doubleYMS4}) is rewritten as 
\bea
&&\mathcal{A}^{\,\text{dYMS}}\big(\big.+,\pmb{\sigma},-||\mathsf{G}\,\big|\,+,\pmb{\rho},-||\W{\mathsf{G}}\,\big)\Label{Eq:doubleYMS422}\\
&=&\Sl_{\small\{A_1, A_2, ..., A_I\}}\Sl_{\small\W{\mathcal{T}}}
\Sl_{\substack{\small {\mathcal{F}}_1,{\mathcal{F}}_2,...,{\mathcal{F}}_I\\ \small \W{\mathcal{F}}_1,\W{\mathcal{F}}_2,...,\W{\mathcal{F}}_I}}\,\begin{minipage}{4cm}  \includegraphics[width=4cm]{PartialIntegrand2} \end{minipage}\times\begin{minipage}{3cm}  \includegraphics[width=3cm]{PartialIntegrand1} \end{minipage}\times\begin{minipage}{4cm} \includegraphics[width=4cm]{PartialIntegrand3} \end{minipage},\nonumber
\eea
	where the right topologies and subgraphs are constructed according to the topology rules and subgraph rules in \cite{Xie:2024pro}. Compared with (\ref{Eq:doubleYMS4}), summing simultaneously over left and right subgraphs proves {to be} more effective in canceling the nonlocalities of integrands involving more particles. However, to illustrate the treatment of nonlocalities and the extraction of vertices more clearly and simply, we omit the detailed construction rules of topologies and subgraphs, and instead give explicit examples starting from (\ref{Eq:doubleYMS4}) with the graphic rules introduced in \secref{sec:GraphicRule}. These distinct construction rules are essentially equivalent.

In the expressions (\ref{Eq:doubleYMS4}) or (\ref{Eq:doubleYMS422}) for dYMS partial integrands, the nonlocalities associated with the left coefficients have already been canceled out. Nevertheless, there still exist Lorentz contractions between two right subgraphs which are separated by linear propagators in the LPFD. These right nonlocalities prevent us from extracting quadratic propagators using (\ref{Eq:partial}). 

In the coming sections, we show that {\it once the right locality has been fulfilled, the dYMS partial integrand is naturally expressed by attaching subcurrents to the linear propagator line through local multi-point vertices. Such a structure implies a cyclic sum in the full integrand, hence results in a quadratic propagator form of the dYMS integrand. When the expressions of dYMS integrands with quadratic propagators are substituted into (\ref{Eq:EYMDoubleYMS1}) and  (\ref{Eq:GRDoubleYMS1}), we obtain the quadratic propagator forms of EYM and GR integrands. It is worth clarifying that the coefficients (i.e., the trace of  field strength tensors) in the result of EYM and GR still involve nonlocalities along the loop. Nevertheless, the localities of dYMS are sufficient for extracting the quadratic propagator forms of EYM and GR. We leave the remaining localization of EYM and GR for future work.}

\section{Localization and quadratic propagators of double YMS with $|\mathsf{W}|=0$}\label{sec:W0}

In this section, we localize the dYMS partial integrand (\ref{Eq:doubleYMS}) and then construct the quadratic propagator form of the full integrand (\ref{Eq:doubleYMSIntegrand}) in the case of $|\mathsf{W}|=0$. The starting point of our discussion is the expression (\ref{Eq:doubleYMS4}), which is already local for the {\it left coefficients}. To show the localization of the {\it right part}, we provide concrete calculations on cancellation of nonlocal terms in dYMS partial integrands with $\mathsf{X}=\{x_1\}$, $\mathsf{Y}=\{y_1,y_2\}$, $\mathsf{Z}=\{z_1,z_2\}$ and $\mathsf{X}=\{x_1\}$, $\mathsf{Y}=\{y_1,y_2\}$, $\mathsf{Z}=\{z_1,z_2,z_3\}$.  The two localization strategies we use here were first presented in \cite{Xie:2024pro}, but the cancellation details are more complicated in the present case. After this localization, one can always find terms in partial integrands that are related by cyclic permutations of external particles. Thus, according to (\ref{Eq:partial}), the sum of all partial integrands gives the full integrand with quadratic propagators. It is straightforward to extend the examples to general dYMS partial and full integrands with $|\mathsf{W}|=0$. We finally present a general formula for this case.  In the general formula, which has the same form as that in (\ref{Eq:GenResultNew}), the effective currents are generalized to incorporate both left and right coefficients and permutations generated by left and right graphs. Correspondingly, two new vertices as the right duals of the vertices in (\ref{Eq:reviewvertex}) arise to provide the right Lorentz index and contract with the effective currents. We show that the results given in this section reduce to those in \secref{sec:2.3YMSreview} when $\mathsf{Z}$ is empty.


\subsection{Localizations demonstrated by integrands with  $\mathsf{X}=\{x_1\}$, $\mathsf{Y}=\{y_1,y_2\}$, $\mathsf{Z}=\{z_1,z_2\}$}
In this subsection, we show localizations of dYMS partial integrand  (\ref{Eq:doubleYMS4}) with  $\mathsf{X}=\{x_1\}$, $\mathsf{Y}=\{y_1,y_2\}$ and $\mathsf{Z}=\{z_1,z_2\}$. The sets of scalars and gluons in the left half integrand are respectively $\{x_1,z_1,z_2\}$ and $\{y_1,y_2\}$, while those of scalars and gluons in the right half integrand are  $\{x_1,y_1,y_2\}$ and $\{z_1,z_2\}$. We fix the permutations $\pmb{\sigma}$ and $\pmb{\rho}$ in  (\ref{Eq:doubleYMS4}) as  $(x_1z_1z_2)$ and $(x_1y_1y_2)$, and fix the  left and right reference orders as $y_1\prec y_2$ and $z_1\prec z_2$. We emphasize that the choices of permutations in the half integrands and the choices of reference orders do not affect our discussion.

\subsubsection{Localization-1}
Let us begin with the following term given by graphic rule for the left partition $\{x_1,y_1\text{-}y_2,z_1,z_2\}$ when $\mathcal{A}^{\,\text{dYMS}}\big(\big.+,x_1,z_1,z_2,-||\{y_1,y_2\}\,\big|\,+,x_1,y_1,y_2,-||\{z_1,z_2\}\,\big)$ is expressed by (\ref{Eq:doubleYMS4})
\bea
\mathrm{T}_1&=&\left(-{1\over 2}\right)\begin{minipage}{2.4cm}\includegraphics[width=2.4cm]{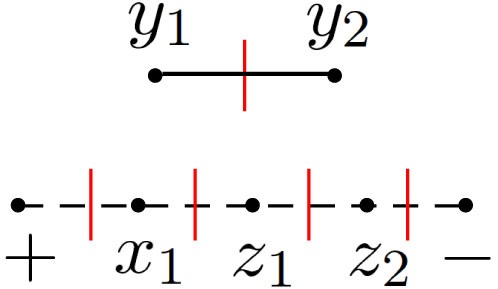} \end{minipage}\times \begin{minipage}{2.3cm}  \includegraphics[width=2.3cm]{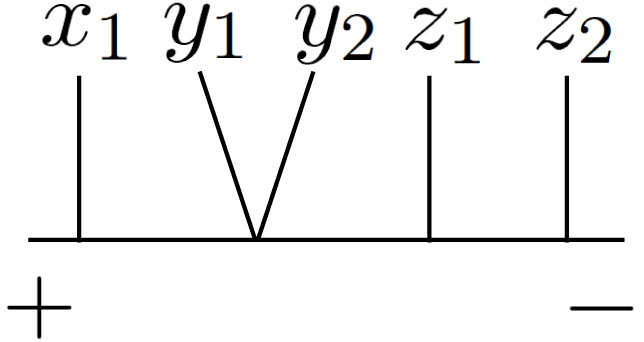} \end{minipage}\times\begin{minipage}{2.45cm}  \includegraphics[width=2.45cm]{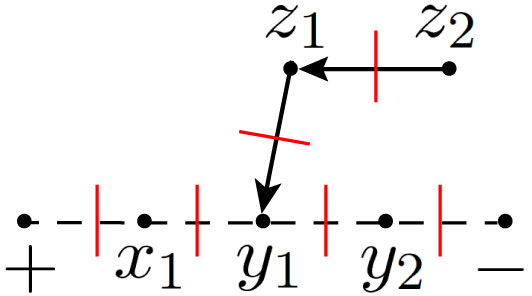} \end{minipage}\Label{Eq:NewW0EG11}\nn
&=&\left(-{1\over 2}\right)(\epsilon_{y_1}\cdot\epsilon_{y_2}){{\phi_{x_1|x_1}\phi_{y_1|y_1}\phi_{y_2|y_2}\phi_{z_1|z_1}\phi_{z_2|z_2}\over l^2 s_{x_1,l}s_{x_1y_1y_2,l}s_{x_1y_1y_2z_1,l}}}(\W\epsilon_{z_1}\cdot k_{y_1})(\W\epsilon_{z_2}\cdot k_{z_1}).
\eea
In this term, we do not have nonlocality associated with the {\it left coefficient} $\epsilon_{y_1}\cdot\epsilon_{y_2}$, as indicated by the $\epsilon\cdot\epsilon$ line in the left graph, since the external particles $y_1$ and $y_2$ are attached to the linear propagator line via a local four-point vertex. However, in the {\it right coefficient}, the factors $\W{\epsilon}_{z_1}\cdot k_{y_1}$ and $\W{\epsilon}_{z_2}\cdot k_{z_1}$ given by the two $\epsilon\cdot k$ lines in the right graph contribute nonlocalities since the $z_1$, $y_1$, and $z_2$, $z_1$ are separated by the linear propagators ${1\over s_{x_1y_1y_2,l}}$ and ${1\over s_{x_1y_1y_2z_1,l}}$, respectively. 

{\it The key technique to treat these nonlocalities is to collect other related nonlocal terms so that the sum of all these terms produces a local result.} We call this treatment {\it localization-1}. Particularly, when all nonlocal terms related to the $\mathrm{T}_1$ in (\ref{Eq:NewW0EG11}) are collected, we get
\bea
\mathrm{T}_1\xrightarrow{\text{Localization}-1}&&\left(-{1\over 2}\right)\begin{minipage}{2.4cm}\includegraphics[width=2.4cm]{NewEQW0EG1L1} \end{minipage}\times \begin{minipage}{2.3cm}  \includegraphics[width=2.3cm]{NewEQW0EG1D1} \end{minipage}\times\begin{minipage}{2.45cm}  \includegraphics[width=2.45cm]{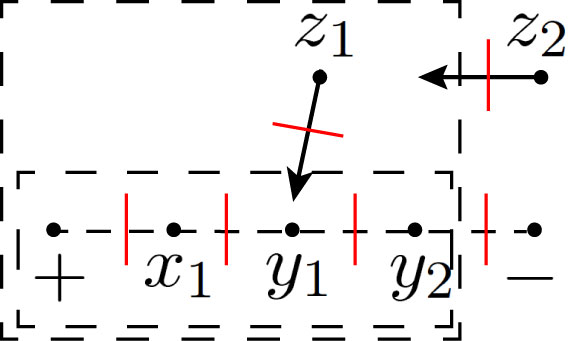} \end{minipage}\Label{Eq:NewW0EG12}\nn
&=&\left(-{1\over 2}\right)(\epsilon_{y_1}\cdot\epsilon_{y_2}){{\phi_{x_1|x_1}\phi_{y_1|y_1}\phi_{y_2|y_2}\phi_{z_1|z_1}\phi_{z_2|z_2}\over l^2 s_{x_1,l}s_{x_1y_1y_2,l}s_{x_1y_1y_2z_1,l}}}(\W\epsilon_{z_1}\cdot X_{z_1})(\W\epsilon_{z_2}\cdot X_{z_2}).
\eea
Here, the notation $\mathrm{T}_1\xrightarrow{\text{Localization}-1}$ means we have collected all nonlocal terms associated with $T_1$: these terms are expressed by the rhs. on the first line. The boxed regions in the right graph of the above equation mean that we sum over all graphs where the $\epsilon\cdot k$ lines $\W{\epsilon}_{z_1}\cdot k_{i}$ and  $\W{\epsilon}_{z_2}\cdot k_{j}$ end at a node in the corresponding boxed region. The explicit expression of the first line is given by the second line, in which  $X^{\mu}_{z_1}=l^{\mu}_{x_1y_1y_2}=l^{\mu}+k_{x_1}^{\mu}+k_{y_1}^{\mu}+k_{y_2}^{\mu}$, $X^{\mu}_{z_2}=l^{\mu}_{x_1y_1y_2z_1}=l^{\mu}+k_{x_1}^{\mu}+k_{y_1}^{\mu}+k_{y_2}^{\mu}+k_{z_1}^{\mu}$.  Correspondingly, $X^{\mu}_{z_1}$ and $X^{\mu}_{z_2}$ are just the momenta of linear propagators that are attached to $z_1$ and $z_2$ from left in the LPFD. 
Thus, the expression (\ref{Eq:NewW0EG12}) is now a local one because there does not exist contraction between subgraphs separated by propagators.

The local expression of the second line in (\ref{Eq:NewW0EG12}) can be reexpressed as LPFD of dYMS
\bea
{1\over l^2}\,J_{(3)}[x_1]\,{1\over s_{x_1,l}}\,J_{(4)}[y_1,y_2]\,{1\over s_{x_1y_1y_2,l}}\,J_{(3)}[z_1]\,{1\over s_{x_1y_1y_2z_1,l}}\,J_{(3)}[z_2]=\begin{minipage}{2.9cm}  \includegraphics[width=2.9cm]{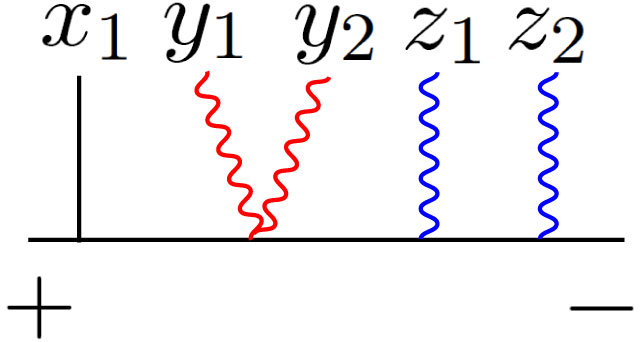} \end{minipage},\Label{Eq:NewFW0EG1}
\eea
where we have introduced local vertex structures as follows:
\begin{itemize}
\item $J_{(3)}[x_1]=\phi_{x_1|x_1}=1$ denotes the structure where the external line $\phi_{x_1|x_1}$ is attached to the propagator line through a 3$x$-vertex
\bea
\begin{minipage}{1.5cm}  \includegraphics[width=1.5cm]{3xVertex} \end{minipage}=1.\Label{Eq:3xVertex}
\eea
The solid lines stand for $\mathsf{X}$ elements. Since the linear propagators in LPFD play as scalar propagators, the $x_2$, $x_3$ in the above vertex correspond to the linear propagators attached to $x_1$ from left and right in \eqref{Eq:NewFW0EG1}.

\item Each $J_{(3)}[z_i]=\big(\W{\epsilon}^{\,\W\mu}_{z_i}\phi_{z_i|z_i}\big)X^{\W\mu}_{z_i}$ ($i=1,2$) in \eqref{Eq:NewFW0EG1} introduces a 2$x$-1$z$ vertex
\bea
\begin{minipage}{1.5cm}  \includegraphics[width=1.5cm]{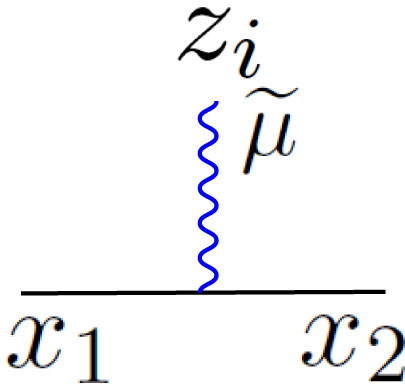} \end{minipage}= k_{x_1}^{\W\mu},\Label{Eq:2x1zVertex}
\eea
by which the external particle $z_i$ is attached to the propagator line. Here, the external particle from the $\mathsf{Z}$ set is denoted by  a blue wave line.
 The $k_{x_1}^{\W\mu}$ becomes the momentum $X^{\W\mu}_{z_i}$ of the loop propagator attached to $z_i$ from left, where ${\W\mu}$ denotes the right Lorentz index.

\item The $J_{(4)}[y_1,y_2]=\left(-{1\over 2}\right)(\epsilon^{\mu}_{y_1}\phi_{y_1|y_1})\eta_{\mu\nu}(\epsilon^{\nu}_{y_2}\phi_{y_2|y_2})$ in \eqref{Eq:NewFW0EG1} introduces a 2$x$-2$y$ four-point vertex
\bea
\begin{minipage}{1.5cm}  \includegraphics[width=1.5cm]{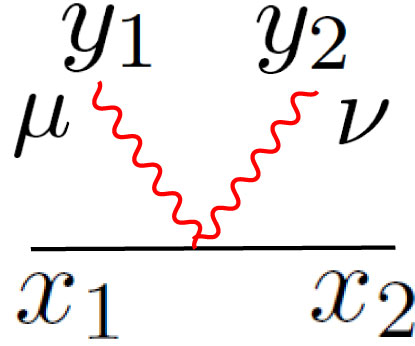} \end{minipage}= \left(-{1\over 2}\right)\eta^{\mu\nu},\Label{Eq:2x2yVertex}
\eea
where the red wave lines stand for $\mathsf{Y}$ elements. The Lorentz indices $\mu$, $\nu$ are the left indices. Noting that the weights of $y_1$ and $y_2$ in this vertex satisfy $y_1\prec y_2$, we do not have such  a 2$x$-2$y$ vertex by exchanging $y_1$ and $y_2$.
\end{itemize}

\begin{figure}
\centering
\includegraphics[width=0.55\textwidth]{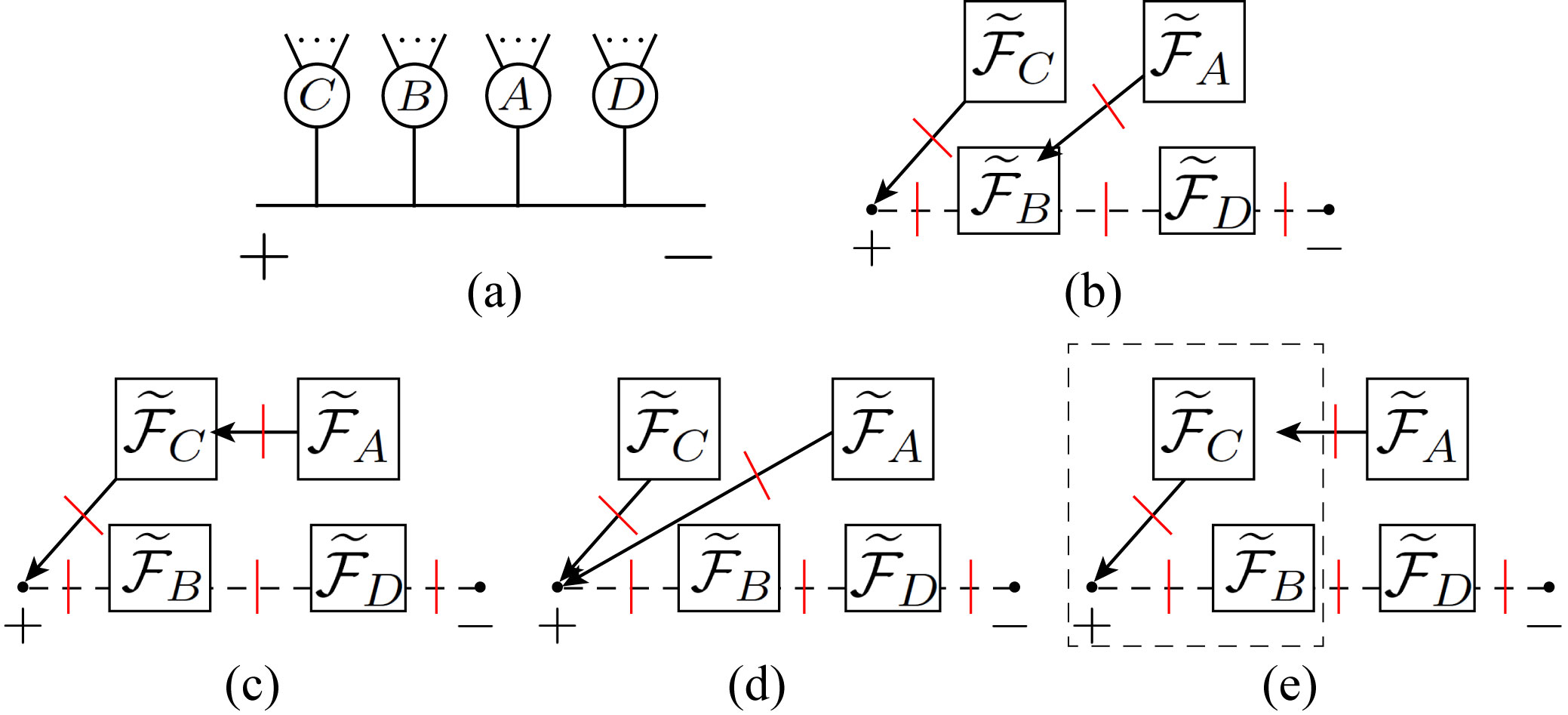}
\caption{Typical right graphs with appropriate LPFD (a) that can apply localization-1}
\label{Fig:GenLoc11}
\end{figure}

{\bf General feature of localization-1:} In general, the localization-1 for right graphs can be applied when the following two conditions are satisfied: (i)  A right subgraph $\W{\mathcal{F}}_A$ is connected to another right subgraph $\W{\mathcal{F}}_B$ which is located below $\W{\mathcal{F}}_A$, via an arrowed $\epsilon\cdot k$ or $k\cdot k$ line pointing towards $\W{\mathcal{F}}_B$, as shown in \figref{Fig:GenLoc11} (b). (ii)  If $\W{\mathcal{F}}_A$ is removed from the full graph, the remaining part is a graph that exists in a fewer-point integrand where all particles in $A$ are deleted from the particle set.
 Once these two conditions are satisfied, we can always find other related right graphs \figref{Fig:GenLoc11} (c) and (d), in which the same subgraph $\W{\mathcal{F}}_A$ is connected to other subgraphs, whose elements locate on the left of the subset $A$ in the LPFD \figref{Fig:GenLoc11} (a) via the same type of line. When collecting all these terms together,  we get \figref{Fig:GenLoc11} (e) and a  $\mathcal{C}[\W{\mathcal{F}}_A]\cdot X_{A}$ factor, which implies a local {\it 2$x$-1$z$ vertex}.

\subsubsection{Localization-2}

In some situations, a nonlocal right graph may not satisfy the general conditions for localization-1. 
We have to introduce {\it localization-2} to cancel these nonlocalities. Now let us demonstrate this with examples.

{\bf Example-1}~~When $\mathcal{A}^{\,\text{dYMS}}\big(\big.+,x_1,z_1,z_2,-||\{y_1,y_2\}\,\big|\,+,x_1,y_1,y_2,-||\{z_1,z_2\}\,\big)$ is expanded according to (\ref{Eq:doubleYMS4}), there exists such a nonlocal term corresponding to the partition $\{x_1,y_1,y_2,z_1,z_2\}$
\bea
\mathrm{T}_1&=&~~~\begin{minipage}{2.4cm}  \includegraphics[width=2.4cm]{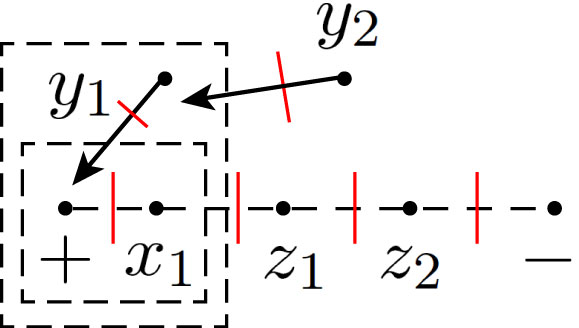} \end{minipage}\times \begin{minipage}{2.3cm}  \includegraphics[width=2.3cm]{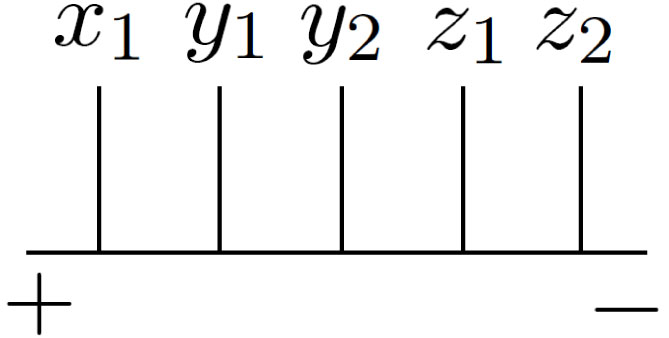} \end{minipage}\times\begin{minipage}{2.3cm}  \includegraphics[width=2.3cm]{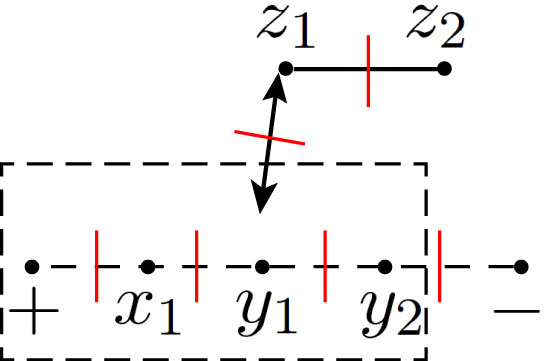} \end{minipage}\Label{Eq:NewW0EG21}\nn
&=&\left(\epsilon_{y_1}\cdot X_{y_1}\right)\left(\epsilon_{y_2}\cdot X_{y_2}\right){\phi_{x_1|x_1}\phi_{y_1|y_1}\phi_{y_2|y_2}\phi_{z_1|z_1}\phi_{z_2|z_2}\over l^2 s_{x_1,l} s_{x_1y_1,l} s_{x_1y_1y_2,l}s_{x_1y_1y_2z_1,l}}\left(\W\epsilon_{z_1}\cdot\W\epsilon_{z_2}\right)\left(-k_{z_1}\cdot X_{z_1}\right).
\eea
In the right graph of the above expression, the nonlocalities associated with the $k\cdot k$ lines between $z_1$ and nodes in the boxed region have been collected to produce a factor $\left(-k_{z_1}\cdot X_{z_1}\right)$. After this step, the contraction between $\W\epsilon^{\W\mu}_{z_1}$ and $\W\epsilon^{\W \mu}_{z_2}$ (given by the $\epsilon\cdot\epsilon$ line) are still separated by the linear propagator ${1\over s_{x_1y_1y_2z_1,l}}$, thus this induces another nonlocality into the right graph. Apparently, this term does not satisfy the conditions for localization-1. Such a structure requires a new approach.

To cancel the nonlocality between $z_1$ and $z_2$, we note that the node $z_1$ is accompanied by a so-called {\it X-pattern} \cite{Xie:2024pro} that brings a $-k_{
z_1}\cdot X_{z_1}$ factor  
\bea
-k_{z_1}\cdot X_{z_1}=\left(-{1\over 2}\right)\big(\,s_{x_1y_1y_2z_1,l}-s_{x_1y_1y_2,l}-k_{z_1}^2\,\big)=\left(-{1\over 2}\right)\big(\,s_{x_1y_1y_2z_1,l}-s_{x_1y_1y_2,l}\,\big),\Label{Eq:Xpattern1}
\eea
where the on-shell condition of $z_1$ was applied. Using (\ref{Eq:Xpattern1}), we split (\ref{Eq:NewW0EG21}) into the following two terms
\bea
\mathrm{T}_1
&=&\begin{minipage}{2.4cm}  \includegraphics[width=2.4cm]{NewEQW0EG2L1} \end{minipage}\times\left(-{1\over 2}\right)\left[\,\begin{minipage}{2.45cm}  \includegraphics[width=2.45cm]{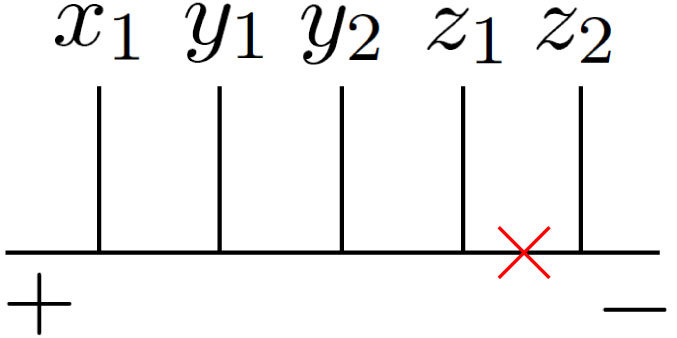} \end{minipage}-\begin{minipage}{2.45cm}  \includegraphics[width=2.45cm]{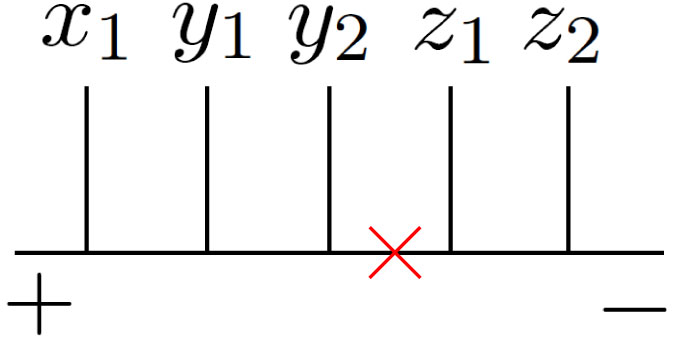} \end{minipage}\,\right]\times\begin{minipage}{2.3cm} \includegraphics[width=2.3cm]{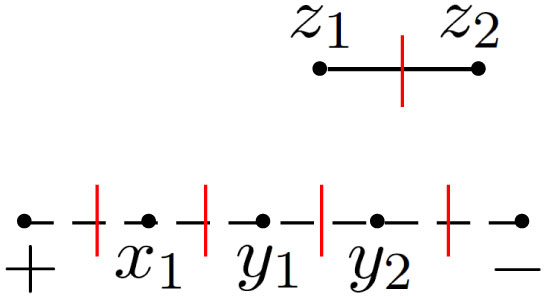} \end{minipage}\nn
&=&\mathrm{T}_{1\text{A}}-\mathrm{T}_{1\text{B}},\Label{Eq:NewW0EG22}
\eea
in which the red crosses mean the propagators between $z_1$, $z_2$ and $y_2$, $z_1$ in the first and the second LPFDs have been removed. The first term $\mathrm{T}_{1\text{A}}$ is now local since $z_1$ and $z_2$ are not separated by any propagator in the LPFD.  The second term $\mathrm{T}_{1\text{B}}$ still exhibits nonlocality. 

To show a further cancellation of the nonlocal term $\mathrm{T}_{1\text{B}}$, we have to investigate the other two terms $\mathrm{T}_2$ and $\mathrm{T}_3$ which have different left graphs and different LPFDs from those of $\mathrm{T}_1$. More explicitly, term $\mathrm{T}_2$ with respect to the partition  $\{x_1,y_1,y_2\text{-}z_1,z_2\}$ is given by 
\bea
\mathrm{T}_2&=&\begin{minipage}{2.4cm}  \includegraphics[width=2.4cm]{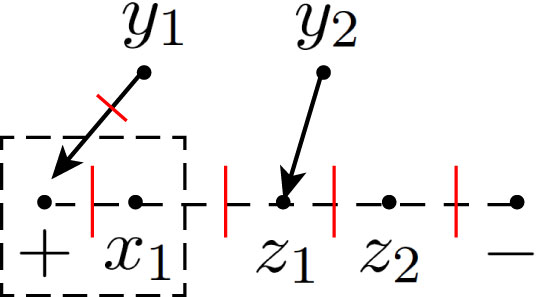} \end{minipage}\times\begin{minipage}{2.45cm}  \includegraphics[width=2.45cm]{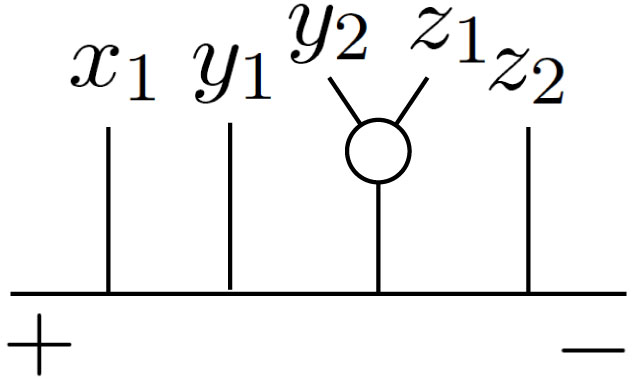} \end{minipage}\times\begin{minipage}{2.3cm} \includegraphics[width=2.3cm]{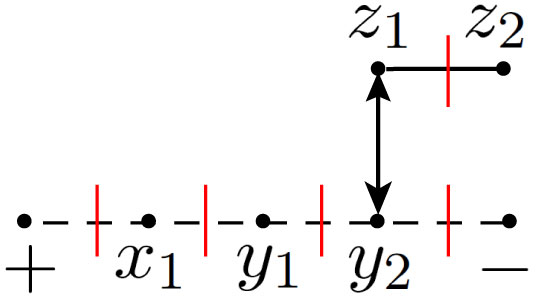} \end{minipage}\nn
&=&\begin{minipage}{2.4cm}  \includegraphics[width=2.4cm]{NewEQW0EG2L2} \end{minipage}\times\left(-{1\over 2}\right)\left[\,\begin{minipage}{2.45cm}  \includegraphics[width=2.45cm]{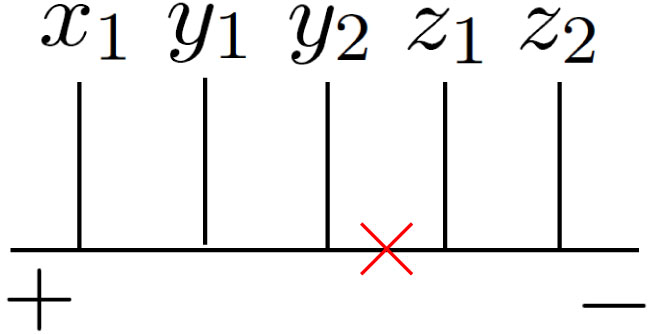} \end{minipage}-\begin{minipage}{2.45cm}  \includegraphics[width=2.45cm]{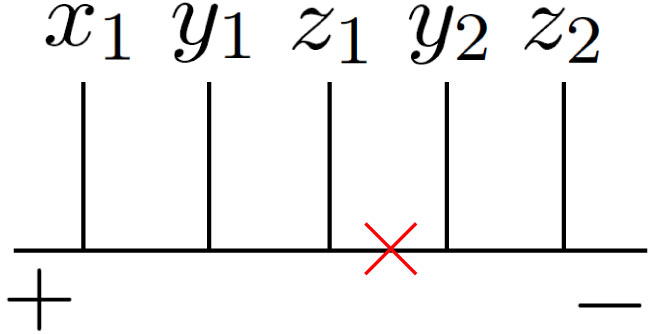} \end{minipage}\,\right]\times\begin{minipage}{2.3cm} \includegraphics[width=2.3cm]{NewEQW0EG2R2} \end{minipage}\nn
&=&\begin{minipage}{2.4cm}  \includegraphics[width=2.4cm]{NewEQW0EG2L2} \end{minipage}\times\left(-{1\over 2}\right)\,(-1)\begin{minipage}{2.45cm}  \includegraphics[width=2.45cm]{NewEQW0EG2D6} \end{minipage}\,\times\begin{minipage}{2.3cm} \includegraphics[width=2.3cm]{NewEQW0EG2R2} \end{minipage}\Label{Eq:NewW0EG23}
\eea
where, we note that the subcurrent involving $y_2$ and $z_1$ on the first line, accompanied by the right factor $-k_{z_1}\cdot k_{y_2}$, agrees with the LHS of the off-shell BCJ relation (\ref{Eq:OffBCJ1}). This structure is called {\it BCJ-pattern} \cite{Xie:2024pro}. According to the off-shell BCJ relation (\ref{Eq:OffBCJ1}), we have 
\bea
 \phi_{z_1y_2|y_2z_1}(-k_{y_2}\cdot k_{z_1})=\left(-\frac{1}{2}\right)\left(\phi_{z_1|y_2}\phi_{y_2|z_1}-\phi_{z_1|z_1}\phi_{y_2|y_2}\right)=\left(-\frac{1}{2}\right)(-1)\phi_{z_1|z_1}\phi_{y_2|y_2}.
\eea
For the first equality, the current with two particles $z_1$ and $y_2$ is divided into two single particle currents. For the second equality, we have used the fact $\phi_{a|b}=0$ ($a\neq b$). Hence, the first line of \eqref{Eq:NewW0EG23} turns into the third line.

Term $\mathrm{T}_3$ is a term that contains an X-pattern associated with $z_1$ and corresponds to the partition $\{x_1,y_1,z_1,y_2,z_2\}$
\bea
\mathrm{T}_3&=&\begin{minipage}{2.4cm}  \includegraphics[width=2.4cm]{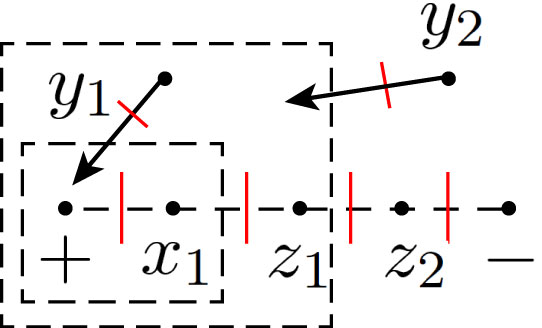} \end{minipage}\times\begin{minipage}{2.45cm}  \includegraphics[width=2.45cm]{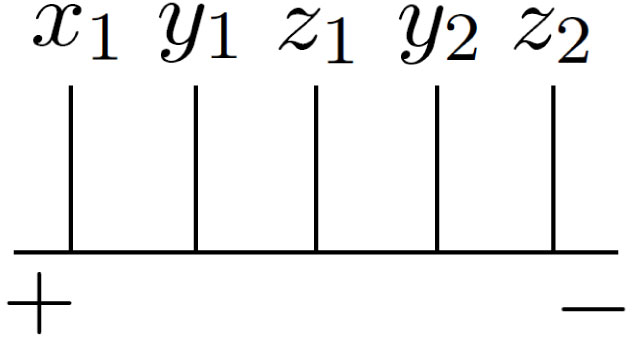} \end{minipage}\times\begin{minipage}{2.3cm} \includegraphics[width=2.3cm]{NewEQW0EG2R1} \end{minipage}\nn
%
%
&=&\left(-{1\over 2}\right)\left[\,\begin{minipage}{2.4cm}  \includegraphics[width=2.4cm]{NewEQW0EG2L1} \end{minipage}\times\begin{minipage}{2.45cm}  \includegraphics[width=2.45cm]{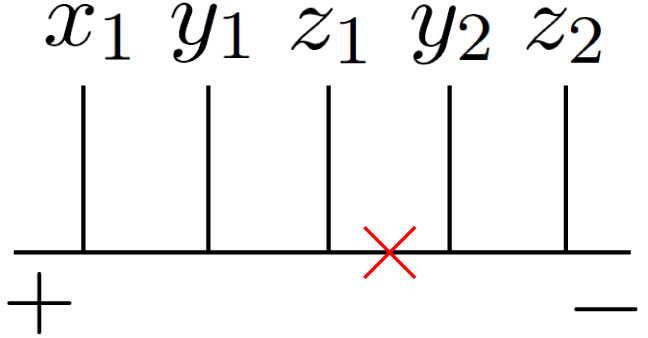} \end{minipage}+\begin{minipage}{2.4cm}  \includegraphics[width=2.4cm]{NewEQW0EG2L2} \end{minipage}\times\begin{minipage}{2.45cm}  \includegraphics[width=2.45cm]{NewEQW0EG2D8} \end{minipage}\right.\nn
&&\left.~~~~~~~~~~~~~~~~~~~~~~~~~~~~~~~~~~~~~~~~~~~~~~~~~~~~~-\begin{minipage}{2.4cm}  \includegraphics[width=2.4cm]{NewEQW0EG2L3} \end{minipage}\times\begin{minipage}{2.45cm}  \includegraphics[width=2.45cm]{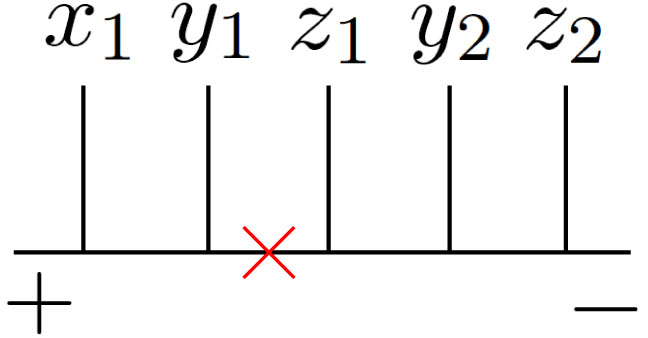} \end{minipage}\,\right]\times\begin{minipage}{2.3cm} \includegraphics[width=2.3cm]{NewEQW0EG2R2} \end{minipage}\nn
&=&\mathrm{T}_{3\text{A}}+\mathrm{T}_{3\text{B}}-\mathrm{T}_{3\text{C}},\Label{Eq:NewW0EG24}
\eea
where we have applied the property (\ref{Eq:Xpattern1}) to X-pattern associated with $z_1$. On the second line of (\ref{Eq:NewW0EG24}), the left coefficient is rewritten as a sum of two left graphs according to the definition. 

{It is obvious that the nonlocal term  $\mathrm{T}_{1\text{B}}$ (\ref{Eq:NewW0EG22}) cancels agaist  $\mathrm{T}_{3\text{A}}$ in (\ref{Eq:NewW0EG24}), while the $\mathrm{T}_{2}$ in (\ref{Eq:NewW0EG23}) cancel with  $\mathrm{T}_{3\text{B}}$ in (\ref{Eq:NewW0EG24}). Therefore, the sum of  $\mathrm{T}_1$, $\mathrm{T}_2$ and $\mathrm{T}_3$ gives rise to}
\bea
\mathrm{T}_1+\mathrm{T}_2+\mathrm{T}_3=\mathrm{T}_{1\text{A}}-\mathrm{T}_{3\text{C}}.\Label{Eq:W0EG16}
\eea
Though $\mathrm{T}_{3\text{C}}$ still has nonlocality, it can be canceled following a similar way  to that for $\mathrm{T}_{1\text{B}}$, and so on. The $\mathrm{T}_{1}$ thus finally introduces the following local expression 
\bea
\mathrm{T}_{1}\xrightarrow{\text{Localization}-2}{1\over l^2}{1\over s_{x_1,l}}(\epsilon_{y_1}\cdot l_{x_1}){1\over s_{x_1y_1,l}}(\epsilon_{y_2}\cdot l_{x_1y_1}){1\over s_{x_1y_1y_2,l}}\left(-{1\over 2}\right)\left(\W\epsilon_{z_1}\cdot\W\epsilon_{z_2}\right){1\over s_{x_1y_1y_2z_1z_2,l}},
\eea
which is already local. Similarly to (\ref{Eq:NewFW0EG1}), the above equation introduces vertex structures as follows
\bea
{1\over l^2}\,J_{(3)}[x_1]\,{1\over s_{x_1,l}}\,J_{(3)}[y_1]\,{1\over s_{x_1y_1,l}}\,J_{(3)}[y_2]\,{1\over s_{x_1y_1y_2,l}}\,J_{(4)}[z_1,z_2]=\begin{minipage}{2.7cm}  \includegraphics[width=2.7cm]{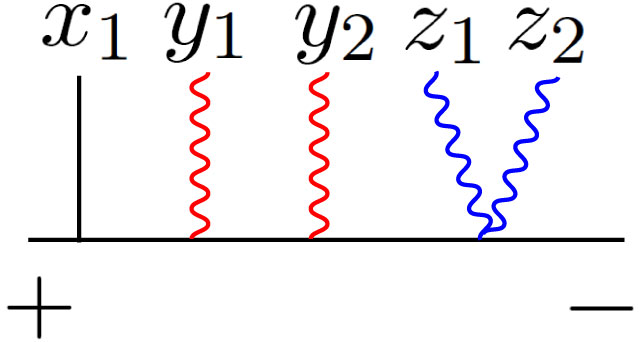} \end{minipage},\Label{Eq:NewFW0EG2D1}
\eea
where $J_{(3)}[x_1]$ is just a $3x$-vertex structure which has already defined in the previous example. The $J_{(3)}[y_i]$ ($i=1,2$) and the $J_{(4)}[z_1,z_2]$ induce new vertices:
\begin{itemize}
\item The $J_{(3)}[y_i]=\left(\epsilon^{\mu}_{y_i}\phi_{y_i|y_i}\right) X^{\mu}_{y_i}$ ($i=1,2$) in (\ref{Eq:NewFW0EG2D1}) defines a $2x$-$1y$ vertex
\bea
\begin{minipage}{1.5cm}  \includegraphics[width=1.5cm]{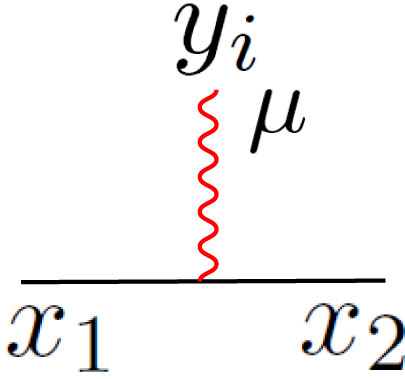} \end{minipage}= k_{x_1}^{\mu},\Label{Eq:2x1yVertex}
\eea
\item  Each $ J_{(4)}[z_1,z_2]=\left(-{1\over 2}\right)\left({\W\epsilon}^{\,\W\mu}_{z_1}\phi_{z_1|z_1}\right)\eta_{\W\mu \W\nu}\left({\W\epsilon}^{\,\W\nu}_{z_2}\phi_{z_2|z_2}\right) $ introduces a $2x$-$2z$ vertices
\bea
\begin{minipage}{1.5cm}  \includegraphics[width=1.5cm]{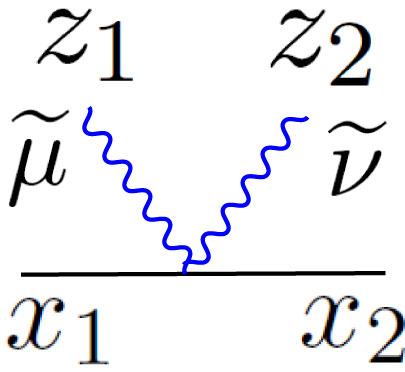} \end{minipage}= \left(-{1\over 2}\right)\eta^{\W\mu\W\nu},\Label{Eq:2x2zVertex}
\eea
\end{itemize}
where the weights of $z_1$ and $z_2$ satisfy $z_1\prec z_2$.  It is worth noting that this vertex can be obtained from (\ref{Eq:2x2yVertex}) by swapping tilded and untilded degrees of freedom, including $y_i \leftrightarrow z_i$. Hence, this exchange symmetry of the four-point vertices (\ref{Eq:2x2yVertex}) and (\ref{Eq:2x2zVertex}) provides a consistent cross-check of their distinct derivations.

From this example, we have shown: (i) The nonlocalities of  X- and BCJ-patterns cancel with each other. (ii) Terms with X-patterns in which  $z_1$ is not adjacent to $z_2$ in the LPFD and terms with BCJ-patterns have to vanish. (iii)  The only surviving term is the one where $z_1$ is adjacent to $z_2$ in the LPFD, and this term contributes a $2x$-$2z$ vertex.  (iv) Although the left graph is already local, it participates in the cancellation of nonlocalities associated with the right graphs. Now we introduce another example to support this statement.

{\bf Example-2}~~Now we consider the following term corresponding to the partition $\{x_1,y_1\text{-}y_2,z_1,z_2\}$ 
\bea
\mathrm{T}_4&=&\left(-{1\over 2}\right)\begin{minipage}{2.4cm}\includegraphics[width=2.4cm]{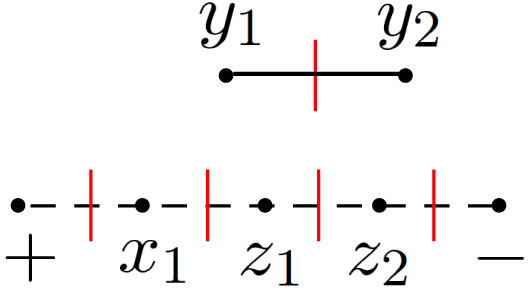} \end{minipage}\times \begin{minipage}{2.45cm}  \includegraphics[width=2.45cm]{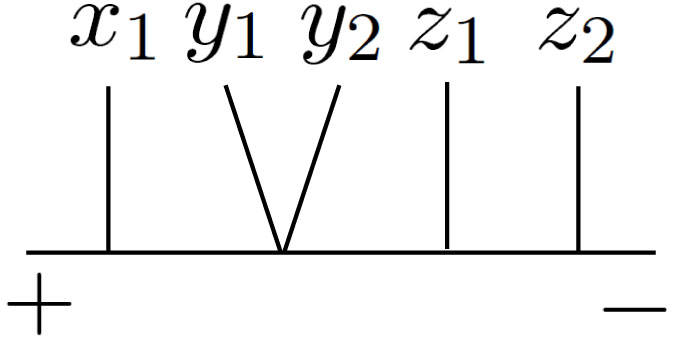} \end{minipage}\times\begin{minipage}{2.3cm}  \includegraphics[width=2.3cm]{NewEQW0EG2R1} \end{minipage}\Label{Eq:NewW0EG31}\nn
&=&\left(-{1\over 2}\right)\begin{minipage}{2.4cm}\includegraphics[width=2.4cm]{NewEQW0EG3L1} \end{minipage}\times\left(-{1\over 2}\right)\left[\,\begin{minipage}{2.45cm}  \includegraphics[width=2.45cm]{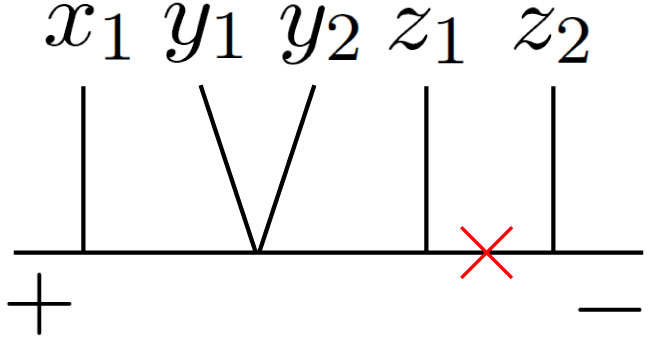} \end{minipage}-\begin{minipage}{2.45cm}  \includegraphics[width=2.45cm]{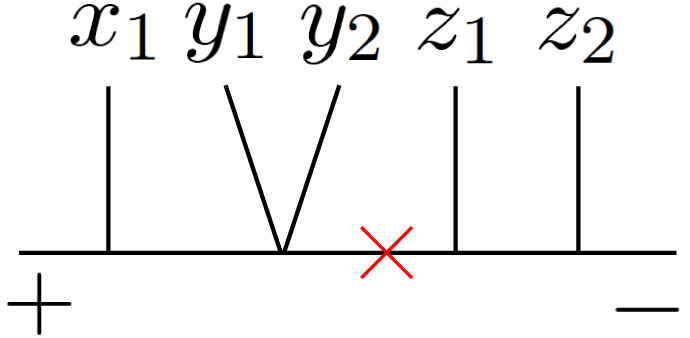} \end{minipage}\,\right]\times\begin{minipage}{2.3cm} \includegraphics[width=2.3cm]{NewEQW0EG2R2} \end{minipage}\nn
&&\xrightarrow {Localization-2}~~{1\over l^2}{1\over s_{x_1,l}}\left(-{1\over 2}\right)(\epsilon_{y_1}\cdot \epsilon_{y_2}){1\over s_{x_1y_1y_2,l}}\left(-{1\over 2}\right)(\W\epsilon_{z_1}\cdot\W\epsilon_{z_2}){1\over s_{x_1y_1y_2z_1z_2,l}},
\eea
where the property (\ref{Eq:Xpattern1}) of X-pattern has been applied. The first term on the second line is local, but the second is not. This remaining nonlocality acquires a further cancellation against a term which contains 
an X-pattern corresponding to the partition $\{x_1,z_1,y_1\text{-}y_2,z_2\}$  
\bea
&&\left(-{1\over 2}\right)\begin{minipage}{2.4cm}\includegraphics[width=2.4cm]{NewEQW0EG3L1} \end{minipage}\times \begin{minipage}{2.45cm}  \includegraphics[width=2.45cm]{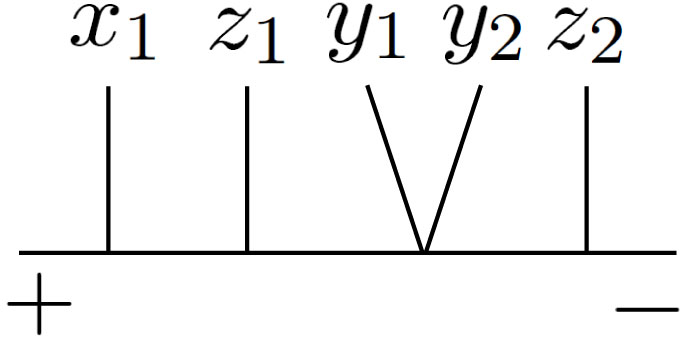} \end{minipage}\times\begin{minipage}{2.3cm}  \includegraphics[width=2.3cm]{NewEQW0EG2R3} \end{minipage}\nn
&=&\left(-{1\over 2}\right)\begin{minipage}{2.4cm}\includegraphics[width=2.4cm]{NewEQW0EG3L1} \end{minipage}\times\left(-{1\over 2}\right)\left[\,\begin{minipage}{2.45cm}  \includegraphics[width=2.45cm]{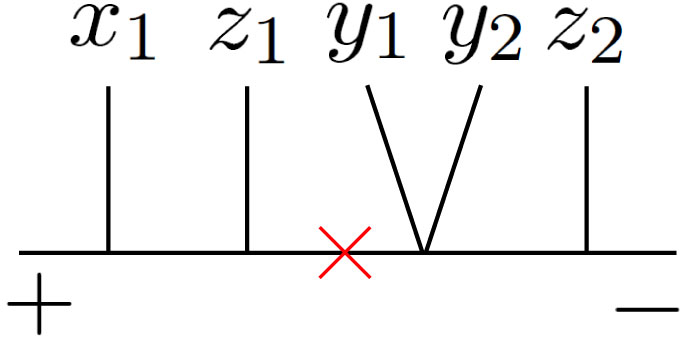} \end{minipage}-\begin{minipage}{2.45cm}  \includegraphics[width=2.45cm]{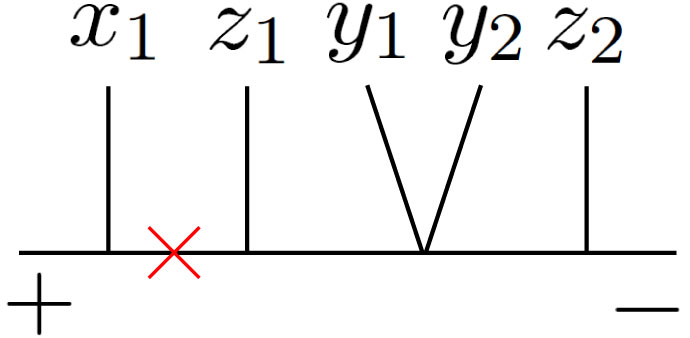} \end{minipage}\,\right]\times\begin{minipage}{2.3cm} \includegraphics[width=2.3cm]{NewEQW0EG2R2} \end{minipage}.
\eea
The second term further cancels with a term containing  a BCJ-pattern. Therefore, the only surviving term, i.e., the third line of $\mathrm{T}_4$ in (\ref{Eq:NewW0EG31}), provides a local expression
\bea
{1\over l^2}\,J_{(3)}[x_1]\,{1\over s_{x_1,l}}\,J_{(4)}[y_1,y_2]\,{1\over s_{x_1y_1y_2,l}}\,J_{(4)}[z_1,z_2]=\begin{minipage}{2.7cm}  \includegraphics[width=2.7cm]{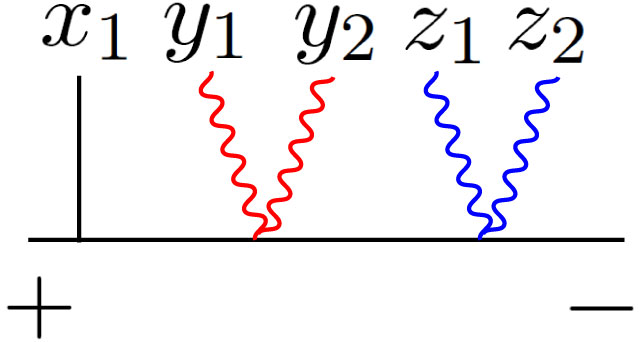} \end{minipage},\Label{Eq:NewFW0EG3D1}
\eea
where the vertices have already been defined in the previous examples.

{\bf General feature of localization-2}~~~The general feature of localization-2 is related to X-patterns and BCJ-patterns. Specifically, as pointed out in \cite{Xie:2024pro}, the nonlocalities associated with X- and BCJ-patterns must cancel with each other. After this cancellation, all terms of the BCJ-patterns 
have been canceled out, while most terms of X-patterns are also canceled. The only surviving terms have the following feature: {\it Assuming that the subgraph $\W{\mathcal{F}}_A$ contributes an X-pattern and the subgraph $\W{\mathcal{F}}_B$ above $\W{\mathcal{F}}_A$ involves the highest-weight node on this chain, the subsets $A$ and $B$ must be adjacent to each other in the LPFD, and the propagator between them is deleted. Since there exists a contraction between $\W{\mathcal{F}}_A$  and $\W{\mathcal{F}}_B$ {which} are not separated by any propagator, this remaining term is a local one.}

\subsection{More on X- and BCJ-patterns}

To eliminate the nonlocalities for more complicated situations, we generalize the X-patterns and the BCJ-patterns, as shown in \cite{Xie:2024pro}.

 We say a graph $\mathcal{F}$ has an {\it X-pattern} if this graph involves a factor $k_A\cdot X_A$, where $k^{\mu}_A$ denotes the total momentum of particles in the subgraph $\mathcal{F}_A$. The $X^{\mu}_A$ denotes the momentum of the propagator attached to the subcurrent corresponding to $A$ from left in the LPFD. Such structure occurs when we collect all graphs where a node in $\mathcal{F}_A$ is connected to those subgraphs, say  $\mathcal{F}_{B_1},...,\mathcal{F}_{B_i}$ corresponding to subsets leaving to the left of the subset $A$ in the LPFD:
\bea
&\!\!\!\!\!&\left[\begin{minipage}{3cm}\includegraphics[width=3cm]{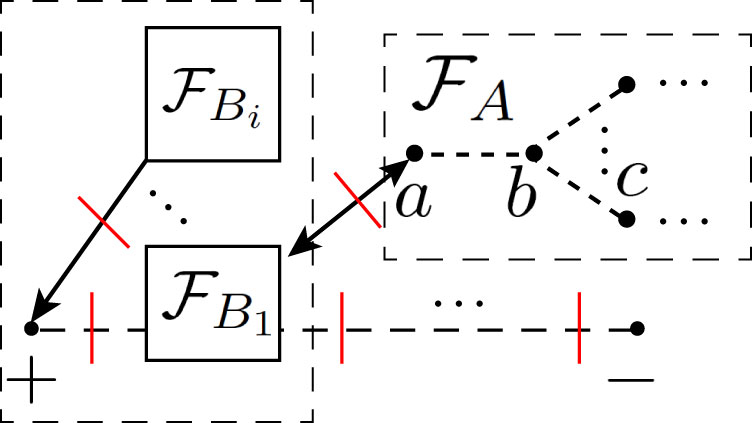} \end{minipage}\times \begin{minipage}{2.1cm}  \includegraphics[width=2.1cm]{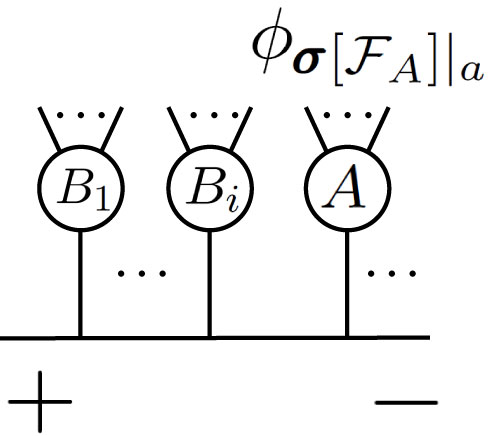} \end{minipage}+(-1)\begin{minipage}{3cm}\includegraphics[width=3cm]{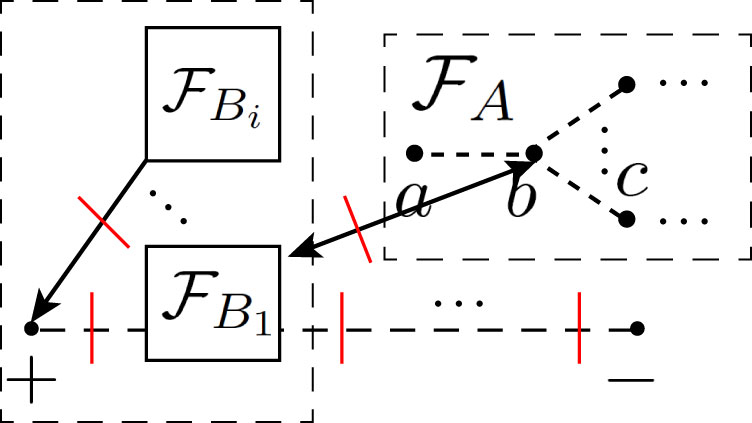} \end{minipage}\times \begin{minipage}{2.1cm}  \includegraphics[width=2.1cm]{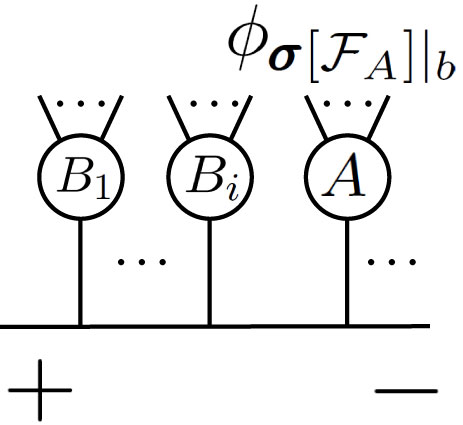} \end{minipage}\right.\nn
&\!\!\!\!\!&~~~~~~\left.+\begin{minipage}{3cm}\includegraphics[width=3cm]{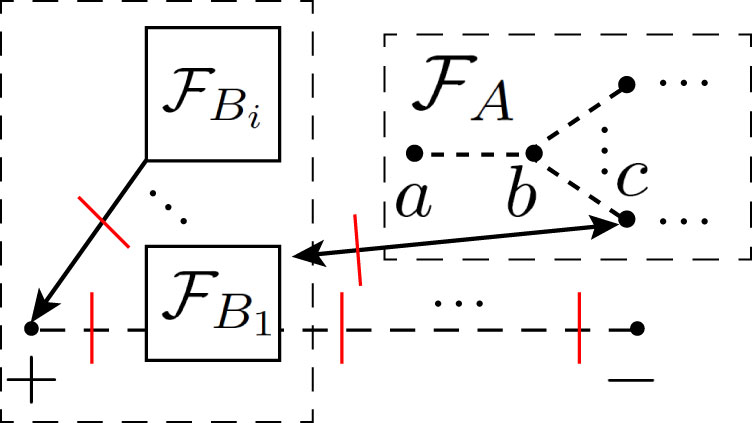} \end{minipage}\times \begin{minipage}{2.1cm}  \includegraphics[width=2.1cm]{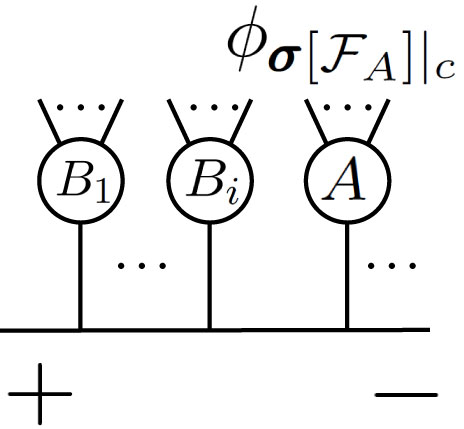} \end{minipage}+...\right]=(-k_{A}\cdot X_A)\begin{minipage}{3cm}\includegraphics[width=3cm]{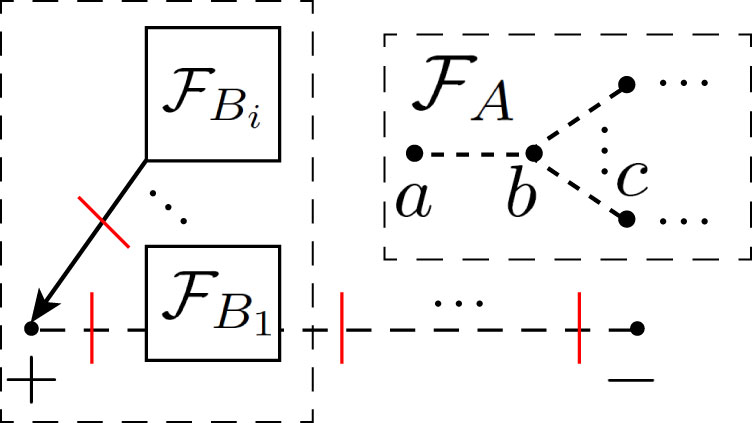} \end{minipage}\times \begin{minipage}{2.1cm}  \includegraphics[width=2.1cm]{GenXPatternD1a} \end{minipage}\!\!,
\eea
where the RHS produces a factor $k_{A}\cdot X_A=k_A\cdot (l+k_{B_1}+...+k_{B_i})$, agrees with the definition of X-pattern. As pointed out in \cite{Xie:2024pro}, once a graph involves an X-pattern, one can apply the following property 
\bea
k_A\cdot X_A={1\over2}\left[(k_A+ X_A)^2-X_A^2-k_A^2\right]={1\over2}\left[\big((k_A+ X_A)^2-l^2\big)-\big(X_A^2-l^2\big)-k_A^2\right]\,,\Label{Eq:XCoefficient}
\eea
which further implies the following identity
\bea
(-k_A\cdot X_A)\times\begin{minipage}{2.0cm}  \includegraphics[width=2.0cm]{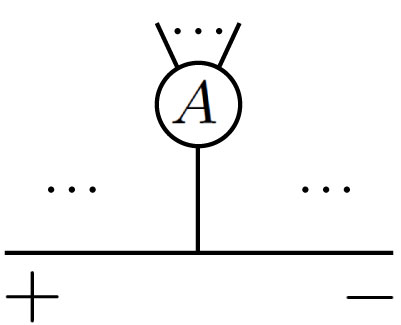} \end{minipage}=\left(-{1\over 2}\right)\left[\,\begin{minipage}{2.1cm}  \includegraphics[width=2.0cm]{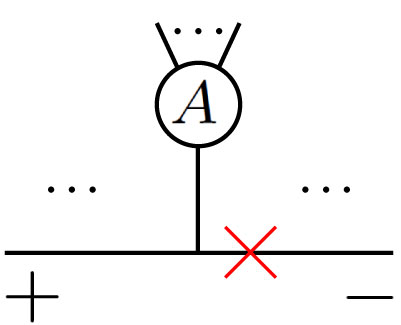} \end{minipage}-\begin{minipage}{2.0cm}  \includegraphics[width=2.0cm]{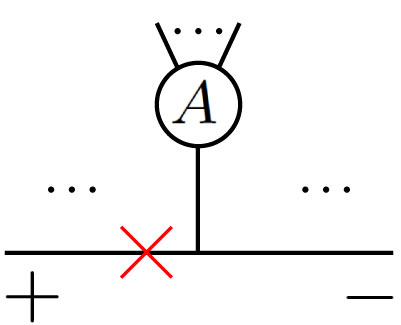} \end{minipage}-\begin{minipage}{2.0cm}  \includegraphics[width=2.0cm]{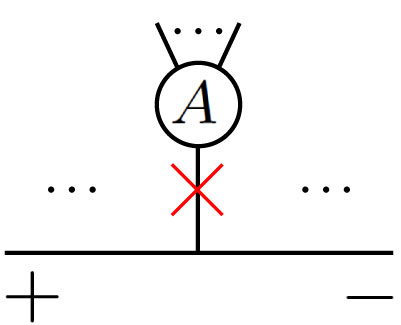} \end{minipage}\,\right].\Label{Eq:PropertyXPattern}
\eea
In the first term and the second term of the above property, the linear propagators attached to the subcurrent $A$ from left and right are deleted, respectively. In the third term, the propagator of the tree-level subcurrent is removed. Apparently, \eqref{Eq:Xpattern1} is the special case when $A$ involves only one on-shell element.

A general BCJ-pattern involves a subcurrent which can be expressed as the lhs. of the off-shell BCJ relation. According to (\ref{Eq:OffBCJ1}), it is reduced as  
\bea
&\!\!\!\!\!\!\!\!\!\!&\begin{minipage}{1.8cm}\includegraphics[width=1.8cm]{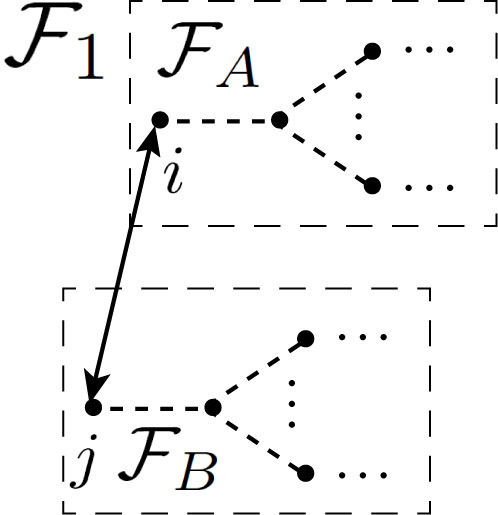} \end{minipage}\times \begin{minipage}{1cm}  \includegraphics[width=1cm]{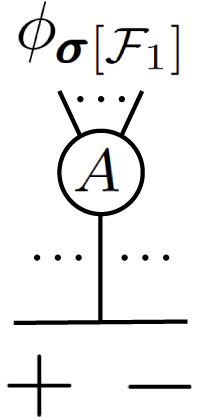} \end{minipage}+(-1)\begin{minipage}{1.8cm}\includegraphics[width=1.8cm]{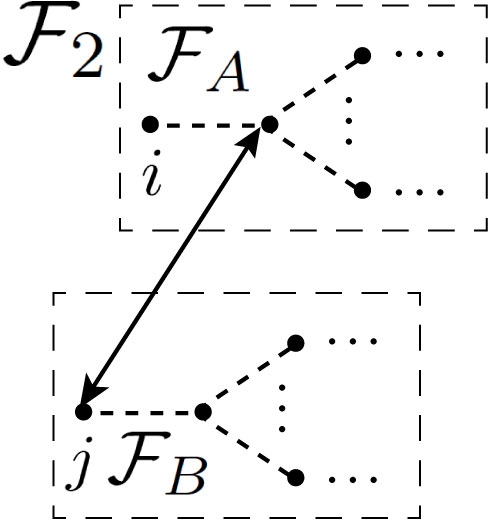} \end{minipage}\times \begin{minipage}{1cm}  \includegraphics[width=1cm]{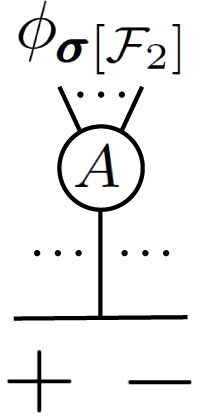} \end{minipage}+...+\begin{minipage}{1.8cm}\includegraphics[width=1.8cm]{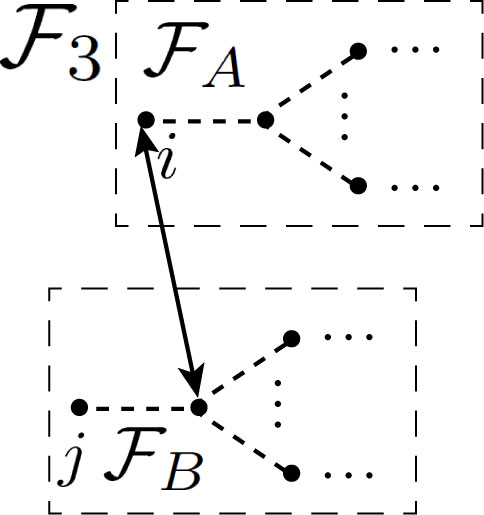} \end{minipage}\times \begin{minipage}{1cm}  \includegraphics[width=1cm]{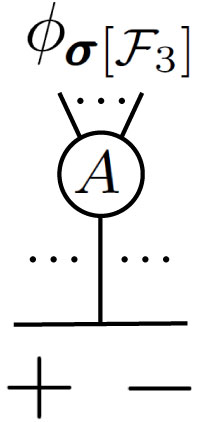} \end{minipage}+(-1)\begin{minipage}{1.8cm}\includegraphics[width=1.8cm]{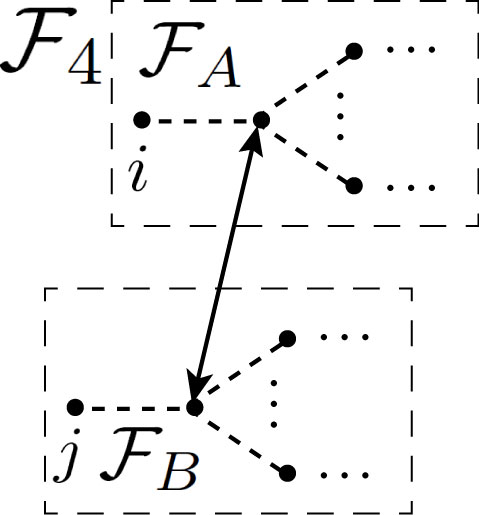} \end{minipage}\times \begin{minipage}{1cm}  \includegraphics[width=1cm]{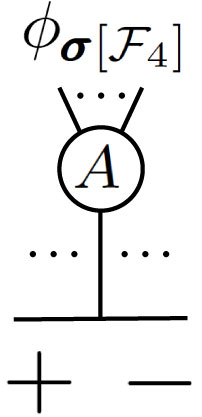} \end{minipage}+...\nn
&\!\!\!\!\!=\!\!\!\!\!&~~~{1\over 2}\left[\,\begin{minipage}{2.1cm}  \includegraphics[width=2.1cm]{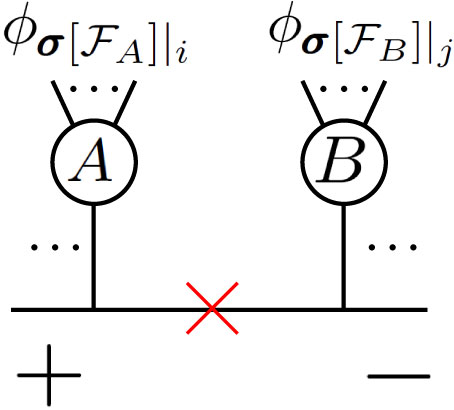} \end{minipage}-\begin{minipage}{2.1cm}  \includegraphics[width=2.1cm]{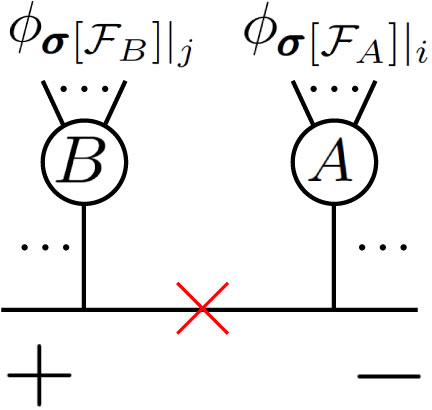} \end{minipage}\,\right],\Label{Eq:PropertyBCJPattern}
\eea
which is also expressed by deleting linear propagators in the LPFDs.

\subsection{More general subgraphs demonstrated by $\mathsf{X}=\{x_1\}$, $\mathsf{Y}=\{y_1,y_2\}$, $\mathsf{Z}=\{z_1,z_2,z_3\}$}\label{sec:W0VertexStructure}
In the previous examples, each subset in a partition involves only one element. However, a general partition may contain subsets with more than one particle. The resulting graphs in this case have more complicated structures according to the graphic rule. In this subsection, we show that the localization-1 and -2 are also effective in these more general situations, and the resulting vertex structures are directly obtained through replacing the polarizations by subcurrents. In the following, we clarify this point by example with  $\mathsf{X}=\{x_1\}$, $\mathsf{Y}=\{y_1,y_2\}$, $\mathsf{Z}=\{z_1,z_2,z_3\}$. The left and right permutations are respectively supposed to be $(x_1z_1z_2z_3)$ and $(x_1y_1y_2)$, while the left and right reference orders are fixed as $y_1\prec y_2$ and $z_1\prec z_2\prec z_3$.

{\bf Example-1}~~Let us consider the partition $\{x_1, \{y_1,y_2\},\{z_1,z_2\},z_3\}$ which contains two subsets $\{y_1,y_2\}$ and $\{z_1,z_2\}$ with more than one element. We just focus on the cases with the right factor $\W\epsilon_{z_1}\cdot\W \epsilon_{z_2}$. The first term under consideration is 
\bea
\begin{minipage}{3.2cm}  \includegraphics[width=3.2cm]{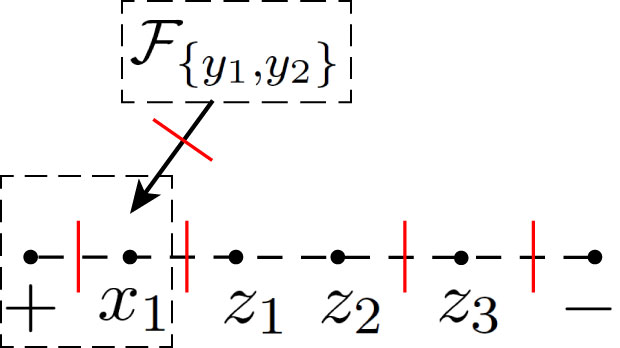} \end{minipage}\times\begin{minipage}{2.45cm}  \includegraphics[width=2.45cm]{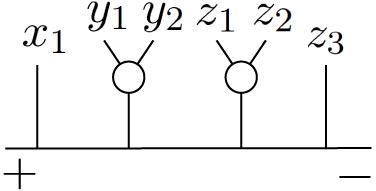} \end{minipage}\times\begin{minipage}{2.7cm} \includegraphics[width=2.7cm]{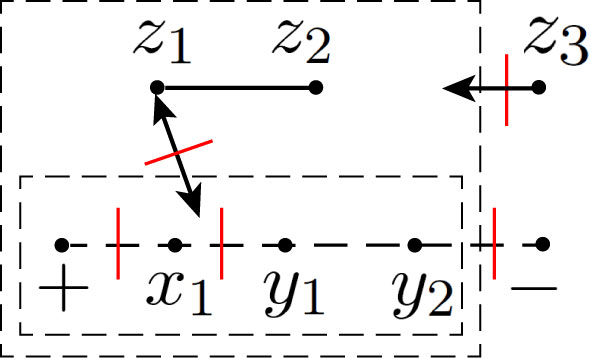} \end{minipage},\Label{Eq:W0EG41}
\eea
where $\mathcal{F}_{\{y_1,y_2\}}$ stands for the collection of all possible {\it left} subgraphs with nodes $y_1$ and $y_2$, recalling that the left graphs do not bring any nonlocality into the full expression. For the right graph, according to localization-1, any nonlocality associated with a factor $(\W\epsilon_{z_1}\cdot\W \epsilon_{z_2})(-k_{z_1}\cdot k_i)$ ($i=,x_1,y_1,y_2$) and that associated with $(\W\epsilon_{z_1}\cdot\W \epsilon_{z_2})(-k_{z_1}\cdot l)$ can be collected into a local expression, with a factor  $(\W\epsilon_{z_1}\cdot\W \epsilon_{z_2})(-k_{z_1}\cdot X_{\{z_1,z_2\}})=(\W\epsilon_{z_1}\cdot\W \epsilon_{z_2})(-k_{z_1}\cdot (l+k_{x_1}+k_{y_1}+k_{y_2}))$. Similarly for those with subgraph $z_3$. Therefore, the expression (\ref{Eq:W0EG41}) is  local and plays as a term 
of 
\bea
{1\over l^2}\,J_{(3)}[x_1]\,{1\over s_{x_1,l}}\,J_{(3)}[y_1,y_2]\,{1\over s_{x_1y_1y_2,l}}\,J_{(3)}[z_1,z_2]\,{1\over s_{x_1y_1y_2z_1z_2,l}}\,J_{(3)}[z_3]=\begin{minipage}{2.9cm}  \includegraphics[width=2.7cm]{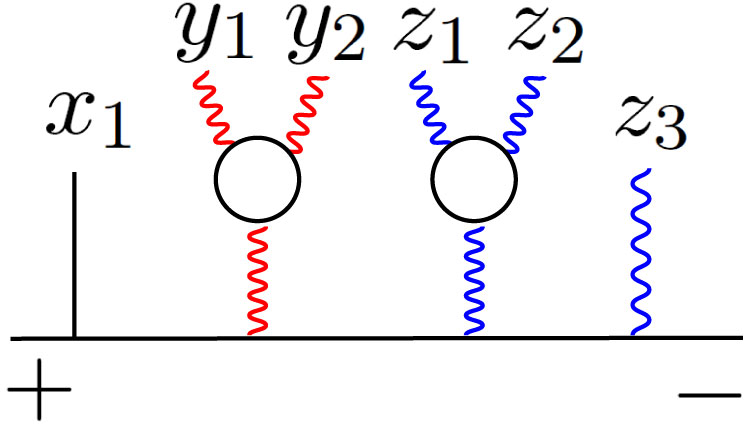} \end{minipage},\Label{Eq:NewFW0EG4D1}
\eea
in which the subcurrents 
\bea
T^{\mu}(y_1,y_2)&=&\Sl_{\mathcal{F}_{\{y_1,y_2\}}}\,\Sl_{\pmb{\sigma}[\mathcal{F}_{\{y_1,y_2\}}]}\mathcal{C}^{\mu}\big[\mathcal{F}_{\{y_1,y_2\}}\big]\,\phi_{\pmb{\sigma}[\mathcal{F}_{\{y_1,y_2\}}]|y_1y_2},\nn
T^{\W\mu}(z_1,z_2)&=&(\W\epsilon_{z_1}\cdot\W\epsilon_{z_2})(-k^{\mu}_{z_1})\phi_{z_1z_2|z_1z_2}
\eea 
with more than one element are attached to the loop line via a 2$x$-1$y$ vertex (\ref{Eq:2x1yVertex}) and a 2$x$-1$z$ vertex (\ref{Eq:2x1zVertex}), respectively.

{\bf Example-2}~~Another example is given by the following term
\bea
\mathrm{T}_1&=&\begin{minipage}{3.2cm}  \includegraphics[width=3.2cm]{EQW0EG3L1} \end{minipage}\times\begin{minipage}{2.45cm}  \includegraphics[width=2.45cm]{EQW0EG3D1} \end{minipage}\times\left[\,\begin{minipage}{2.7cm} \includegraphics[width=2.7cm]{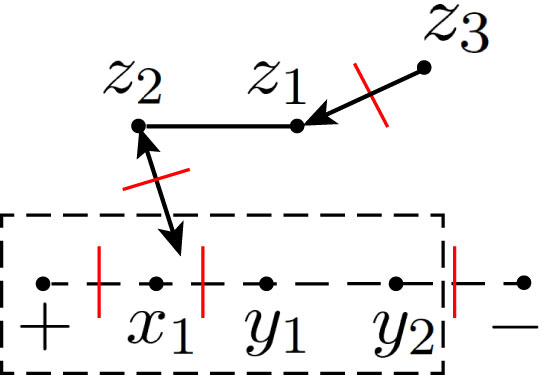} \end{minipage}-\begin{minipage}{2.7cm} \includegraphics[width=2.7cm]{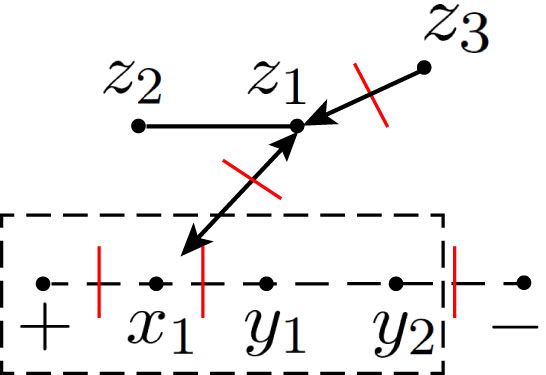} \end{minipage}\right]\nn
&=&\begin{minipage}{3.2cm}  \includegraphics[width=3.2cm]{EQW0EG3L1} \end{minipage}\times\left(-{1\over 2}\right)\left[\,\begin{minipage}{2.45cm}  \includegraphics[width=2.45cm]{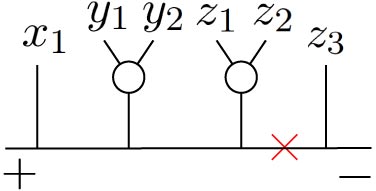} \end{minipage}-\begin{minipage}{2.45cm}  \includegraphics[width=2.45cm]{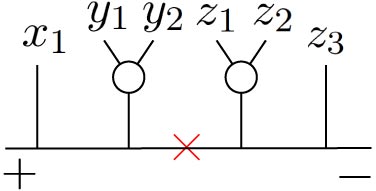} \end{minipage}-\begin{minipage}{2.45cm}  \includegraphics[width=2.45cm]{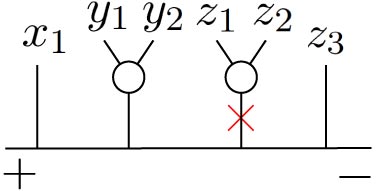} \end{minipage}\,\right]\times\begin{minipage}{2.5cm} \includegraphics[width=2.5cm]{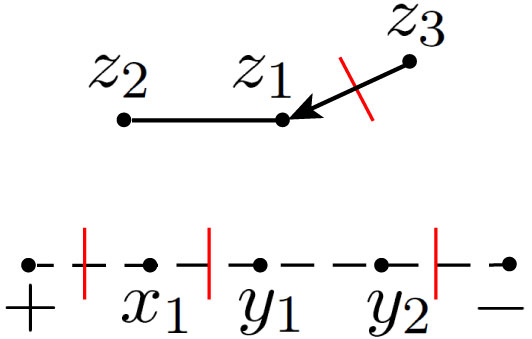} \end{minipage}\nn
&=&\mathrm{T}_{1\text{A}}-\mathrm{T}_{1\text{B}}-\mathrm{T}_{1\text{C}},\Label{Eq:W0EG51}
\eea
which contains a nonlocality associated with the subgraph $\mathcal{F}_{\{z_1,z_2\}}$. This nonlocality cannot be treated by localization-1; we have to consider localization-2 since an X-pattern is involved in the right graph. When the property (\ref{Eq:PropertyXPattern}) is applied, the first line of (\ref{Eq:W0EG51}) turns into the second.  Note that the right graph of the second term in the first line is, in fact, cannot be produced by the graphic rules; we subtract this spurious graph in order to complete the X-pattern. Indeed, the same spurious graph has already been added in (\ref{Eq:W0EG41}) to enable applying localization-1, so the spurious graphs leave no contribution for the full amplitude \footnote{A systematic discussion on spurious graphs can be found in \cite{Xie:2024pro}.}. The term $\mathrm{T}_{1\text{A}}$ is already local because the propagator between the subsets $\{z_1,z_2\}$ and $\{z_3\}$ has been removed. On the contrary, $\mathrm{T}_{1\text{B}}$ and $\mathrm{T}_{1\text{C}}$ still have nonlocality.
{\it According to the definition of Berends-Giele current (\ref{Eq:BScurrent}),  the Feynman diagram of $\mathrm{T}_{1\text{C}}$  further splits as}
\bea
\begin{minipage}{2.45cm}  \includegraphics[width=2.45cm]{EQW0EG3D4} \end{minipage}=\begin{minipage}{2.45cm}  \includegraphics[width=2.45cm]{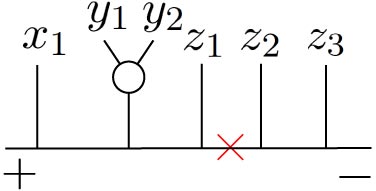} \end{minipage}-\begin{minipage}{2.45cm}  \includegraphics[width=2.45cm]{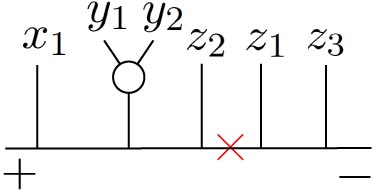} \end{minipage},\Label{Eq:SplitT1C}
\eea
in which the second term vanishes due to the left permutation $(x_1z_1z_2z_3)$.

According to localization-2, we need to look for other terms with X- and BCJ-patterns that cancel with (\ref{Eq:W0EG51}). One term with a BCJ-pattern is the following
\bea
\mathrm{T}_2&=&\begin{minipage}{3.2cm}  \includegraphics[width=3.2cm]{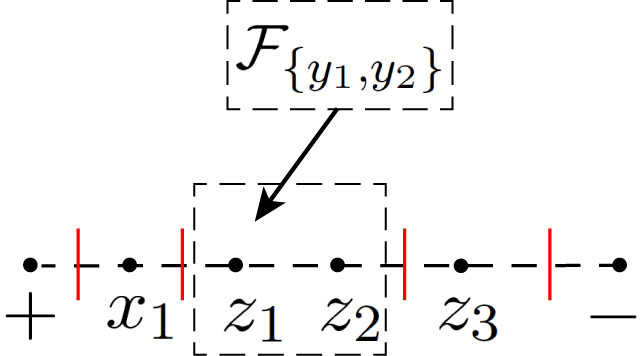} \end{minipage}\times\begin{minipage}{2.45cm}  \includegraphics[width=2.45cm]{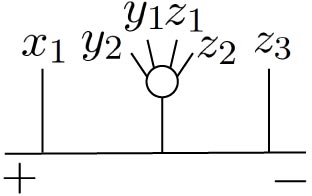} \end{minipage}\times\begin{minipage}{2.7cm} \includegraphics[width=2.7cm]{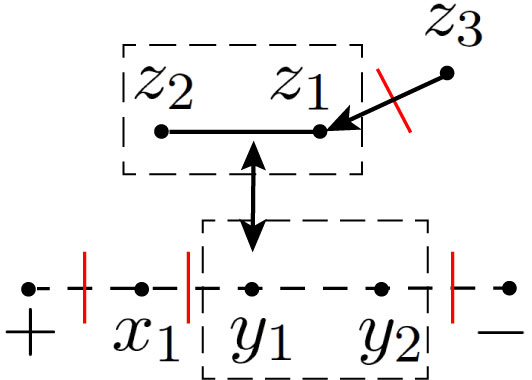} \end{minipage}\nn
&=&\begin{minipage}{3.2cm}  \includegraphics[width=3.2cm]{EQW0EG3L2} \end{minipage}\times\left(-{1\over 2}\right)\left[\,\begin{minipage}{2.45cm}  \includegraphics[width=2.45cm]{EQW0EG3D3} \end{minipage}-\begin{minipage}{2.45cm}  \includegraphics[width=2.45cm]{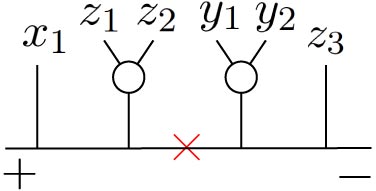} \end{minipage}\,\right]\times\begin{minipage}{2.7cm} \includegraphics[width=2.7cm]{EQW0EG3R3} \end{minipage}\nn
&=&\mathrm{T}_{2\text{A}}-\mathrm{T}_{2\text{B}},
\eea
where we have applied the property (\ref{Eq:PropertyBCJPattern}) of BCJ-pattern by subtracting corresponding spurious graphs implicitly contained in the right graph on the first line. Term $\mathrm{T}_{2\text{A}}$ equals to $0$ since the left permutation of elements in the subset $\{y_1,y_2,z_1,z_2\}$ is of the form $(z_1...)$.
Another term $\mathrm{T}_3$ contains an X-pattern
\bea
\mathrm{T}_3&=&\begin{minipage}{3.2cm}  \includegraphics[width=3.2cm]{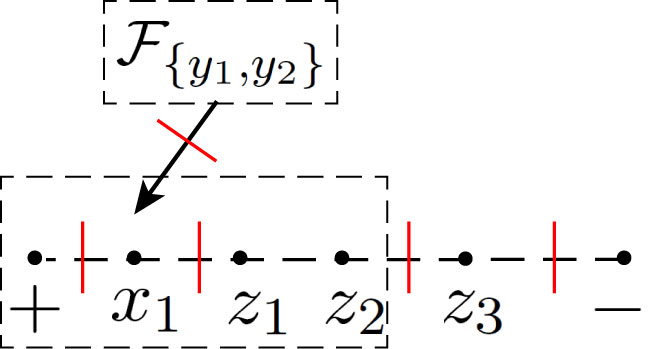} \end{minipage}\times\begin{minipage}{2.45cm}  \includegraphics[width=2.45cm]{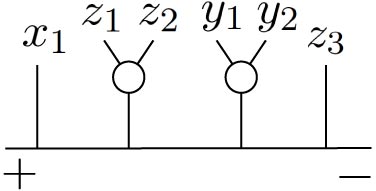} \end{minipage}\times\begin{minipage}{2.7cm} \includegraphics[width=2.7cm]{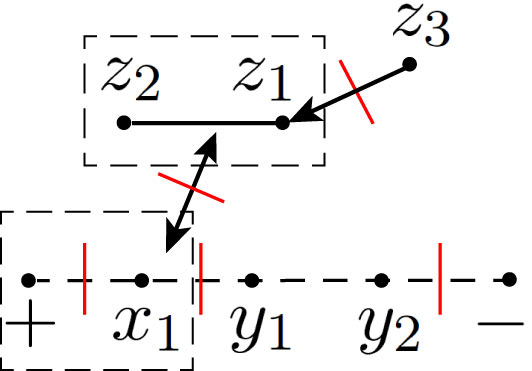} \end{minipage}\nn
&=&\begin{minipage}{3.2cm}  \includegraphics[width=3.2cm]{EQW0EG3L3} \end{minipage}\times\left(-{1\over 2}\right)\left[\,\begin{minipage}{2.45cm}  \includegraphics[width=2.45cm]{EQW0EG3D30} \end{minipage}-\begin{minipage}{2.45cm}  \includegraphics[width=2.45cm]{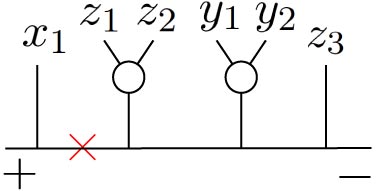} \end{minipage}-\begin{minipage}{2.45cm}  \includegraphics[width=2.45cm]{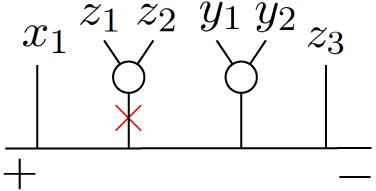} \end{minipage}\,\right]\times\begin{minipage}{2.5cm} \includegraphics[width=2.5cm]{EQW0EG3R3} \end{minipage}\nn
&=&\mathrm{T}_{3\text{A}}-\mathrm{T}_{3\text{B}}-\mathrm{T}_{3\text{C}}.
\eea
Apparently, $\mathrm{T}_{3\text{A}}=\mathrm{T}_{1\text{B}}+\mathrm{T}_{2\text{B}}$, each of  $\mathrm{T}_{1\text{C}}$ and $\mathrm{T}_{3\text{C}}$ cancels with another X-pattern which comes from a further division of the subset $\{z_1,z_2\}$. After cancellation, the only surviving term in $\mathrm{T}_1$, $\mathrm{T}_2$ and $\mathrm{T}_3$ is the $\mathrm{T}_{1\text{A}}$ which is local and has the explicit expression
\bea
&&{1\over l^2}{1\over s_{x_1,l}}\left[\mathcal{C}[\mathcal{F}_{\{y_1,y_2\}}]\cdot X_{\{y_1,y_2\}}\phi_{\pmb\alpha[\mathcal{F}_{\{y_1,y_2\}}]|y_1y_2}\right]{1\over s_{x_1y_1y_2,l}}\left[\left(-{1\over 2}\right)\left(\W\epsilon_{z_2}\cdot \W\epsilon_{z_1}\right)\left(-\W\epsilon_{z_3}\cdot k_{z_1}\right)\phi_{z_1z_2|z_1z_2}\right].
\eea
This just contributes to the following local result
\bea
&&{1\over l^2}\,J_{(3)}[x_1]\,{1\over s_{x_1,l}}\,J_{(3)}[y_1,y_2]\,{1\over s_{x_1y_1y_2,l}}\,J_{(4)}[z_1,z_2,z_3]=\begin{minipage}{2.9cm}  \includegraphics[width=2.7cm]{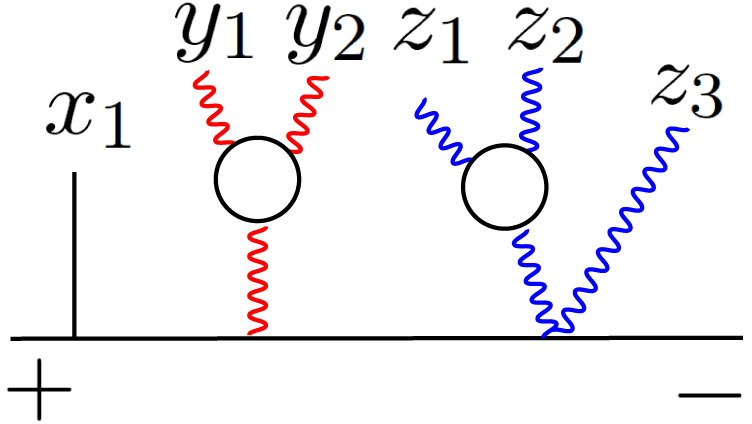} \end{minipage},\Label{Eq:NewFW0EG5D1}
\eea
where the subcurrent $T^{\mu}(y_1,y_2)$ is contracted with a $2x$-$1y$ vertex, while the subcurrents $T^{\W\mu}(z_1,z_2)$ and $T^{\mu}(z_3)$ are attached to a $2x$-$2z$ vertex.

{\bf Comments on the general feature~~} From this example, we see that the cancellation of nonlocalities associated with subgraphs containing more nodes also follows localization-1 and -2. The local result has the general feature where a single external particle is replaced by a {\it multi-particle effective subcurrent} of the same type. Note that the relative reference order between subcurrents {(i.e., the relative order of the highest-weight nodes of the subcurrents in the reference order)} must be preserved in this replacement.

\subsection{General local formula for partial and full integrands with $|\mathsf{W}|=0$}
According to the previous examples and comments, a general local partial integrand $\mathcal{A}^{\,\text{dYMS},\,|\mathsf{W}|=0}$ has the following form
\bea
\mathcal{A}^{\,\text{dYMS},\,|\mathsf{W}|=0}=\,\Sl_{\{A_1A_2...A_I\}}{1\over l^2}\,J[A_1]\, {1\over s_{A_1,l}}\,J[A_2]\cdots\,{1\over s_{A_1...A_{I-1},l}}\,J[A_I], \Label{Eq:doubleYMSW0Partial}
\eea
where {$\{A_1A_2...A_I\}$ are possible partitions that preserve both permutations of left and right scalars}. Each $J[A_i]$ ($i=1,...,I$) is defined as follows:
\begin{itemize} 

\item If $A_i$ contains only elements in $\mathsf{Y}$ or $\mathsf{Z}$,
\bea
J[A_i]&=&J_{(3)}[A_i]+J_{(4)}[A_i]\nn
&=&{T}(A_i)\cdot X_{A_i}+\left(-{1\over 2}\right)\left[\Sl_{\{A_{i1},A_{i2}\}}T(A_{i1})\cdot  T (A_{i2})\right],\Label{Eq:doubleYMSW0Final1}
\eea
where $A_i=\mathsf{Y}_i$ or $A_i=\mathsf{Z}_i$. The subcurrents $T^{\mu}(\mathsf{Y}_i)$ and $T^{\W\mu}(\mathsf{Z}_i)$ are the effective currents respectively defined by 
\bea
T^{\mu} (\mathsf{Y}_i)&=&\Sl_{\mathcal{F}_{i}}\Sl_{\pmb{\sigma}[\mathcal{F}_{i}]}\mathcal{C}^{\mu}[\mathcal{F}_{i}]\phi_{\pmb{\sigma}[\mathcal{F}_{i}]|\pmb{\sigma}[\W{\mathcal{F}}_{i}] }\,,\nn
T^{\W\mu} (\mathsf{Z}_i)&=&\Sl_{\W{\mathcal{F}}_{i}}\Sl_{\pmb{\sigma}[\W{\mathcal{F}}_{i}]}\phi_{\pmb\sigma[\mathcal{F}_{i}]|\pmb{\sigma}[\W{\mathcal{F}}_{i}]} \mathcal{C}^{\W\mu}[\W{\mathcal{F}}_{i}], \Label{Eq:doubleYMSW0Final2}
\eea
 where the expression for $T^{\mu}(\mathsf{Y}_i)$ is the same as that of the effective current $T^{\mu}(A_i)$ given in (\ref{Eq:EffectiveCurrent}), and $T^{\W\mu}(\mathsf{Z}_i)$ can be obtained from $T^{\mu}(A_i)$ by substituting the left graph and left Lorentz index with their right counterparts. Thus, the local structure $J[A_i]$ in (\ref{Eq:doubleYMSW0Final1}) is either the same as or the right dual of that in (\ref{Eq:GenResultNewJ2}).


\item For other cases, $A_i$ in general can be expressed as $\mathsf{X}_i\cup\mathsf{Y}_i\cup \mathsf{Z}_i$, and $J[A_i]$ is given by 
\bea
J[A_i]&=&J_{(3)}[A_i]=\Sl_{\mathcal{F}_i,\W{\mathcal{F}}_i}\Sl_{\pmb{\sigma}[\mathcal{F}_i],\pmb{\sigma}[\W{\mathcal{F}}_i]}\mathcal{C}\left[\mathcal{F}_i\right]\phi_{\pmb{\sigma}[\mathcal{F}_i]|\pmb{\sigma}[\W{\mathcal{F}}_i]}\mathcal{C}[\W{\mathcal{F}}_i],\Label{Eq:doubleYMSW0PartialFinal3}
\eea
where $\mathcal{F}_i$ and  $\W{\mathcal{F}}_i$ respectively denote the graphs in which $\mathsf{X}_i\cup \mathsf{Z}_i$ elements and $\mathsf{X}_i\cup \mathsf{Y}_i$ elements play as the roots. The above expression for $J_{(3)}[A_i]$ reduces to $T(A_i)$ in (\ref{Eq:GenResultNewJ1}) and (\ref{Eq:EffectiveCurrent}) when $\mathsf{Z}_i$ is empty. 

\end{itemize}
For any local partial integrand, one can always find other partial integrands which are related by cyclic permutations in (\ref{Eq:doubleYMSIntegrand}). As a result of (\ref{Eq:partial}), the partial integrand (\ref{Eq:doubleYMSW0Partial}) induces a full integrand {expressed by} quadratic propagators
\bea
\mathcal{I}\cong\,\Sl_{\{A_1A_2...A_I\}}{1\over l^2}\,J[A_1]\, {1\over l^2_{A_1}}\,J[A_2]\cdots\,{1\over l^2_{A_1...A_{I-1}}}\,J[A_I]. \Label{Eq:doubleYMSW0Final}
\eea

Now let us summarize helpful observations on calculations of dYMS partial integrands with $|\mathsf{W}|=0$:
\begin{itemize}
 \item [(i)] Local vertex structures directly arise from partitions in which each subset contains only one element, through localization-1 and -2.
 
\item  [(ii)] More complicated structures, where nontrivial effective currents are attached to vertices, are obtained via replacing a single particle by a {multi-particle} effective current in a proper way. 
 
\item  [(iii)] Once a local LPFD is obtained, one can also find terms related by cyclic permutations, hence the local partial integrands with linear propagators sum into a local full integrand with quadratic propagators.

\end{itemize}

To study the localization of partial integrands with more $\mathsf{W}$ elements, we may encounter new types of effective currents. The next section introduces a general discussion of effective currents, which will be helpful in the remaining sections.

\section{General effective currents and double-copy structure inside}\label{sec:5}
As shown in the final local results (\ref{Eq:doubleYMSW0Partial}) and (\ref{Eq:doubleYMSW0Final}) of $|\mathsf{W}|=0$, for a given partition, each subset $A_i$ is associated with $J[A_i]$ which is further expressed by contracting effective currents (\ref{Eq:doubleYMSW0Final2}) and (\ref{Eq:doubleYMSW0PartialFinal3}) to a three-point or four-point vertex. 
In general, when we study integrands with at least one $\mathsf{W}$ element, the local results are also expressed via $J[A_i]$, but more complicated vertices and more complicated effective currents arise when the number of $\mathsf{W}$ elements increases. In this section, we extend the effective currents (\ref{Eq:doubleYMSW0Final2})  and (\ref{Eq:doubleYMSW0PartialFinal3}) to more generic structures that have double-copy structures and will be encountered in the coming discussions. The convention of notations for Lorentz contractions of these effective currents is also established.

A general {\it effective current is expressed by a combination of BS currents, where the coefficients and permutations are defined by the left and/or right graphs.}  According to different particle species in the particle set, the effective currents can be classified as follows 
\bea
T^{\mu \W\mu}\left(A^{\text{YM}\otimes\text{YM}}_i\right)&\equiv& \Sl_{\mathcal{F}^{\text{YM}},\,\W{\mathcal{F}}^{\text{YM}}}\,\Sl_{\pmb{\sigma}[\mathcal{F}^{\text{YM}}],\,\pmb{\sigma}[\W{\mathcal{F}}^{\text{YM}}]} \mathcal{C}^{\mu}\big[\mathcal{F}^{\text{YM}}\big]\,\phi_{\pmb{\sigma}[\mathcal{F}^{\text{YM}}]\,\big|\,\pmb{\sigma}[\W{\mathcal{F}}^{\text{YM}}]}\,\mathcal{C}^{\W\mu}\big[\W{\mathcal{F}}^{\text{YM}}\big],\Label{Eq:J1}\\
T^{\mu}\left(A^{\text{YM}\otimes\text{YMS}}_i\right)&\equiv& \Sl_{\mathcal{F}^{\text{YM}},\,\W{\mathcal{F}}^{\text{YMS}}}\,\Sl_{\pmb{\sigma}[\mathcal{F}^{\text{YM}}],\,\pmb{\sigma}[\W{\mathcal{F}}^{\text{YMS}}]} \mathcal{C}^{\mu}\big[\mathcal{F}^{\text{YM}}\big]\,\phi_{\pmb{\sigma}[\mathcal{F}^{\text{YM}}]\,\big|\,\pmb{\sigma}[\W{\mathcal{F}}^{\text{YMS}}]}\,\mathcal{C}\big[\W{\mathcal{F}}^{\text{YMS}}\big],\Label{Eq:J2}\\
T^{\W\mu}\left(A^{\text{YMS}\otimes\text{YM}}_i\right)&\equiv& \Sl_{\mathcal{F}^{\text{YMS}},\,\W{\mathcal{F}}^{\text{YM}}}\,\Sl_{\pmb{\sigma}[\mathcal{F}^{\text{YMS}}],\,\pmb{\sigma}[\W{\mathcal{F}}^{\text{YM}}]} \mathcal{C}\big[\mathcal{F}^{\text{YMS}}\big]\,\phi_{\pmb{\sigma}[\mathcal{F}^{\text{YMS}}]\,\big|\,\pmb{\sigma}[\W{\mathcal{F}}^{\text{YM}}]}\,\mathcal{C}^{\W\mu}\big[\W{\mathcal{F}}^{\text{YM}}\big],\Label{Eq:J3}\\
T\left(A^{\text{YMS}\otimes\text{YMS}}_i\right)&\equiv& \Sl_{\mathcal{F}^{\text{YMS}},\,\W{\mathcal{F}}^{\text{YMS}}}\,\Sl_{\pmb{\sigma}[\mathcal{F}^{\text{YMS}}],\,\pmb{\sigma}[\W{\mathcal{F}}^{\text{YMS}}]} \mathcal{C}\big[\mathcal{F}^{\text{YMS}}\big]\,\phi_{\pmb{\sigma}[\mathcal{F}^{\text{YMS}}]\,\big|\,\pmb{\sigma}[\W{\mathcal{F}}^{\text{YMS}}]}\,\mathcal{C}\big[\W{\mathcal{F}}^{\text{YMS}}\big],\Label{Eq:J4}
\eea
where $\mathcal{F}_{A_i}$ and $\W{\mathcal{F}}_{A_i}$ are abbreviated as $\mathcal{F}$ and $\W{\mathcal{F}}$, respectively. In each of the above expressions, we summed over graphs  $\mathcal{F}$ and $\W{\mathcal{F}}$ which are generated by the graphic rule for the left and right half  integrands in (\ref{Eq:ILR}). The permutations  $\pmb{\sigma}[\mathcal{F}]$ and $\pmb{\sigma}[\W{\mathcal{F}}]$ defined by $\mathcal{F}$ and $\W{\mathcal{F}}$ were also summed over. According to the left (or right) half integrands, graphs $\mathcal{F}$ (or $\W{\mathcal{F}}$) are classified into the following two categories:
\begin{itemize}
\item [(i)] {\it The graph $\mathcal{F}$ (or $\W{\mathcal{F}}$) contains only gluons in the left (or right) half integrand.} The graphs are denoted by $\mathcal{F}^{\text{YM}}$ (or $\W{\mathcal{F}}^{\text{YM}}$) and the particles come from $\mathsf{Y}\cup \mathsf{W}$ (or $\mathsf{Z}\cup \mathsf{W}$). The coefficient defined by the graph has one left (or right) Lorentz index $\mu$ (or $\W\mu$).

\item [(ii)] {\it The graph $\mathcal{F}$ (or $\W{\mathcal{F}}$) contains scalars in the left (or right) half integrand.} The graphs are denoted by $\mathcal{F}^{\text{YMS}}$ (or $\W{\mathcal{F}}^{\text{YMS}}$), while the scalars and gluons come from $\mathsf{X}\cup \mathsf{Z}$ ($\mathsf{X}\cup\mathsf{Y}$) and  $\mathsf{Y}\cup \mathsf{W}$ ($\mathsf{Z}\cup \mathsf{W}$), respectively. The coefficient defined by such a graph does not have a Lorentz index.
\end{itemize}
 When both left and right graphs are taken into account, we arrive at four types of possible effective currents carrying different Lorentz indices, as shown by (\ref{Eq:J1})-(\ref{Eq:J4}). Each current in (\ref{Eq:doubleYMSW0Final2}), which only has a left or a right Lorentz index, is the boundary case of (\ref{Eq:J2}) or (\ref{Eq:J3}) when the YMS part only contains scalars. The current (\ref{Eq:doubleYMSW0PartialFinal3}) agrees with (\ref{Eq:J4}). In the special case when $A_i$ contains a single particle, (\ref{Eq:J1})-(\ref{Eq:J4}) are just the single particle external lines
 \bea
 T^{\mu \W\mu}\left(\{w_i\}^{\text{YM}\otimes\text{YM}}\right)&=&\epsilon^{\mu}_{w_i}\phi_{w_i|w_i}\epsilon^{\W\mu}_{w_i}, ~~~~T^{\mu}\left(\{y_i\}^{\text{YM}\otimes\text{YMS}}\right)=\epsilon^{\mu}_{y_i}\phi_{y_i|y_i},\nn
 T^{\W\mu}\left(\{z_i\}^{\text{YMS}\otimes\text{YM}}\right)&=&\phi_{z_i|z_i}\W\epsilon^{\W\mu}_{z_i},~~~~~~~~~\,T\left(\{x_i\}^{\text{YMS}\otimes\text{YMS}}\right)=\phi_{x_i|x_i}.
 \eea
Notice that the single particle BS current $\phi_{a|a}$ gives a trivial factor $1$.

The expressions of effective currents (\ref{Eq:J1})-(\ref{Eq:J4}) possess a significant {\it double-copy structure}: Each effective current is a combination of BS currents. The combination coefficients are factorized into two copy coefficients, which correspond to left and right graphs depending on theories (YM or YMS). When the possible left (right) graphs contributing to the same left (right) permutations are collected, the expressions (\ref{Eq:J1})-(\ref{Eq:J4}) are rearranged as 
\bea
T\left(A^{\mathsf{t}\otimes\W{\mathsf{t}}}\right)&\equiv& \Sl_{\pmb{\sigma},\,\pmb{\rho}}\,{N}^{\,\mathsf{t}}_{\pmb{\sigma}}\,\phi_{\pmb{\sigma}\,|\,\pmb{\rho}}\,{N}^{\,\W{\mathsf{t}}}_{\pmb{\rho}},\Label{Eq:DoubleCopy}
\eea
where the superscripts $\mathsf{t}$ and $\W{\mathsf{t}}$ denote theories, i.e., YM and/or YMS, $\pmb{\rho}$ and $\pmb{\sigma}$ are possible permutations on the left and the right sides. The ${N}^{\,\mathsf{t}}_{\pmb{\sigma}}$ (${N}^{\,\W{\mathsf{t}}}_{\pmb{\rho}}$) refers to off-shell numerator for the permutation $\pmb{\sigma}$ ($\pmb{\rho}$) of external particles
\bea
{N}^{\,\mathsf{t}}_{\pmb{\sigma}}=\Sl_{\mathcal{F}^{\,\mathsf{t}}_{\pmb{\sigma}}} \mathcal{C}\big[\mathcal{F}^{\,\mathsf{t}}_{\pmb{\sigma}}\big],~~~~~{N}^{\,\W{\mathsf{t}}}_{\pmb{\rho}}=\Sl_{\W{\mathcal{F}}^{\,\W{\mathsf{t}}}_{\pmb{\rho}}} \mathcal{C}\big[\W{\mathcal{F}}^{\,\W{\mathsf{t}}}_{\pmb{\rho}}\big].\Label{Eq:Numerators}
\eea
In the above expressions, $\mathcal{F}^{\,\mathsf{t}}_{\pmb{\sigma}}$ and $\W{\mathcal{F}}^{\,\W{\mathsf{t}}}_{\pmb{\rho}}$ are the graphs which are consistent with the permutations $\pmb\sigma$ and $\pmb\rho$, respectively. Possible Lorentz indices are implied by the theories $\mathsf{t}$ and $\W{\mathsf{t}}$. The off-shell numerators provided in (\ref{Eq:Numerators}) are closely related to the BCJ numerators \cite{Britto:2004ap,Britto:2005fq} at tree-level.

 Now let us clarify the relationship between the double-copy expression (\ref{Eq:Numerators}) for {\it tree-level effective currents} and the standard BCJ double copy \cite{Bern:2008qj,Bern:2010ue} for {\it tree amplitudes}. In a standard BCJ double-copy formula, the color-dressed YM tree amplitude is written as a sum over trivalent diagrams. For each diagram, the denominator is given by the inverse of propagators, while the numerator involves a color factor and a kinematic factor. These kinematic factors (BCJ numerators) are required to satisfy the same Jacobi identities with the corresponding color ones. As pointed out in \cite{Britto:2004ap,Britto:2005fq},  once such kinematic factors of YM have been obtained,  one can replace the color factors by another copy of such kinematic factors to get a GR amplitude. A deformed form of the BCJ double copy is the following. By applying the Jacobi identities on both copies of BCJ numerators, one can express a tree-level GR amplitude as a combination of tree-level BS amplitudes (propagator matrices) associated with two copies of BCJ numerators corresponding to half ladder diagrams. These half-ladder BCJ numerators serve as a basis of tree-level BCJ numerators. The double-copy formula (\ref{Eq:Numerators}) for effective currents already has an expression in terms of BS currents. Thus, in the on-shell limit, it expresses an on-shell (GR if the theories on both sides are YM) amplitude in terms of BS amplitudes, with two copies of kinematic coefficients. As demonstrated in \cite{Wu:2021exa}, this is equivalent\footnote{More precisely, upto certain U(1)-decoupling identities, symmetrizations over gravitons and an arrangement into nested commutators. Detail can be found in \cite{Wu:2021exa}.} to the deformed BCJ double copy with half ladder BCJ numerators.
 
At one-loop level, the integrands derived in the previous section and in the coming sections are generally expressed as a sum of terms which are constructed by inserting the double-copy effective currents to the (quadratic propagator) loop via local vertices. Such an expression of one-loop integrand serves as an intermediate form between (i) those  straightforwardly generated from tree amplitudes via forward limit \cite{He:2016mzd,He:2017spx,Geyer:2017ela}, which contains linear propagators, and (ii) the expected one-loop BCJ form  which results in a full expansion formula in terms of quadratic propagator expressed BS integrands (i.e., the one-loop version of deformed BCJ form).

 {\it Recursion for the effective currents}~~Since the effective subcurrents are expressed by the graphic rule which is a result of the recursive expansion formula \cite{Fu:2017uzt}, these currents can naturally be expressed recursively: Each effective current is given by a combination of those with fewer (left or right) gluons but more scalars. More explicitly, assuming $a$ is the highest-weight gluon in the left part of $\mathsf{A}_i$, we have 
\bea
T^{\mu}\left(A_i^{\text{YM}\otimes\,\W{\mathsf{t}}}\right)&\equiv& T^{\mu}\left(\emptyset||\mathsf{G}_i\left|\pmb\rho_i||\W{\mathsf{G}}_i\right.\right)\nn
&=&\Sl_{\{a,i_1,...,i_j\}\in\mathsf{G}}\left(\epsilon_a\cdot F_{i_1}\cdot...\cdot F_{i_j}\right)^{\mu}\underbrace{T\left(i_j,i_{j-1},...,i_1,a||\mathsf{G}_i\setminus\{i_j,i_{j-1},...,i_1,a\}\left|\pmb\rho_i||\W{\mathsf{G}}_i\right.\right)}_{T\left(A_i^{\text{YMS}\otimes\,\W{\mathsf{t}}}\right)},\Label{Eq:RecursionCurrent1}\\
T\left(A_i^{\text{YMS}\otimes\,\W{\mathsf{t}}}\right)&\equiv& T\left(\pmb\sigma_i||\mathsf{G}_i\left|\pmb\rho_i||\W{\mathsf{G}}_i\right.\right)\nn
&=&\Sl_{\{a,i_1,...,i_j\}\in\mathsf{G}}\left(\epsilon_a\cdot F_{i_1}\cdot...\cdot F_{i_j}\cdot Y_{i_j}\right)\nn
&&~~~~~~~~~~\times\underbrace{T\left(\sigma_1,\{\sigma_2,...,\sigma_r\}\shuffle\{i_j,...,i_1,a\}||\mathsf{G}_i\setminus\{i_j,i_{j-1},...,i_1,a\}\left|\pmb\rho_i||\W{\mathsf{G}}_i\right.\right)}_{T\left(A_i^{\text{YMS}\otimes\,\W{\mathsf{t}}}\right)}.\Label{Eq:RecursionCurrent2}
\eea
When exchanging the roles of the left and the right, we also get the recursive relations for the right side. In the first relation, the left set contains only gluons and the effective current is expanded in terms of those with fewer left gluons and more left scalars (i.e., $i_j,i_{j-1},...,i_1,a$ in each term), no matter whether the right side is YMS or YM. In the second relation, the effective current where the left part is YMS, is expanded in terms of those with fewer left gluons but more left scalars. This expansion is also independent of the choice of $\W{\mathsf{t}}$. The $Y^{\mu}_{i_j}$ denotes the total momentum of the particles living on the left of $i_j$ in $(\sigma_1,\{\sigma_2,...,\sigma_r\}\shuffle\{i_j,...,i_1,a\})$\footnote{Here, we use a new notation $Y^{\mu}_{i_j}$ that is different form the $X^{\mu}_{i_j}$ defined before. The reason is that if we do the recursion iteratively, there may be other gluons inserted into the left side of $i_j$, but the $Y^{\mu}_{i_j}$  does not include the momenta of these gluons.}. When applying these expansions iteratively till that both left and right gluons are turned into scalars,  the effective currents return to the double-copy formulas (\ref{Eq:J1})-(\ref{Eq:J4}). This recursive form of the effective current is crucial, as it allows the current to be computed without any knowledge of the graph rules.

{\it $U(1)$-decoupling identities of the effective currents}~~Tree-level effective currents $T^{\mu}\left(\emptyset||\mathsf{G}\left|\pmb\rho||\W{\mathsf{G}}\right.\right)$ or $T\left(\pmb\sigma||\mathsf{G}\left|\pmb\rho||\W{\mathsf{G}}\right.\right)$ defined by (\ref{Eq:DoubleCopy}) (or equivalently, by (\ref{Eq:RecursionCurrent1}) and (\ref{Eq:RecursionCurrent2})) satisfy the following relations:
\bea
T^{\mu}\left(\emptyset||\mathsf{G}\left|\pmb\rho||\W{\mathsf{G}}\right.\right)+\text{cyc}(\pmb\rho)&=&0,\nn
T\left(\pmb\sigma||\mathsf{G}\left|\pmb\rho||\W{\mathsf{G}}\right.\right)+\text{cyc}(\pmb\sigma)&=&T\left(\pmb\sigma||\mathsf{G}\left|\pmb\rho||\W{\mathsf{G}}\right.\right)+\text{cyc}(\pmb\rho)=0. \Label{Eq:LRU1ID}
\eea
The lhs. of these identities is expressed by the cyclic sum of the left or right scalars. Although there are in general four types of particles in the effective currents, the vanishing of the cyclic sum, e.g., $T\left(\pmb\sigma||\mathsf{G}\left|\pmb\rho||\W{\mathsf{G}}\right.\right)+\text{cyc}(\pmb\sigma)$, is independent of $\mathsf{G}$, $\W{\mathsf{G}}$ and the right scalars. Essentially, these identities are guaranteed by the antisymmetry of the BS vertices. One may understand the off-shell line for the current as a $U(1)$ particle, thus these identities are just the off-shell version of the $U(1)$-decoupling identity for amplitudes.

{\it Cancellation of tadpole diagrams}~~A consequent result of the  $U(1)$-decoupling identities (\ref{Eq:LRU1ID}) is the cancellation of tadpole diagrams that are divergent (before taking the loop integral) when the momentum conservation for external particles is considered. In a nonsupersymmetric theory, the one-loop integrand obtained from forward limit is accompanied by divergences (before integrating over the loop momentum) which have to be regularized. In the framework of CHY formula, divergences are regularized in a gauge invariance way by dropping out certain singular solution contributions, as shown in \cite{He:2015yua,Cachazo:2015aol}. A crucial observation  \cite{He:2015yua} on BS integrand is the following. For singular solutions to scattering equations, the cyclic sum over external particles results in a $U(1)$-decoupling identity for PT factors such that the full CHY integrand does not diverge. On the Feynman diagram side, the cyclic sum permits one to cancel all tadpole diagrams (which are divergent under forward limit, due to momentum conservation of external particles) in BS. This cancellation of divergent tadpole diagrams is generalized to EYM and GR integrands. To see this, we consider the following term with divergent tadpole diagrams for example:
\bea
{1\over l^2}\left[\,T\left(\pmb\sigma||\mathsf{G}\left|\pmb\rho||\W{\mathsf{G}}\right.\right)+\text{cyc}(\pmb\sigma)\,\right],
\eea
where $T\left(\pmb\sigma||\mathsf{G}\left|\pmb\rho||\W{\mathsf{G}}\right.\right)$ is a tree-level effective current involving all external particles.
Recalling that each term is a combination of BS currents which contain a propagator ${1\over P^2}$, where $P^{\mu}$ denotes the total momentum of all external particles. Each term thus diverges due to momentum conservation. However,  if we do not apply momentum conservation before taking the sum over cyclic permutations, these divergent tadpole terms cancel out after summation due to the $U(1)$-decoupling identity (\ref{Eq:LRU1ID}). With this cancellation mechanism, one finds that the tadpole contributions of the EYM integrands (\ref{Eq:EYMDoubleYMS1}) with at least two external gluons have to vanish. The only possible surviving tadpole contribution of the GR integrand comes from the $(D-2)^2$ term in (\ref{Eq:GRDoubleYMS1}) (and also EYM integrand where all external particles are gravitons), which supplies multipoint vertex structures inserted to the loop (but do not contain the divergent propagator $1\over P^2$)\footnote{Bubble diagrams, in which one of the tree structures on the loop is a single particle, also diverge due to momentum conservation and on-shell condition. However, we can regularize this by Minahaning \cite{Minahan:1987ha,Berg:2016wux,Berg:2016fui} when setting the total momentum to a vector $P^{\mu}$ such that $P\cdot v=0$ ($v^{\mu}\neq k^{\mu}_i$ for all external particle $i$). All localization techniques can be applied under this regularization.}.

When the effective currents are connected to possible vertex structures, Lorentz indices carried by the currents must be contracted out. In general,  such a contraction satisfies {\it (i) A left (right) Lorentz index is contracted to another left (right) index, i.e.,  there does not exist a contraction between left and right indices.  (ii) Each current may be contracted with another current or a momentum.} In the coming  discussion, we will briefly write contractions between effective currents  as shown by the following examples
\bea
T^{\alpha}( \mathsf{Y}_1\cup \mathsf{W}_1)T^{\alpha\W{\beta}}(\mathsf{W}_2)T^{\W{\beta}}( \mathsf{Z}_3\cup \mathsf{W}_3)&=&T( \mathsf{Y}_1\cup \mathsf{W}_1)\cdot T(\mathsf{W}_2)\cdot T( \mathsf{Z}_3\cup \mathsf{W}_3),\nn
T^{\alpha}( \mathsf{Y}_1\cup \mathsf{W}_1)T^{\alpha\W{\mu}}(\mathsf{W}_2)T^{\nu\W{\mu}}(\mathsf{W}_3)T^{{\nu}}( \mathsf{Y}_4\cup \mathsf{W}_4)&=&T( \mathsf{Y}_1\cup \mathsf{W}_1)\cdot T(\mathsf{W}_2)\cdot T(\mathsf{W}_3)\cdot T( \mathsf{Y}_4\cup \mathsf{W}_4),\nn
T^{\W\mu}( \mathsf{Z}_1\cup \mathsf{W}_1)T^{\alpha\W{\mu}}(\mathsf{W}_2)T^{\alpha\W{\nu}}(\mathsf{W}_3)T^{{\W\nu}}( \mathsf{Z}_4\cup \mathsf{W}_4)&=&T( \mathsf{Z}_1\cup \mathsf{W}_1)\cdot T(\mathsf{W}_2)\cdot T(\mathsf{W}_3)\cdot T(\mathsf{Z}_4\cup \mathsf{W}_4),\nn
T^{\alpha}( \mathsf{Y}_1\cup \mathsf{W}_1)T^{\alpha\W{\mu}}(\mathsf{W}_2)T^{\nu\W{\mu}}(\mathsf{W}_3)\,{k_i}^{{\nu}}&=&T( \mathsf{Y}_1\cup \mathsf{W}_1)\cdot T(\mathsf{W}_2)\cdot T(\mathsf{W}_3)\cdot k_i,\nn
k_i^{\alpha}T^{\alpha\W{\beta}}(\mathsf{W}_1)T^{\mu\W{\beta}}(\mathsf{W}_2)T^{\mu\W{\nu}}(\mathsf{W}_3)\,{k_j}^{\W{\nu}}&=&k_i\cdot T(\mathsf{W}_1)\cdot T(\mathsf{W}_2)\cdot T(\mathsf{W}_3)\cdot k_{\,\W j},\nn
k_{i}^{\W\alpha}T^{\beta\W{\alpha}}(\mathsf{W}_1)T^{\beta\W{\mu}}(\mathsf{W}_2)\,k^{\W\mu}_{j}&=&k_{\,\W i}\cdot T(\mathsf{W}_1)\cdot T(\mathsf{W}_2)\cdot k_{\,\W j}.
\eea
In the above expressions, the subset $A_i$ in each current for fixed $\mathsf{t}$ and $\W{\mathsf{t}}$ is explicitly expressed by the particle species, for instance $T^{\alpha}\big(\,\mathsf{A}_1^{\text{YM}\otimes\text{YMS}}\big)=T^{\alpha}( \mathsf{Y}_1\cup \mathsf{W}_1)$, where the superscripts are supressed. On the rhs. of each line, the contraction between a pair of Lorentz indices is implied by a dot. One can deduce the full form on the lhs. from the expression on the rhs. In particular, if there exists a current with one Lorentz index, i.e., $T^{\alpha}( \mathsf{Y}_1\cup \mathsf{W}_1)$ or $T^{\W\mu}( \mathsf{Z}_1\cup \mathsf{W}_1)$,  one can always deduce whether the index comes from the left or right half integrand, by the elements in the current. Consequently, the full contraction of Lorentz indices is uniquely fixed. If an expression is given by contractions between two momenta and currents with two Lorentz indices, e.g., the last two lines, one cannot deduce the contraction of Lorentz indices directly. Thus, we use $k_i$ and $k_{\,\W j}$ to imply that the corresponding Lorentz indices come from the left and the right half integrands, respectively.

Having defined the effective currents and the Lorentz contractions among them,  we investigate the localizations of partial integrands with $|\mathsf{W}|=1,2,3$ and then provide the quadratic propagator forms of the full integrand.

\section{ Double YMS with $|\mathsf{W}|=1$}\label{sec:W1}

For a partial integrand with $\mathsf{W}=\{w_1\}$, the local vertices (\ref{Eq:3xVertex}), (\ref{Eq:2x1zVertex}), (\ref{Eq:2x2yVertex}), (\ref{Eq:2x1yVertex}) and (\ref{Eq:2x2zVertex}), which were previously constructed in the $|\mathsf{W}|=0$ case, can also be found in the localized partial integrands. In addition, the particle $w_1$ introduces new local vertices into the partial integrands. In this section, we construct these new  vertices associated with $w_1$ by explicit examples, and then provide the general formula of the local partial integrand and the quadratic propagator expressed integrand.


\subsection{Localizations and vertices demonstrated by $\mathsf{X}=\{x_1\}$, $\mathsf{Y}=\{y_1\}$, $\mathsf{Z}=\{z_1\}$, $\mathsf{W}=\{w_1\}$}
In this example, we study the localization of dYMS partial integrands with $\mathsf{X}=\{x_1\}$, $\mathsf{Y}=\{y_1\}$, $\mathsf{Z}=\{z_1\}$ and $\mathsf{W}=\{w_1\}$. 
Permutations of the left and right PT factors are fixed as $(x_1z_1)$ and $(x_1y_1)$, while the left and right reference orders are respectively supposed to be $y_1\prec w_1$ and $z_1\prec w_1$.  

%
%

%
\subsubsection{$2x$-$1w$ vertex}

At first, we consider the following term with the partition $\{x_1,z_1,y_1,w_1\}$
%
%
%
%
\bea
\mathrm{T}_1&=&\begin{minipage}{2.2cm}  \includegraphics[width=2.2cm]{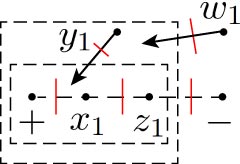} \end{minipage}\times\begin{minipage}{2.2cm}  \includegraphics[width=2.2cm]{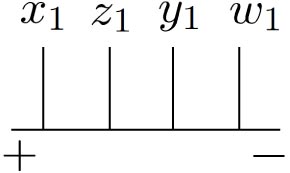} \end{minipage}\times\begin{minipage}{2.2cm} \includegraphics[width=2.2cm]{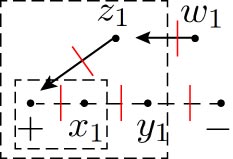} \end{minipage}\Label{Eq:NewW1EG1}\nn
&\rightarrow&{1\over l^2}\,\phi_{x_1|x_1}\,{1\over s_{x_1,l}}\,(\W\epsilon_{z_1}\cdot l_{x_1})\phi_{z_1|z_1}\,{1\over s_{x_1z_1,l}}\,(\epsilon_{y_1}\cdot l_{x_1z_1})\phi_{y_1|y_1}\,{1\over s_{x_1z_1y_1,l}}\,(l_{x_1z_1y_1}\cdot\epsilon_{w_1}\cdot l_{x_1z_1y_1})\phi_{w_1|w_1},
\eea
where the relevant nonlocal terms have been assembled into  a local expression on the first line. The polarization tensor for $w_1$ on the second line 
is presented as $\epsilon_{w_1}^{\mu\W\mu}\equiv \epsilon_{w_1}^{\mu}\W \epsilon_{w_1}^{\W\mu}$. This term induces a local $2x$-$1w$ vertex 
\bea
\begin{minipage}{1.5cm}  \includegraphics[width=1.5cm]{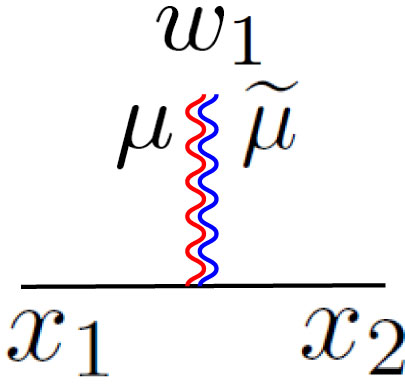} \end{minipage}= k_{x_1}^{\mu}k_{x_1}^{\W\mu},\Label{Eq:2x1wVertex}
\eea
in which $k_{x_1}^{\mu}$ and $k_{x_1}^{\W\mu}$ are both the momentum of the scalar line $x_1$ but are contracted with the left and right Lorentz indices, separately. The local result (\ref{Eq:NewW1EG1}) then  contributes to the partial integrand as:
\bea
{1\over l^2}\,J_{(3)}[x_1]\,{1\over s_{x_1,l}}\,J_{(3)}[z_1]\,{1\over s_{x_1z_1,l}}\,J_{(3)}[y_1]\,{1\over s_{x_1z_1y_1,l}}\,J_{(3)}[w_1]=\begin{minipage}{2.9cm}  \includegraphics[width=2.7cm]{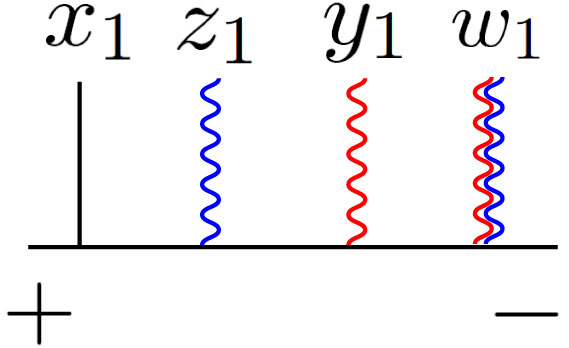} \end{minipage}\Label{Eq:NewFW1EG1D1}
\eea
\subsubsection{$2x$-$1y$-$1w$ and $2x$-$1z$-$1w$ vertices}
The second example is shown by the following term with partition $\{x_1,z_1,y_1\text{-}w_1\}$ which introduces a four-point vertex by the left graph
\bea
\mathrm{T}_1&=&\left(-{1\over 2}\right)\begin{minipage}{2.2cm} \includegraphics[width=2.2cm]{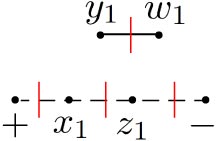} \end{minipage}\times\begin{minipage}{2.2cm}  \includegraphics[width=2.2cm]{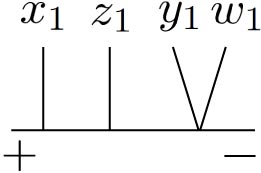} \end{minipage}\times\begin{minipage}{2.2cm} \includegraphics[width=2.2cm]{EQW1EG1R3} \end{minipage}\nn
&=&{1\over l^2}\phi_{x_1|x_1}{1\over s_{x_1,l}}(\W\epsilon_{z_1}\cdot l_{x_1})\phi_{z_1|z_1}{1\over s_{x_1z_1,l}}\left(-{1\over 2}\right)\big[\epsilon_{y_1}\cdot\epsilon_{w_1}\cdot(l_{x_1z_1}+k_{y_1})\big]\phi_{y_1|y_1}\phi_{w_1|w_1},\Label{EQ:W1EG1FourVertex1}
\eea
where the localization-1 have been implied by the boxed regions and  a {\it $2x$-$1y$-$1w$ vertex} (with $w_1$) has been introduced
\bea
\begin{minipage}{1.5cm}  \includegraphics[width=1.5cm]{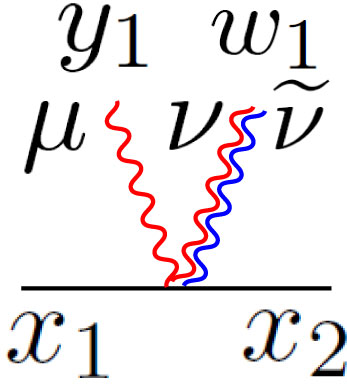} \end{minipage}=\left(-{1\over 2}\right) \eta^{\mu\nu}\left(k_{x_1}^{\W\nu}+k_{y_1}^{\W\nu}\right).\Label{Eq:2x1y1wVertex}
\eea
The expression (\ref{EQ:W1EG1FourVertex1}) finally plays as the following local term
\bea
{1\over l^2}\,J_{(3)}[x_1]\,{1\over s_{x_1,l}}\,J_{(3)}[z_1]\,{1\over s_{x_1z_1,l}}\,J_{(4)}[y_1,w_1]=\begin{minipage}{2.9cm}  \includegraphics[width=2.7cm]{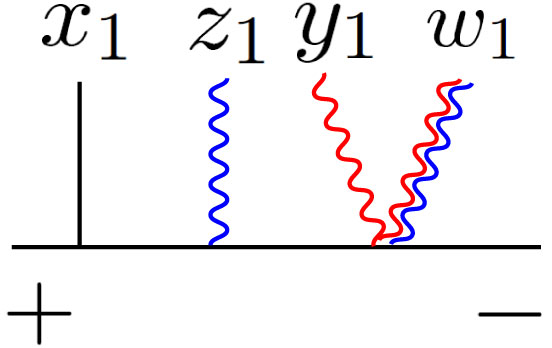} \end{minipage}.\Label{Eq:NewFW1EG2D1}
\eea

{\it Since both half integrands of dYMS are given by the left half integrand of YMS, the local partial integrand can be equivalently obtained by exchanging the roles of the left and right.} As a result, a  {\it $2x$-$1z$-$1w$ vertex} must also be induced 
\bea
\begin{minipage}{1.5cm}  \includegraphics[width=1.5cm]{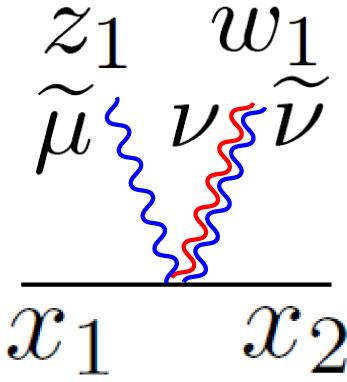} \end{minipage}= \left(-{1\over 2}\right)\eta^{\W\mu\W\nu}\left(k_{x_1}^{\nu}+k_{z_1}^{\nu}\right).\Label{Eq:2x1z1wVertex}
\eea
This vertex must also be constructed without changing the roles of left and right. To see this, we look at the following term with partition $\{x_1,y_1,z_1,w_1\}$
\bea
\mathrm{T}_2&&=\begin{minipage}{2.2cm}  \includegraphics[width=2.2cm]{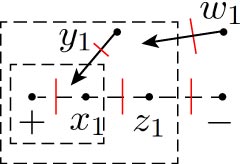} \end{minipage}\times\begin{minipage}{2.2cm}  \includegraphics[width=2.2cm]{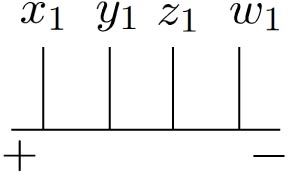} \end{minipage}\times\begin{minipage}{2.2cm} \includegraphics[width=2.2cm]{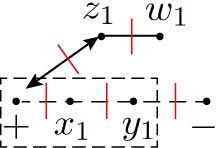} \end{minipage}\nn
&&\xrightarrow{\text{Localization-2}}\begin{minipage}{2.2cm}  \includegraphics[width=2.2cm]{EQW1EG2L1} \end{minipage}\times\left(-{1\over 2}\right)\begin{minipage}{2.2cm}  \includegraphics[width=2.2cm]{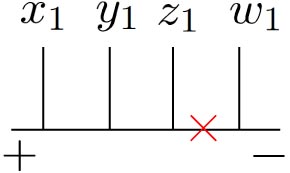} \end{minipage}\times\begin{minipage}{2.2cm} \includegraphics[width=2.2cm]{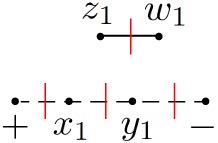} \end{minipage}\nn
&\rightarrow&{1\over l^2}{1\over s_{x_1,l}}(\epsilon_{y_1}\cdot l_{x_1}){1\over s_{x_1y_1,l}}\left(-{1\over 2}\right)\big[(l_{x_1y_1}+k_{z_1})\cdot\epsilon_{w_1}\cdot \W\epsilon_{z_1}\big],\Label{EQ:W1EG1FourVertex2}
\eea
which plays as the following local term
\bea
{1\over l^2}\,J_{(3)}[x_1]\,{1\over s_{x_1,l}}\,J_{(3)}[y_1]\,{1\over s_{x_1z_1,l}}\,J_{(4)}[z_1,w_1]=\begin{minipage}{2.9cm}  \includegraphics[width=2.7cm]{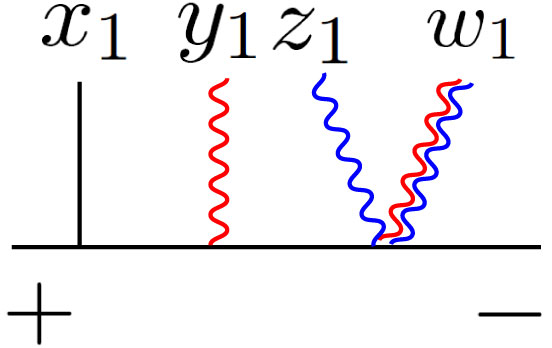} \end{minipage},\Label{Eq:NewFW1EG2D2}
\eea
involving the $2x$-$1z$-$1w$ vertex (\ref{Eq:2x1z1wVertex}).

\subsubsection{$2x$-$1y$-$1z$-$1w$ vertex}

Last, we turn to a term with partition  $\{x_1,z_1,y_1\text{-}w_1\}$
\bea
\mathrm{T}_1&&=\left(-{1\over 2}\right)\begin{minipage}{2.2cm}  \includegraphics[width=2.2cm]{EQW1EG1L2} \end{minipage}\times\begin{minipage}{2.2cm}  \includegraphics[width=2.2cm]{EQW1EG1D2} \end{minipage}\times\begin{minipage}{2.2cm} \includegraphics[width=2.2cm]{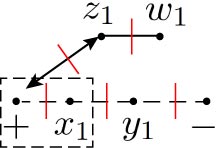} \end{minipage}\nn
&&=\left(-{1\over 2}\right)\begin{minipage}{2.2cm}  \includegraphics[width=2.2cm]{EQW1EG1L2} \end{minipage}\times\left(-{1\over 2}\right)\left[\,\begin{minipage}{2.2cm}\includegraphics[width=2.2cm]{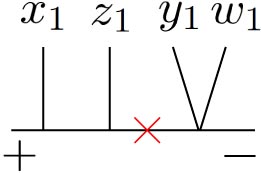} \end{minipage}-\begin{minipage}{2.2cm}\includegraphics[width=2.2cm]{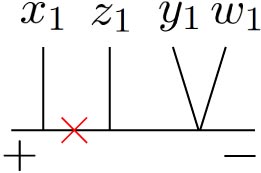} \end{minipage}\,\right]\times\begin{minipage}{2.2cm} \includegraphics[width=2.2cm]{EQW1EG1R2} \end{minipage},\nn
&&\xrightarrow{\text{Localization-2}}\left(-{1\over 2}\right)\begin{minipage}{2.2cm}  \includegraphics[width=2.2cm]{EQW1EG1L2} \end{minipage}\times\left(-{1\over 2}\right)\,\begin{minipage}{2.2cm}\includegraphics[width=2.2cm]{EQW1EG1D3} \end{minipage}\,\times\begin{minipage}{2.2cm} \includegraphics[width=2.2cm]{EQW1EG1R2} \end{minipage}\nn
=&&{1\over l^2}\phi_{x_1|x_1}{1\over s_{x_1,l}}\left(-{1\over 2}\right)^2(\epsilon_{y_1}\cdot\epsilon_{w_1}\cdot\W\epsilon_{z_1})\phi_{z_1|z_1}\phi_{y_1|y_1}\phi_{w_1|w_1}.\Label{EQ:W1EG1FiveVertex1}
\eea
On the first line, there is an X-pattern associated with $z_1$. When the property (\ref{Eq:PropertyXPattern}) of X-pattern is applied, we arrive at the second line. The second term on the second line cancels with another graph which contains the BCJ-pattern where $x_1$ and $z_1$ belong to the same subcurrent according to localization-2. Hence, only the first term survives and is already local. This term induces a {\it $2x$-$1y$-$1z$-$1w$ vertex}
\bea
\begin{minipage}{1.7cm}  \includegraphics[width=1.7cm]{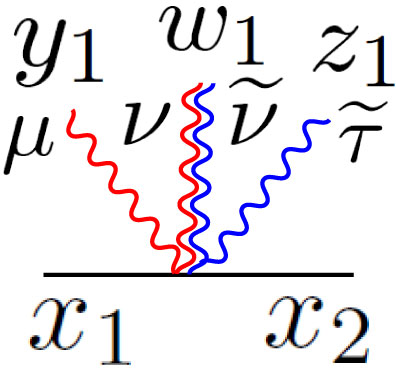} \end{minipage}= \left(-{1\over 2}\right)^2\eta^{\mu\nu}\eta^{\W\nu\W\tau}, \Label{Eq:2x1y1z1wVertex}
\eea
and  the last line of (\ref{EQ:W1EG1FiveVertex1}) contributes the following term to the local partial integrand
\bea
{1\over l^2}\,J_{(3)}[x_1]\,{1\over s_{x_1,l}}\,J_{(5)}[y_1,z_1,w_1]=     \begin{minipage}{2.9cm}  \includegraphics[width=2.7cm]{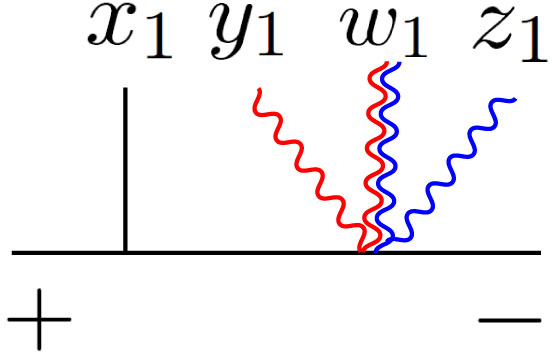} \end{minipage}.\Label{Eq:NewFW1EG3D1}
\eea
%

\subsection{Contracting nontrivial subcurrents with local vertices}
In the previous examples, each external line attached to a vertex is a single particle. Now we show that more complicated structures where nontrivial subcurrents are attached to the vertex structure can be  derived by {\it replacing a single particle with a subcurrent}.

We first consider the following term with partition $\{x_1,z_1,y_1\text{-}\{y_2,w_1\}\}$ which contains a nontrivial subcurrent with $y_2,w_1$ connecting to a $2x$-$2y$ vertex
\bea
\mathrm{T}_1&=&\left(-{1\over 2}\right)\begin{minipage}{2.2cm}  \includegraphics[width=2.2cm]{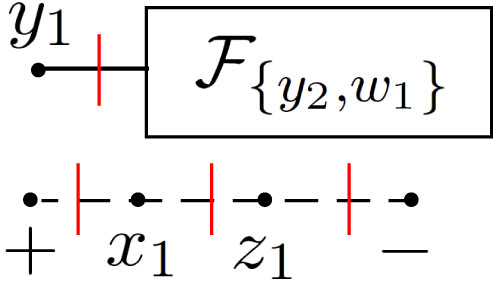} \end{minipage}\times\begin{minipage}{2.2cm}  \includegraphics[width=2.2cm]{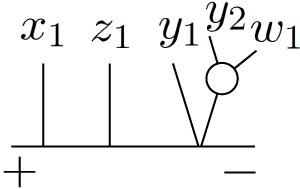} \end{minipage}\times\begin{minipage}{2.4cm} \includegraphics[width=2.4cm]{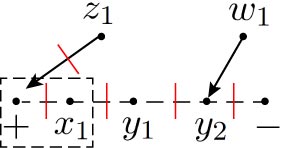} \end{minipage}\nn
&=&{1\over l^2}\phi_{x_1|x_1}{1\over s_{x_1,l}}(\W\epsilon_{z_1}\cdot l_{x_1})\phi_{z_1|z_1}{1\over s_{x_1z_1,l}}\left[\left(-{1\over 2}\right)
\left(\epsilon_{y_1}\phi_{y_1|y_1}\right)\cdot\Big(\mathcal{C}\left[\mathcal{F}_{\{y_2,w_1\}}\right]\phi_{\pmb\sigma\left[\mathcal{F}_{\{y_2,w_1\}}\right]\big|y_2w_1}(\W\epsilon_{w_1}\cdot k_{y_2})\Big)
\right]\nn
&=&\begin{minipage}{3cm}  \includegraphics[width=3cm]{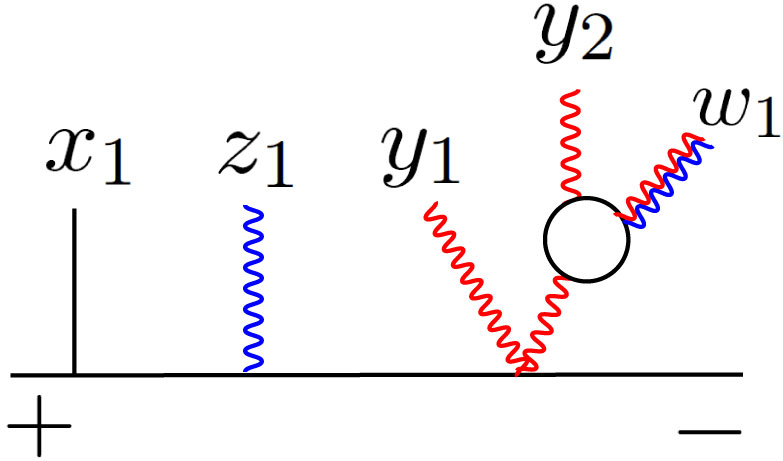} \end{minipage},
\eea
where the first line is already local due to localization-1. The expression on the second line turns into the Feynman diagram on the last line, which involves a $2x$-$2y$ vertex structure
\bea
\left(-{1\over 2}\right)T(y_1)\cdot T(y_2,w_1).
\eea
In the above expression, $T^{\mu}(y_1)=\epsilon^{\mu}_{y_1}\phi_{y_1|y_1}$ is a single-particle subcurrent, while $ T^{\mu}(y_2,w_1)$ is the Berends-Giele current (\ref{Eq:J2}) with both $y_2$ and $w_1$, noting $\mathcal{C}\left[\W{\mathcal{F}}_{\{y_2,w_1\}}\right]=\W\epsilon_{w_1}\cdot k_{y_2}$.  This four-point vertex structure can be given  by replacing $\epsilon^{\mu}_{y_2}$ in the $2x$-$2y$ vertex structure $\left(-{1\over 2}\right)\epsilon_{y_1}\cdot\epsilon_{y_2}$ with $T^{\mu}(y_2,w_1)$.

A more general five-point vertex structure is demonstrated by the following term associated with the partition $\{x_1,\{z_1,z_2\},\{y_1,y_2\}\text{-}w_1\}$
\bea
\mathrm{T}_2&=&\left(-{1\over 2}\right)\begin{minipage}{2.2cm}  \includegraphics[width=2.2cm]{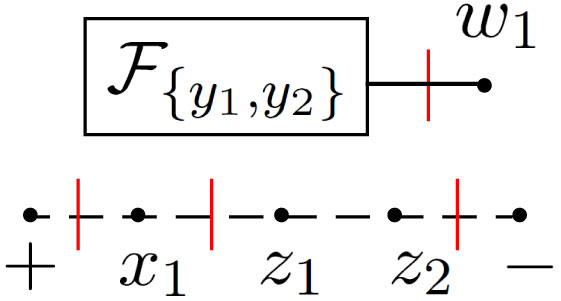} \end{minipage}\times\begin{minipage}{2.2cm}  \includegraphics[width=2.2cm]{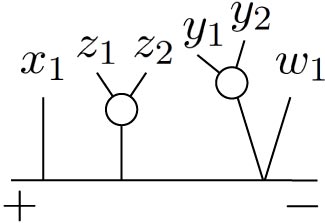} \end{minipage}\times\begin{minipage}{2.4cm} \includegraphics[width=2.4cm]{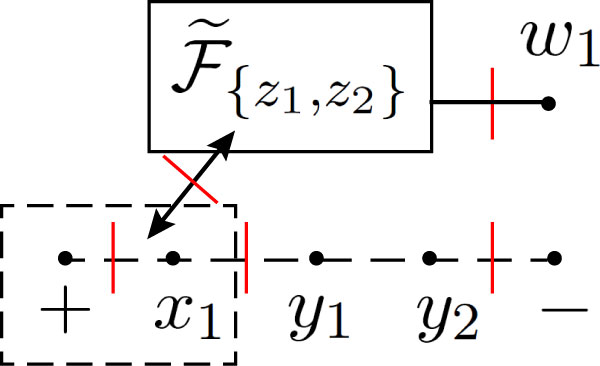} \end{minipage}\nn
&\rightarrow&\left(-{1\over 2}\right)^2\begin{minipage}{2.2cm}  \includegraphics[width=2.2cm]{EQW1EG3L2} \end{minipage}\times\begin{minipage}{2.2cm}  \includegraphics[width=2.2cm]{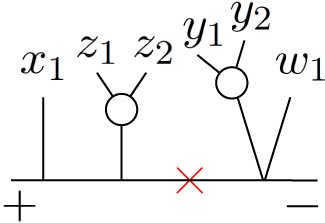} \end{minipage}\times\begin{minipage}{2.4cm} \includegraphics[width=2.4cm]{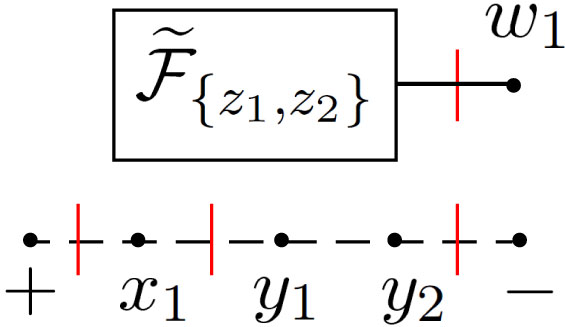} \end{minipage}\nn
&=&{1\over l^2}{1\over s_{x_1,l}}\left(-{1\over 2}\right)^2 T(y_1,y_2)\cdot T(w_1)\cdot T(z_1,z_2)\nn
&=&\begin{minipage}{3cm}  \includegraphics[width=3cm]{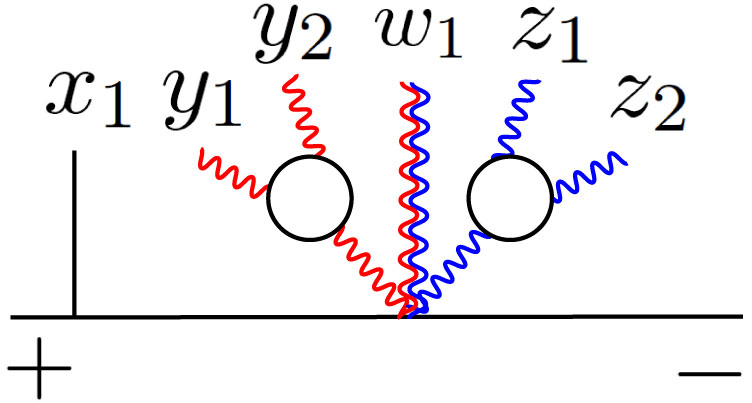} \end{minipage},
\eea
where the first and the second line are given by localization-1 and -2, respectively. The subcurrents of the $2x$-$1y$-$1z$-$1w$ vertex structure are explicitly written as 
\bea
T^{\mu\W\mu}(w_1)&=&\epsilon^{\mu\W\mu}_{w_1}=\epsilon^{\mu}_{w_1}\W{\epsilon}^{\,\W\mu}_{w_1},\nn
T^{\mu}(y_1,y_2)&=&\mathcal{C}^{\mu}[\mathcal{F}_{\{y_1,y_2\}}]\phi_{\pmb\alpha[\mathcal{F}_{\{y_1,y_2\}}]|y_1y_2},\nn
T^{\W\mu}(z_1,z_2)&=&\phi_{z_1z_2|\pmb\alpha[\W{\mathcal{F}}_{\{z_1,z_2\}}]}\mathcal{C}^{\W\mu}[\W{\mathcal{F}}_{\{z_1,z_2\}}],
\eea
according to the definitions in \secref{sec:5}. From this example, we find that in the cases with subcurrents containing more $\mathsf{Y}$ and $\mathsf{Z}$ elements, the five-point vertex also arises. The corresponding vertex structure can be obtained by applying the following replacements to (\ref{Eq:NewFW1EG3D1})
\bea
\epsilon^{\mu}_{y_1}\to T^{\mu}(y_1,y_2),~\epsilon^{\W\mu}_{z_1}\to T^{\W\mu}(z_1,z_2).
\eea

Another example is given by the following term with partition  $\{x_1,\{z_1,z_2\},\{y_1,y_2\}\text{-}w_1\}$
\bea
\mathrm{T}_3&=&\left(-{1\over 2}\right)\begin{minipage}{2.2cm}  \includegraphics[width=2.2cm]{EQW1EG3L2} \end{minipage}\times\begin{minipage}{2.2cm}  \includegraphics[width=2.2cm]{EQW1EG3D2} \end{minipage}\times\begin{minipage}{2.4cm} \includegraphics[width=2.4cm]{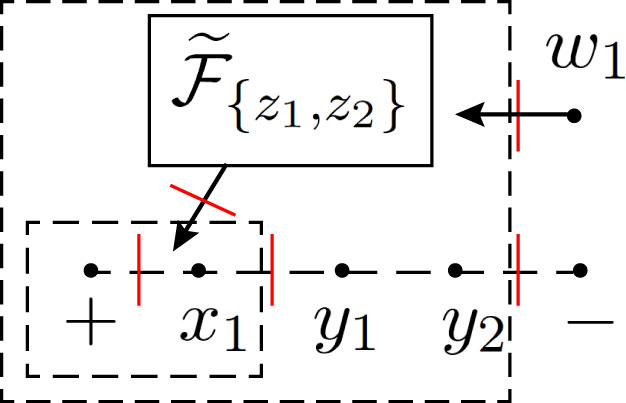} \end{minipage}\nn
&=&{1\over l^2}{1\over s_{x_1,l}}\,T(z_1,z_2)\cdot l_{x_1}\,{1\over s_{x_1z_1z_2,l}}\,\left(-{1\over 2}\right) T(y_1,y_2)\cdot T(w_1)\cdot (l_{x_1z_1z_2}+k_{y_1}+k_{y_2})\nn
&=&\begin{minipage}{3cm}  \includegraphics[width=3cm]{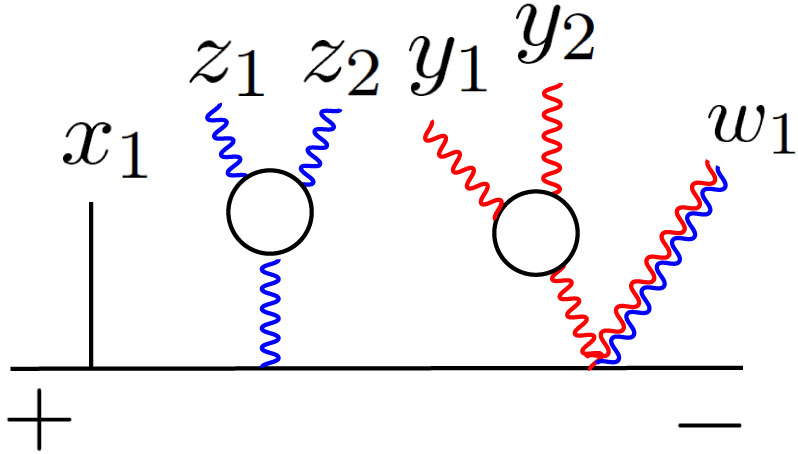} \end{minipage}
\eea
in which the $2x$-$1z$ and $2x$-$1y$-$1w$ vertex structures are given by taking the following replacements on those in (\ref{EQ:W1EG1FourVertex1})
\bea
\epsilon^{\mu}_{y_1}\to T^{\mu}(y_1,y_2),~~{\W\epsilon}^{\,\W\mu}_{z_1}\to T^{\W\mu}(z_1,z_2),~~k^{\mu}_{y_1}\to  k^{\mu}_{y_1}+ k^{\mu}_{y_2},~~k^{\W\mu}_{z_1}\to  k^{\W\mu}_{z_1}+ k^{\W\mu}_{z_2}.
\eea
Note the replacement of ${\W\epsilon}^{\,\W\mu}_{z_1}$ was already introduced in  example of $|\mathsf{W}|=0$ (see \secref{sec:W0VertexStructure}).

From the examples in this section, terms with more complicated partitions do not introduce new vertices into the final local result.
{\it Vertex structures with more complicated currents can be obtained directly from those with single particles as follows: (i)  replace the single particles by currents of the same type that have the identical Lorentz indices, (ii) replace the momentum of each particle by the total momentum of the corresponding current.} This observation will be helpful when investigating more general situations.

\subsection{General local formula for partial and full integrands with $|\mathsf{W}|=1$ }

In general, the double YMS partial integrand (\ref{Eq:doubleYMS}) with $|\mathsf{W}|=1$ is reduced into the local form
\bea
\mathcal{A}^{\,\text{dYMS},|\mathsf{W}|=1}=\,\Sl_{\{A_1A_2...A_I\}}{1\over l^2}\,J[A_1]\, {1\over s_{A_1,l}}\,J[A_2]\cdots\,{1\over s_{A_1...A_{I-1},l}}\,J[A_I]. \Label{Eq:doubleYMSW1Final}
\eea
In the above expression, the local vertex structure $J[A_i]$ for a given subset $A_i$ is defined as
\bea
J[A_i]=J_{(3)}[A_i]+J_{(4)}[A_i]+J_{(5)}[A_i]. \Label{Eq:doubleYMSW1Final1}
\eea
The three-point vertex structures $J_{(3)}[A_i]$ have the following possible forms
\bea
J_{(3)}\left[\,A^{\text{YM}\otimes\text{YM}}_i\,\right]&=& X_{A_i}\cdot T\left(A_i^{\text{YM}\otimes\text{YM}}\right)\cdot X_{\W A_i},\\
J_{(3)}\left[\,A^{\text{YM}\otimes\text{YMS}}_i\,\right]&=&T\left(A_i^{\text{YM}\otimes\text{YMS}}\right)\cdot X_{A_i},\\
J_{(3)}\left[\,A^{\text{YMS}\otimes\text{YM}}_i\,\right]&=&T\left(A_i^{\text{YMS}\otimes\text{YM}}\right)\cdot X_{A_i},\\
J_{(3)}\left[\,A^{\text{YMS}\otimes\text{YMS}}_i\,\right]&=& T\left(A^{\text{YMS}\otimes\text{YMS}}_i\right),
\eea
where the possible particle species in a subset $A^{\mathsf{t}\otimes\W{\mathsf{t}}}_i$ are implied by the superscript $\mathsf{t}\otimes\W{\mathsf{t}}$. For example, the notation $A^{\text{YM}\otimes\text{YM}}_i$ denotes possible subsets whose elements play as gluons for both left and right half integrands. In the case $\mathsf{W}=\{w_1\}$, $A^{\text{YM}\otimes\text{YM}}_i=\{w_1\}$. Another example is that the subset $A^{\text{YM}\otimes\text{YMS}}_i$ can either be $\mathsf{Y}_i$ or  $\mathsf{Y}_i\cup\{w_1\}$. The four-point vertex structures $J_{(4)}[A_i]$ in (\ref{Eq:doubleYMSW1Final1}) are given by
\bea
J_{\text{(4)}}[\mathsf{Y}_i]&=&\left(-{1\over 2}\right)\,\Sl_{\{\mathsf{Y}_{i1}\mathsf{Y}_{i2}\}}\Big[T(\mathsf{Y}_{i1})\cdot T(\mathsf{Y}_{i2})\Big],\\
J_{\text{(4)}}[\mathsf{Z}_i]&=&\left(-{1\over 2}\right)\,\Sl_{\{\mathsf{Z}_{i1}\mathsf{Z}_{i2}\}}\Big[T(\mathsf{Z}_{i1})\cdot T(\mathsf{Z}_{i2})\Big],\\
J_{\text{(4)}}[\mathsf{Y}_i\cup \{w_1\}]&=&\left(-{1\over2}\right)\bigg[\Sl_{\{\mathsf{Y}_{i1}\mathsf{Y}_{i2}\}}T(\mathsf{Y}_{i_1})\cdot T(\mathsf{Y}_{i_2}\cup \{w_1\})+T(\mathsf{Y}_{i})\cdot T(w_1)\cdot (X_{\mathsf{Y}_i\cup \{w_1\}}+k_{\mathsf{Y}_i})\bigg],\\
J_{\text{(4)}}[\mathsf{Z}_i\cup \{w_1\}]&=&\left(-{1\over2}\right)\bigg[\Sl_{\{\mathsf{Z}_{i1}\mathsf{Z}_{i2}\}}T(\mathsf{Z}_{i_1})\cdot T(\mathsf{Z}_{i_2}\cup \{w_1\})+T(\mathsf{Z}_{i})\cdot T(w_1)\cdot (X_{\mathsf{Z}_i\cup \{w_1\}}+k_{\mathsf{Z}_i})\bigg].
\eea
The five-point vertex structure  $J_{(5)}[A_i]$ in (\ref{Eq:doubleYMSW1Final1}) is given by 
\bea
J_{\text{(5)}}\left[\,\mathsf{Y}_i\cup\mathsf{Z}_i\cup\{w_1\}\right]&=&\left(-{1\over 2}\right)^2 \,T\left(\mathsf{Y}_i\right)\cdot T\left(w_1\right)\cdot T\left(\mathsf{Z}_i\right).
\Label{eq:w15vertex}
\eea
This $J_{\text{(5)}}$ arises only when the subset contains $\mathsf{Y}$, $\mathsf{Z}$ and $\mathsf{W}$ elements but does not contain $\mathsf{X}$ element.

Once the nonlocal partial integrand (\ref{Eq:doubleYMS}) has been transformed into the local partial integrand (\ref{Eq:doubleYMSW1Final}), one can always find other terms related by cyclic permutations. Thus, according to (\ref{Eq:partial}), all partial integrands sum into the full integrand with quadratic propagators 
\bea
\mathcal{I}^{\,\text{dYMS},|\mathsf{W}|=1}=\,\Sl_{\{A_1A_2...A_I\}}{1\over l^2}\,J[A_1]\, {1\over l^2_{A_1}}\,J[A_2]\cdots\,{1\over l^2_{A_1...A_{I-1}}}\,J[A_I]. \Label{Eq:doubleYMSW1Final2}
\eea
%

\section{Double YMS with $|\mathsf{W}|=2$}\label{sec:W2}
For  $\mathsf{W}=\{w_1,w_2\}$, all vertices involved in cases of $|\mathsf{W}|=0$ and $|\mathsf{W}|=1$ contribute to the final local partial integrand. In addition, new vertices containing two $\mathsf{W}$ particles emerge after localization. In this section, we show how to extract these new vertices by examples and then provide the general formulas for dYMS partial and full integrands.

\subsection{Six-point vertices}

The five-point vertex structure given by (\ref{eq:w15vertex}) in the $|\mathsf{W}|=1$ case can be explicitly written as
\bea
\left(-{1\over 2}\right)^2 \,T\left(\mathsf{Y}_i\right)\cdot T\left(w_1\right)\cdot T\left(\mathsf{Z}_i\right)=\left(-{1\over 2}\right)^2 T^{\mu}( \mathsf{Y}_i)T^{\mu\W{\mu}}(w_1)T^{\W{\mu}}( \mathsf{Z}_i).
\eea
In the above expression, the effective current $T^{\mu\W{\mu}}(w_1)$ carries left and right Lorentz indices $\mu$ and $\W{\mu}$, which contract with the left and right indices of the other two effective currents $T^{\mu}( \mathsf{Y}_i)$ and $T^{\W{\mu}}( \mathsf{Z}_i)$, respectively. Hence, for the case of $\mathsf{W}=\{w_1,w_2\}$, the analysis of the Lorentz indices suggests the following possible forms of the six-point vertex structures
\bea
T\left(\mathsf{Y}_{i1}\right)\cdot T\left(w_1\right)\cdot T\left(w_2\right)\cdot T\left(\mathsf{Y}_{i2}\right)&=&T^{\alpha}( \mathsf{Y}_{i1})T^{\alpha\W{\beta}}(w_1)T^{\gamma\W{\beta}}(w_2)T^{{\gamma}}( \mathsf{Y}_{i2}),\nn
T\left(\mathsf{Z}_{i1}\right)\cdot T\left(w_1\right)\cdot T\left(w_2\right)\cdot T\left(\mathsf{Z}_{i2}\right)&=&T^{\W{\alpha}}( \mathsf{Z}_{i1})T^{\beta\W{\alpha}}(w_1)T^{\beta\W{\gamma}}(w_2)T^{\W{\gamma}}( \mathsf{Z}_{i2}),\nn
T\left(\mathsf{Y}_{i1}\right)\cdot T\left(w_2\right)\cdot T\left(w_1\right)\cdot T\left(\mathsf{Y}_{i2}\right)&=&T^{\alpha}( \mathsf{Y}_{i1})T^{\alpha\W{\beta}}(w_2)T^{\gamma\W{\beta}}(w_1)T^{{\gamma}}( \mathsf{Y}_{i2}),\nn
T\left(\mathsf{Z}_{i1}\right)\cdot T\left(w_2\right)\cdot T\left(w_1\right)\cdot T\left(\mathsf{Z}_{i2}\right)&=&T^{\W{\alpha}}( \mathsf{Z}_{i1})T^{\beta\W{\alpha}}(w_2)T^{\beta\W{\gamma}}(w_1)T^{\W{\gamma}}( \mathsf{Z}_{i2}).
\Label{eq:possi6vertex}
\eea
In this subsection, we demonstrate through examples that the dYMS partial integrand with $\mathsf{W}=\{w_1,w_2\}$ actually only contains the last two 
six-point vertex structures under the choice of reference order and the left/right permutations in the current paper\footnote{Full calculations and program verification indicate that the case of $\mathsf{W}=\{w_1,w_2\}$ does not involve more complex six-point vertex structures, even though such vertex structures are allowed by the Lorentz index analysis. }. In the appendix, we show that the first two structures in (\ref{eq:possi6vertex}) do not appear in the result.


\subsubsection{$2x$-$2y$-$2w$ vertex}

We now demonstrate how to induce a $2x$-$2y$-$2w$ vertex by localization of  the dYMS partial integrand $\mathcal{A}^{\,\text{dYMS}}\big(\big.+,-||\{y_1,y_2,w_1,w_2\}\,\big|\,+,y_1,y_2,-||\{w_1,w_2\}\,\big)$. The left and right reference orders are respectively chosen as $y_1\prec y_2\prec w_1\prec w_2$ and $w_1\prec w_2$. The right permutation is fixed as $(y_1y_2)$. Let us begin with the following term
\bea
\left(-{1\over 2}\right)\begin{minipage}{2.1cm}  \includegraphics[width=2.1cm]{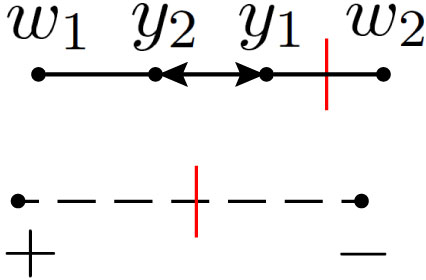} \end{minipage}\times\begin{minipage}{1.6cm}  \includegraphics[width=1.6cm]{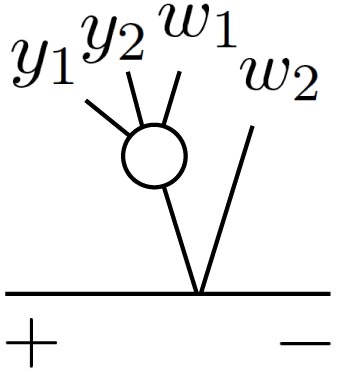} \end{minipage}\times\begin{minipage}{2cm} \includegraphics[width=2cm]{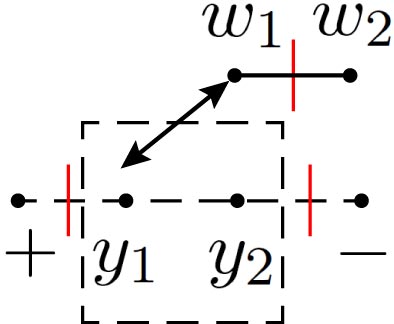} \end{minipage},
\eea
which corresponds to the partition $\{\{y_1,y_2,w_1\}\text{-}w_2\}$. The LPFD together with the right graph involves an off-shell BCJ relation in the subcurrent with $y_1$, $y_2$ and $w_1$. Thus (\ref{Eq:OffBCJ1}) reduces the above expression into 
\bea
\left(-{1\over 2}\right)^2\begin{minipage}{2.1cm}  \includegraphics[width=2.1cm]{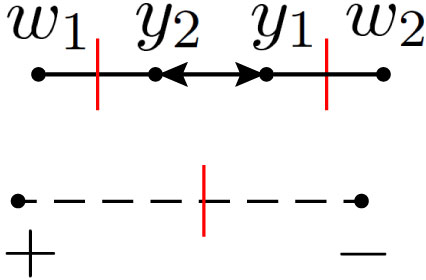} \end{minipage}\times\begin{minipage}{2.2cm}  \includegraphics[width=1.8cm]{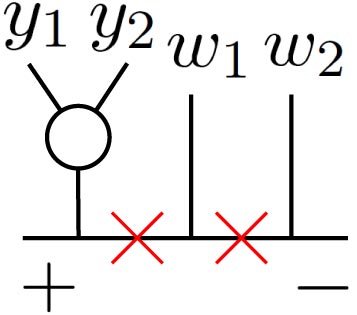} \end{minipage}\times\begin{minipage}{2cm} \includegraphics[width=2cm]{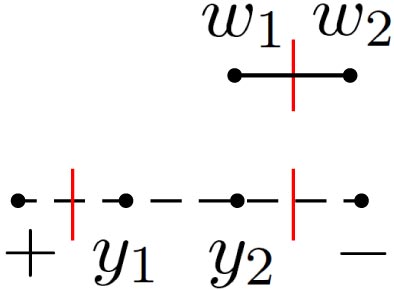} \end{minipage}.
\eea
After this step, one may notice that the subcurrent containing $y_1$ and $y_2$, together with a left coefficient $\left(-k_{y_2}\cdot k_{y_1}\right)$ also matches the lhs. of off-shell BCJ relation (\ref{Eq:OffBCJ1}). This leads to a further reduction
\bea
&&\left(-{1\over 2}\right)^3\begin{minipage}{2.1cm}  \includegraphics[width=2.1cm]{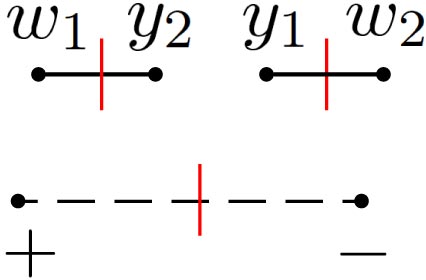} \end{minipage}\times\begin{minipage}{2cm}  \includegraphics[width=2cm]{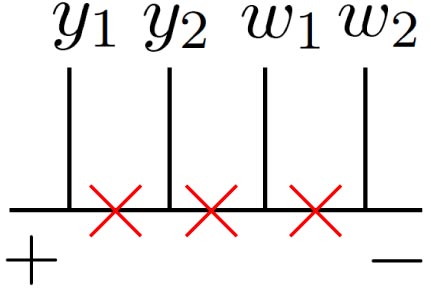} \end{minipage}\times\begin{minipage}{2cm} \includegraphics[width=2cm]{EQW2EG1R2} \end{minipage}\nn
&=&\left(-{1\over 2}\right)^3{1\over l^2}(\epsilon_{y_2}\cdot\epsilon_{w_1})(\epsilon_{y_1}\cdot\epsilon_{w_2})(\W\epsilon_{w_1}\cdot\W\epsilon_{w_2}),\Label{Eq:2x2y2wEG1}
\eea
which induces the following six-point vertex structure
\bea
{1\over l^2}\left(-{1\over 2}\right)^3\,T(y_1)\cdot T(w_2)\cdot T(w_1)\cdot T(y_2)=\,\begin{minipage}{2.3cm}  \includegraphics[width=2.3cm]{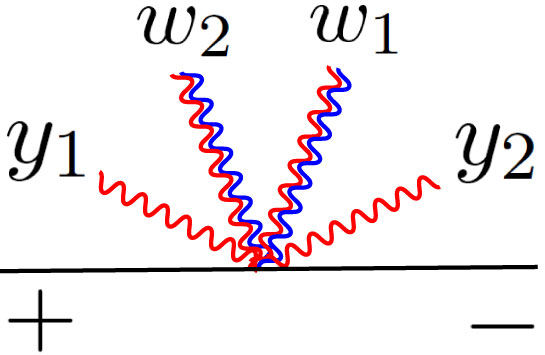} \end{minipage}.
\Label{Eq:2x2y2wVertexStr}
\eea
This term implies the {\it $2x$-$2y$-$2w$ vertex}
\bea
\begin{minipage}{2.3cm}  \includegraphics[width=2.3cm]{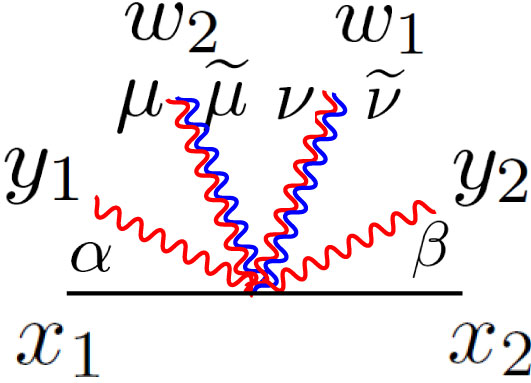} \end{minipage}= \left(-{1\over 2}\right)^3\eta^{\alpha\mu}\eta^{\W\mu\W\nu}\eta^{\nu\beta}. \Label{Eq:2x2y2wVertex}
\eea

{\bf Comments:} One may expect another six-point vertex term $(\epsilon_{y_1}\cdot\epsilon_{w_1})(\epsilon_{y_2}\cdot\epsilon_{w_2})(\W\epsilon_{w_1}\cdot\W\epsilon_{w_2})$ which is obtained by the one (\ref{Eq:2x2y2wEG1}) via exchanging the roles of $y_1$ and $y_2$. However, there exist two other  terms with partitions 
 $\{y_1\text{-}w_1,y_2\text{-}w_2\}$ and $\{\{y_1,w_1\},y_2\text{-}w_2\}$ that cancel with  this six-point vertex term. Since the relative order of $y_1$ and $y_2$ has been fixed as $(y_1y_2)$, there do not exist such terms to cancel with (\ref{Eq:2x2y2wEG1}). That's why we have nonvanishing $(\epsilon_{y_2}\cdot\epsilon_{w_1})(\epsilon_{y_1}\cdot\epsilon_{w_2})(\W\epsilon_{w_1}\cdot\W\epsilon_{w_2})$ structure but do not have $(\epsilon_{y_1}\cdot\epsilon_{w_1})(\epsilon_{y_2}\cdot\epsilon_{w_2})(\W\epsilon_{w_1}\cdot\W\epsilon_{w_2})$. We leave the detail of the cancellation of the latter term in the appendix.

\subsubsection{$2x$-$2z$-$2w$ vertex}

For dYMS partial integrand $\mathcal{A}^{\,\text{dYMS}}\big(\big.+,z_1,z_2,-||\{w_1,w_2\}\,\big|\,+,-||\{z_1,z_2,w_1,w_2\}\,\big)$ with the left and right reference orders $w_1\prec w_2$ and $z_1\prec z_2\prec w_1\prec w_2$, vertices and vertex structures can be obtained straightforwardly via exchanging the roles of the left and the right half integrands. This implies that we have the six-point vertex structure
\bea
\left(-{1\over 2}\right)^3\,T(z_1)\cdot T(w_2)\cdot T(w_1)\cdot T(z_2),
\eea
which replaces $y_1$  and $y_2$ in the vertex structure of (\ref{Eq:2x2y2wVertexStr})  by $z_1$ and $z_2$, respectively. The above structure is related to a $2x$-$2z$-$2w$ vertex. On another hand, consistency requires an alternative approach through the localization of the right coefficients for the right half integrand. To show this, we induce the local  $2x$-$2z$-$2w$ vertex  from the following term with the partition $\{z_1,z_2,w_1\text{-}w_2\}$

\bea
&&\left(-{1\over 2}\right)\begin{minipage}{2cm}  \includegraphics[width=2cm]{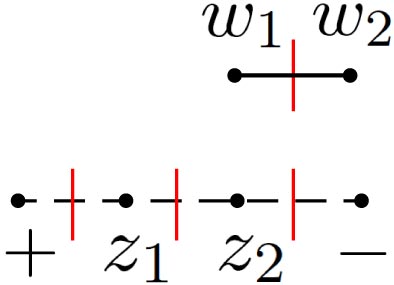} \end{minipage}\times\begin{minipage}{2cm}  \includegraphics[width=2cm]{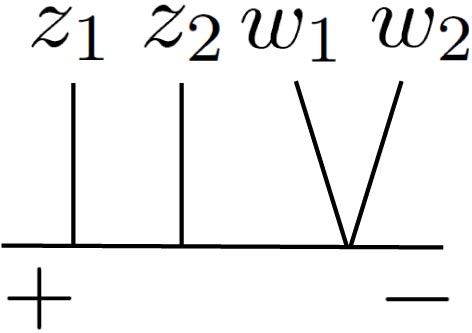} \end{minipage}\times\begin{minipage}{2cm} \includegraphics[width=2cm]{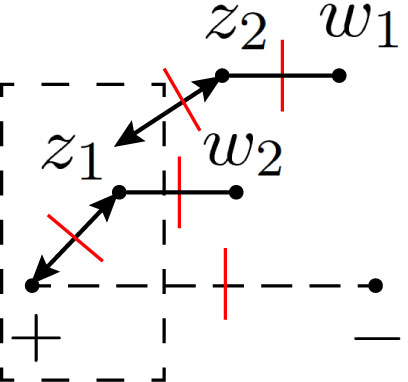} \end{minipage}\nn
&=&\left(-{1\over 2}\right)^3\begin{minipage}{2cm}  \includegraphics[width=2cm]{EQW2EG2L1} \end{minipage}\times\left[\,\begin{minipage}{2cm}  \includegraphics[width=2cm]{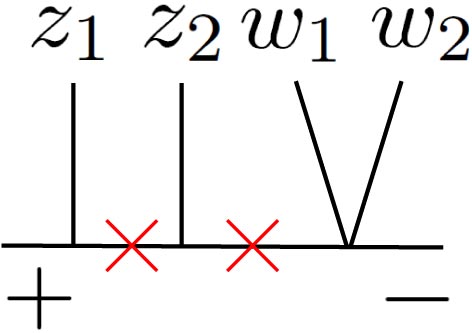} \end{minipage}-s_{z_1,l}\begin{minipage}{2cm}  \includegraphics[width=2cm]{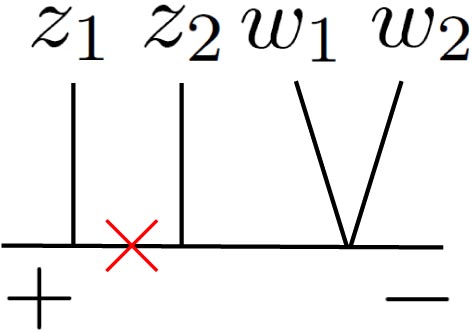} \end{minipage}\,\right]\times\begin{minipage}{2cm} \includegraphics[width=2cm]{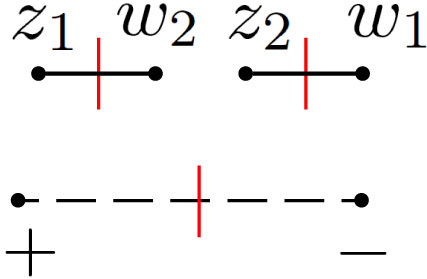} \end{minipage}\Label{EQ:W2EG2SixVertex1}
\eea
where the X-patterns associated with $z_1$ and $z_2$ have been considered. The second term cancels with another term which corresponds to the partition $\{\{z_1,z_2\},w_1\text{-}w_2\}$ and contains a BCJ-pattern
\bea
\left(-{1\over 2}\right)\begin{minipage}{2cm}  \includegraphics[width=2cm]{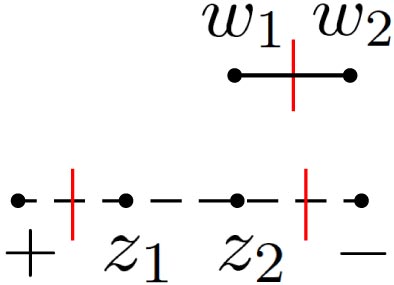} \end{minipage}\times\begin{minipage}{1.8cm}  \includegraphics[width=1.8cm]{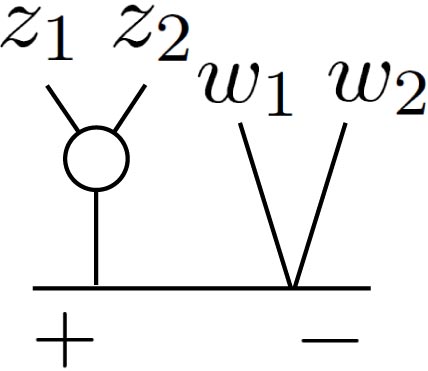} \end{minipage}\times\begin{minipage}{2cm} \includegraphics[width=2cm]{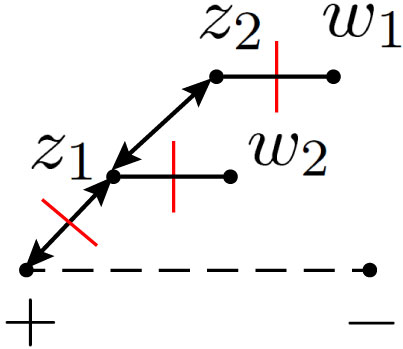} \end{minipage}=\left(-{1\over 2}\right)^3\begin{minipage}{2cm}  \includegraphics[width=2cm]{EQW2EG2L1} \end{minipage}\times s_{z_1,l}\begin{minipage}{2cm}  \includegraphics[width=2cm]{EQW2EG2D3} \end{minipage}\times\begin{minipage}{2cm} \includegraphics[width=2cm]{EQW2EG2R7} \end{minipage}.\nn\Label{EQ:W2EG2SixVertex2}
\eea
The first term in (\ref{EQ:W2EG2SixVertex1}) survives and is explicitly written as 
\bea
{1\over l^2}\left(-{1\over 2}\right)^3\,({\W\epsilon}_{z_1}\cdot{\W\epsilon}_{w_2}) ({\epsilon}_{w_2}\cdot{\epsilon}_{w_1})({\W\epsilon}_{w_1}\cdot{\W\epsilon}_{z_2})={1\over l^2}\left(-{1\over 2}\right)^3\,T(z_1)\cdot T(w_2)\cdot T(w_1)\cdot T(z_2),
\eea
which induces a  {\it $2x$-$2z$-$2w$ vertex}
\bea
\begin{minipage}{2.3cm}  \includegraphics[width=2.3cm]{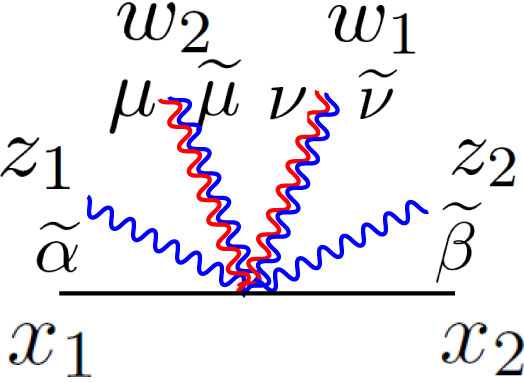} \end{minipage}= \left(-{1\over 2}\right)^3\eta^{\W\alpha\W\mu}\eta^{\mu\nu}\eta^{\W\nu\W\beta}. \Label{Eq:2x2z2wVertex}
\eea

{\bf Comments:} As in the $2x$-$2y$-$2w$ case, the reference order and the left permutation $(z_1z_2)$ forbid the vertex with $({\W\epsilon}_{z_2}\cdot{\W\epsilon}_{w_2}) ({\epsilon}_{w_2}\cdot{\epsilon}_{w_1})({\W\epsilon}_{w_1}\cdot{\W\epsilon}_{z_1})$. The full cancellation of this term is displayed in the appendix.

\subsection{New five-point vertices}
In the previous discussions, we have pointed out that the six-point vertex structures are produced in the $|\mathsf{W}|=2$ case. These vertices are not expected in the partial integrands with $|\mathsf{W}|\leq 1$. In this subsection, we demonstrate that the localization of  $|\mathsf{W}|=2$ partial integrands also introduces new five-point vertices which were not produced by  $|\mathsf{W}|\leq 1$.  The two $\mathsf{W}$ elements belong to two distinct currents which are contracted via the new five-point vertices. Such possible vertex structures are given as follows
\bea
&T\left(\mathsf{Y}_l\right)\cdot T\left(w_i\right)\cdot T\left(w_j\right)\cdot k,&
~~~~~~~~T\left(\mathsf{Z}_l\right)\cdot T\left(w_i\right)\cdot T\left(w_j\right)\cdot k,\nn
&\text{Tr}[T\left(w_i\right)\cdot T\left(w_j\right)]\,T\left(\mathsf{Y}_l\right)\cdot k,&~~~~~~~~\text{Tr}[T\left(w_i\right)\cdot T\left(w_j\right)]\,T\left(\mathsf{Z}_l\right)\cdot k,\nn
&T(\mathsf{Y}_l\cup \{w_{i}\})\cdot T(w_j)\cdot T(\mathsf{Z}_l)&~~~~~~~~T(\mathsf{Y}_l)\cdot T(w_j)\cdot T(\mathsf{Z}_l\cup\{w_i\}),
\eea
where $w_i$, $w_j$ are the two $\mathsf{W}$ elements. The $k^{\mu}$ is the momentum of the loop propagator or the sum of momenta of an effective current. Now we extract these vertex structures by localization of partial integrands.

\subsubsection{$2x$-$1y$-$2w$ vertex}

To produce a $2x$-$1y$-$2w$ vertex, we study the partial integrand with $\mathsf{Y}=\{y_1\}$, $\mathsf{W}=\{w_1,w_2\}$. The left and right reference orders are $y_1\prec w_1\prec w_2$ and $w_1\prec w_2$, respectively. There are three terms contributing to the $2x$-$1y$-$2w$ vertex. We first consider the following term with partition $\{w_1,y_1\text{-}w_2\}$
 \bea
\left(-{1\over 2}\right)\begin{minipage}{1.85cm}  \includegraphics[width=1.85cm]{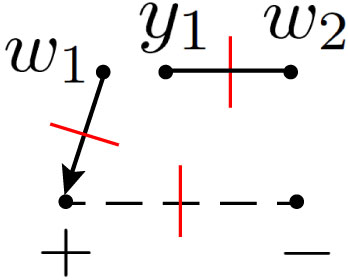} \end{minipage}\times\begin{minipage}{1.8cm}  \includegraphics[width=1.8cm]{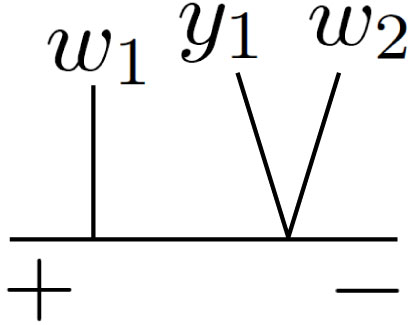} \end{minipage}\times\begin{minipage}{1.9cm} \includegraphics[width=1.9cm]{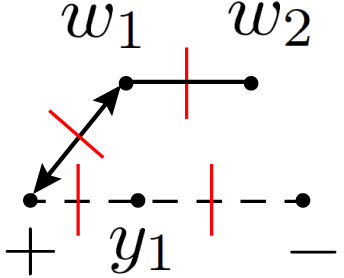} \end{minipage}
&=&\left(-{1\over 2}\right)\begin{minipage}{1.85cm}  \includegraphics[width=1.85cm]{EQW2EG3L1} \end{minipage}\times\begin{minipage}{1.8cm}  \includegraphics[width=1.8cm]{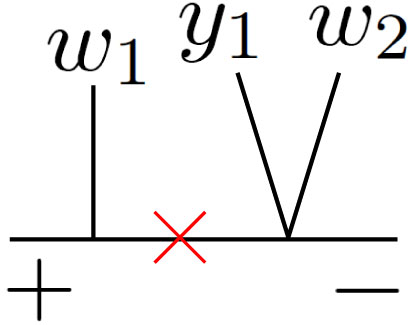} \end{minipage}\times\begin{minipage}{1.9cm} \includegraphics[width=1.9cm]{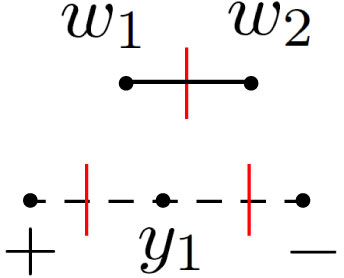} \end{minipage}\nn
&=&{1\over l^2}\left(-{1\over 2}\right)^2 T(y_1)\cdot T(w_2)\cdot T(w_1)\cdot l.\Label{2x1y2wVertex1}
\eea
When the property (\ref{Eq:PropertyXPattern}) of X-pattern and on-shell condition of $w_1$ are applied, we get a  $2x$-$1y$-$2w$ vertex structure on the last line. 

The second term related to a  $2x$-$1y$-$2w$ vertex is given as follows  
  \bea
\left(-{1\over 2}\right)\begin{minipage}{2.1cm}  \includegraphics[width=2.1cm]{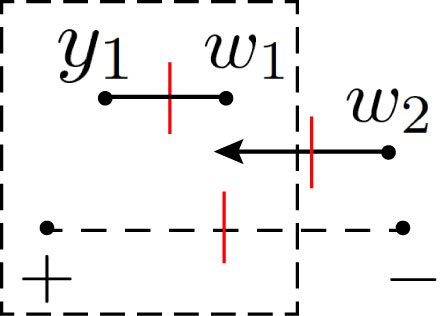} \end{minipage}\times\begin{minipage}{1.7cm}  \includegraphics[width=1.7cm]{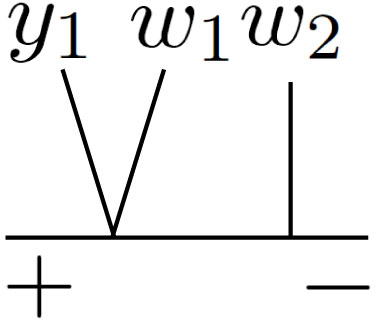} \end{minipage}\times\begin{minipage}{1.9cm} \includegraphics[width=1.9cm]{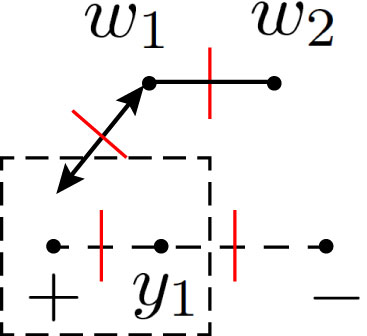} \end{minipage}\to{1\over l^2}\left(-{1\over 2}\right)^2 T(y_1)\cdot T(w_1)\cdot T(w_2)\cdot l_{w_1y_1}.\Label{2x1y2wVertex2}
\eea
The right graph contains an X-pattern associated with $w_1$. When (\ref{Eq:PropertyXPattern}) is applied, it splits into two terms. According to localization-2, one of them cancels with another term containing BCJ-pattern. The surviving term, i.e., the rhs. of above expression contributes to a $2x$-$1y$-$2w$  vertex.

Now we turn to the third term with partition $\{\{y_1,w_1\}\text{-}w_2\}$, which contains three left graphs and a BCJ relation associated with the right coefficient
\bea
&&\left(-{1\over 2}\right)\left[\begin{minipage}{1.8cm}  \includegraphics[width=1.8cm]{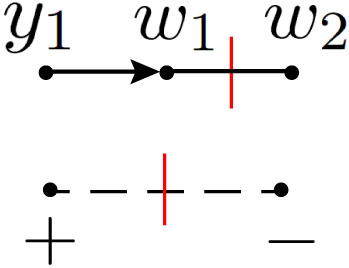} \end{minipage}+\begin{minipage}{1.8cm}  \includegraphics[width=1.8cm]{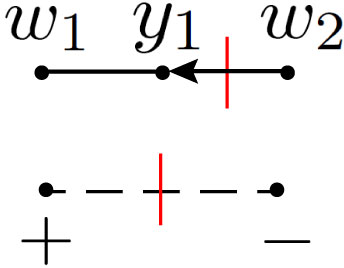} \end{minipage}+\begin{minipage}{1.8cm}  \includegraphics[width=1.8cm]{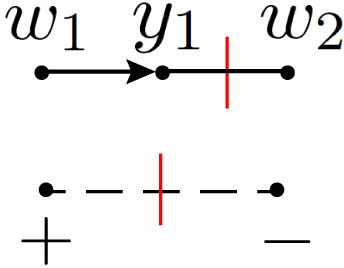} \end{minipage}\right]\times\begin{minipage}{1.5cm}  \includegraphics[width=1.5cm]{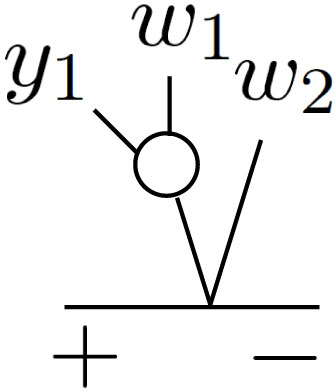} \end{minipage}\times\begin{minipage}{1.7cm} \includegraphics[width=1.7cm]{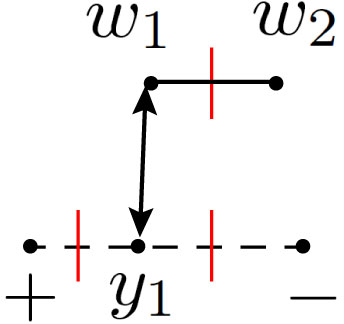} \end{minipage}\Label{2x1y2wVertex3}\\
&\!\!\to\!\!&{1\over l^2}\left(-{1\over 2}\right)^2 \Big[-(\epsilon_{y_1}\cdot k_{w_1})(\epsilon_{w_1}\cdot\epsilon_{w_2})-(\epsilon_{w_2}\cdot k_{y_1})(\epsilon_{w_1}\cdot\epsilon_{y_1})+(\epsilon_{w_1}\cdot k_{y_1})(\epsilon_{y_1}\cdot\epsilon_{w_2})\Big]({\W\epsilon}_{w_2}\cdot {\W\epsilon}_{w_1})\nn
&\!\!=\!\!&{1\over l^2}\left(-{1\over 2}\right)^2 \Big[-\left(T(y_1)\cdot k_{w_1}\right) \text{Tr}[T(w_1)\cdot T(w_2)]-T(y_1)\cdot T(w_1)\cdot T(w_2)\cdot k_{y_1}+T(y_1)\cdot T(w_2)\cdot T(w_1)\cdot k_{y_1}\Big].\nonumber
\eea
On the second line, the off-shell BCJ relation (\ref{Eq:OffBCJ1}) was used. We have rewritten the polarizations as subcurrents with a single particle and expressed $(\epsilon_{w_1}\cdot\epsilon_{w_2})({\W\epsilon}_{w_2}\cdot {\W\epsilon}_{w_1})$ by trace in the first term of the third line.

The terms (\ref{2x1y2wVertex1}), (\ref{2x1y2wVertex2}) and (\ref{2x1y2wVertex3}) together finally introduce a $2x$-$1y$-$2w$ vertex
\bea
\begin{minipage}{2.3cm}  \includegraphics[width=2.3cm]{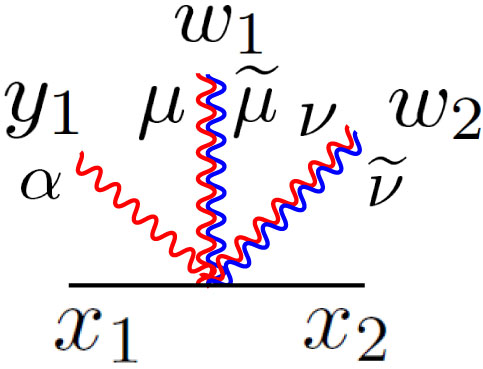} \end{minipage}= \left(-{1\over 2}\right)^2\left[\eta^{\alpha\nu}\eta^{\W\nu\W\mu}(k^{\mu}_{x_1}+k^{\mu}_{y_1})+\eta^{\alpha\mu}\eta^{\W\mu\W\nu}(k^{\nu}_{x_1}+k^{\nu}_{w_1})-\eta^{\mu\nu}\eta^{\W\nu\W\mu}k_{w_1}^{\alpha}\right]. \Label{Eq:2x1y2wVertex}
\eea
The left and right weights of currents attached to this vertex satisfy $y_1\prec w_1\prec w_2$, $w_1\prec w_2$, respectively.

\subsubsection{$2x$-$1z$-$2w$ vertex}

On the dual side, once we exchange the roles of $\mathsf{Y}$ and $\mathsf{Z}$ elements, the $2x$-$1y$-$2w$ vertex (\ref{Eq:2x1y2wVertex}) immediately induces a  $2x$-$1z$-$2w$ vertex. Now we verify this statement by canceling the right nonlocality of the following three terms with respect to the partition $\{z_1,w_1\text{-}w_2\}$
 \bea
&\!\!\!\!\!\!\!&\left(-{1\over 2}\right)\begin{minipage}{1.7cm}  \includegraphics[width=1.7cm]{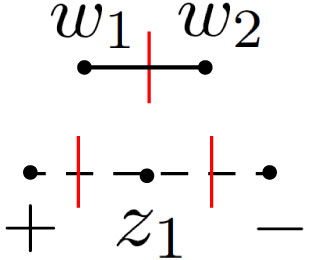} \end{minipage}\times\begin{minipage}{1.65cm}  \includegraphics[width=1.65cm]{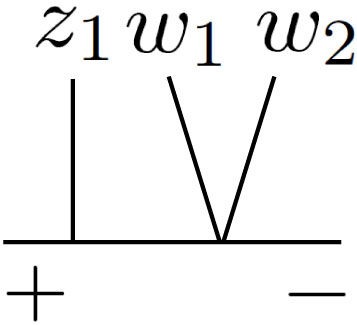} \end{minipage}\times\begin{minipage}{1.6cm} \includegraphics[width=1.6cm]{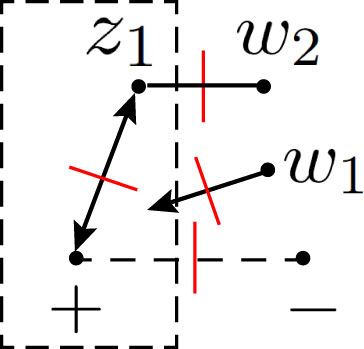} \end{minipage}\to{1\over l^2}\left(-{1\over 2}\right)^2  T(z_1)\cdot T(w_2)\cdot T(w_1)\cdot l_{z_1},\\
&\!\!\!\!\!\!\!&\left(-{1\over 2}\right)\begin{minipage}{1,7cm}  \includegraphics[width=1.7cm]{EQW2EG3L3} \end{minipage}\times\begin{minipage}{1.65cm}  \includegraphics[width=1.65cm]{EQW2EG3D3} \end{minipage}\times\left[\begin{minipage}{1.55cm} \includegraphics[width=1.55cm]{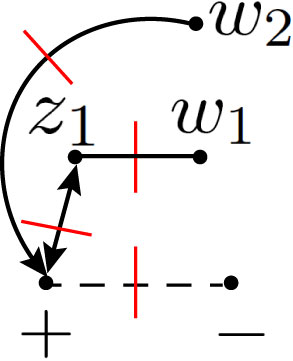} \end{minipage}+\begin{minipage}{1.55cm} \includegraphics[width=1.55cm]{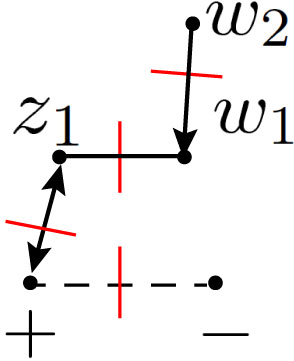} \end{minipage}\right]\!\!\to\!\!{1\over l^2}\left(-{1\over 2}\right)^2  T(z_1)\cdot T(w_1)\cdot T(w_2)\cdot l_{w_1},\\
&\!\!\!\!\!\!\!&\left(-{1\over 2}\right)\begin{minipage}{1.7cm}  \includegraphics[width=1.7cm]{EQW2EG3L3} \end{minipage}\times\begin{minipage}{1.5cm}  \includegraphics[width=1.48cm]{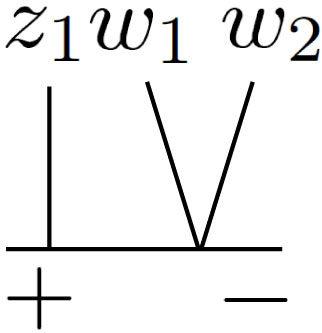} \end{minipage}\times\begin{minipage}{1.9cm} \includegraphics[width=1.9cm]{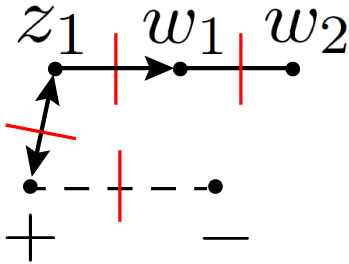} \end{minipage}\!\!\!\!\!\to\!\!{1\over l^2}(-1)\left(-{1\over 2}\right)^2 \left(T(z_1)\cdot k_{w_1}\right) \text{Tr}[T(w_1)\cdot T(w_2)].\eea
These three terms together imply the following $2x$-$1z$-$2w$ vertex
\bea
\begin{minipage}{2.3cm}  \includegraphics[width=2.3cm]{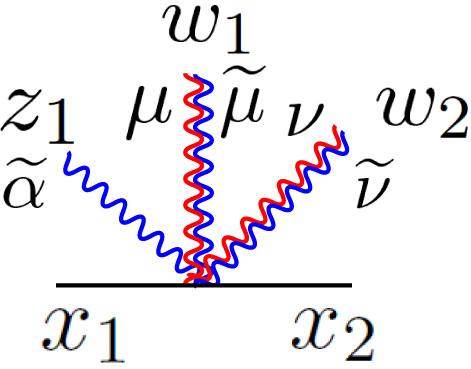} \end{minipage}= \left(-{1\over 2}\right)^2\left[\eta^{\W\alpha\W\nu}\eta^{\nu\mu}(k^{\W\mu}_{x_1}+k^{\W\mu}_{z_1})+\eta^{\W\alpha\W\mu}\eta^{\mu\nu}(k^{\W\nu}_{x_1}+k^{\W\nu}_{w_1})-\eta^{\mu\nu}\eta^{\W\nu\W\mu}k_{w_1}^{\W\alpha}\right],\Label{Eq:2x1z2wVertex}
\eea
where the left and right weights of currents attached to the vertex satisfy $w_1\prec w_2$, $z_1\prec w_1\prec w_2$, respectively. The vertex (\ref{Eq:2x1z2wVertex}) is just the dual of (\ref{Eq:2x1y2wVertex}).

\subsubsection{New $2x$-$1y$-$1z$-$1w$ vertex}

In the partial integrand with $|\mathsf{W}|=1$, we have already constructed a $2x$-$1y$-$1z$-$1w$ vertex (\ref{Eq:2x1y1z1wVertex}), where the weight of the current attached to  $w_1$ is higher than those of $y_1$ and $z_1$. When another $\mathsf{W}$ element is added, a current with both  $\mathsf{Y}$ and  $\mathsf{W}$ ($\mathsf{Z}$ and  $\mathsf{W}$) elements is contracted with the $y_1$ ($z_1$) line. Such a current may have a higher weight than the one contracted to the $w_1$ line. Thus, this situation induces new $2x$-$1y$-$1z$-$1w$ vertices. First, the following term induces a $2x$-$1y$-$1z$-$1w$ vertex structure
\bea
\left(-{1\over 2}\right)\begin{minipage}{2.1cm}  \includegraphics[width=2.1cm]{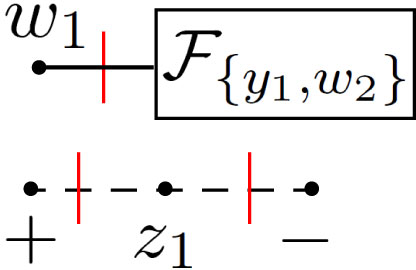} \end{minipage}\times\begin{minipage}{1.8cm}  \includegraphics[width=1.8cm]{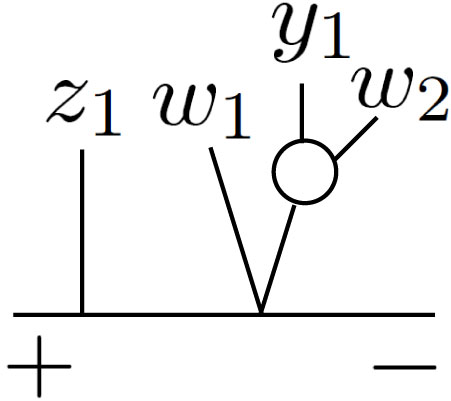} \end{minipage}\times\begin{minipage}{1.8cm} \includegraphics[width=1.8cm]{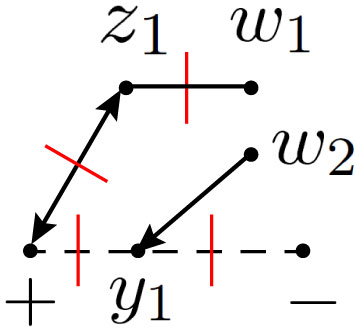} \end{minipage}={1\over l^2}\left(-{1\over 2}\right)^2T(y_1,w_2)\cdot T(w_1)\cdot T(z_1),\Label{Eq:2x1y1z1wVertexTerm1}
\eea
which follows from a similar discussion with (\ref{EQ:W1EG1FiveVertex1}). This term implies a new $2x$-$1y$-$1z$-$1w$ vertex
\bea
\begin{minipage}{1.5cm}  \includegraphics[width=1.5cm]{2x1y1z1wVertex} \end{minipage}= \left(-{1\over 2}\right)^2\eta^{\mu\nu}\eta^{\W\nu\W\tau} ~~~~~~~(w_1\prec y_1, z_1\prec w_1). \Label{Eq:2x1y1z1wVertex1} 
\eea
It is worth pointing out that the form of this vertex is the same as (\ref{Eq:2x1y1z1wVertex}), but the weight of the subcurrent containing $w_1$ is lower than that of the subcurrent with $y_1$. Another term is given by
\bea
\left(-{1\over 2}\right)\begin{minipage}{1.5cm}  \includegraphics[width=1.5cm]{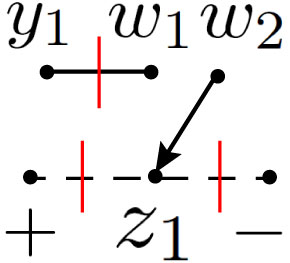} \end{minipage}\times\begin{minipage}{1.8cm}  \includegraphics[width=1.8cm]{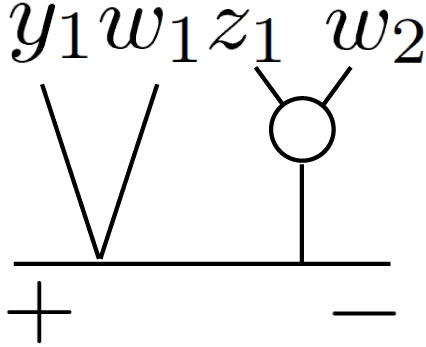} \end{minipage}\times\begin{minipage}{2.15cm} \includegraphics[width=2.15cm]{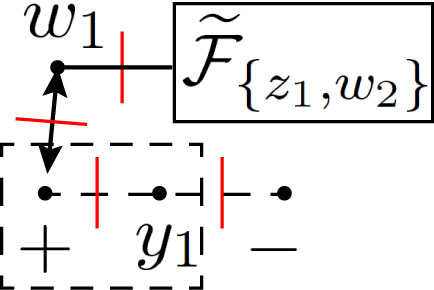} \end{minipage}\to{1\over l^2}\left(-{1\over 2}\right)^2T(z_1,w_2)\cdot T(w_1)\cdot T(y_1),\Label{Eq:2x1y1z1wVertexTerm2}
\eea
where the right graph has an X-pattern associated with $w_1$. When the property (\ref{Eq:PropertyXPattern}) is applied to this X-pattern, the last term in (\ref{Eq:PropertyXPattern}) vanishes due to on-shell condition, the second term cancel with another term with partition $\{\{y_1,w_1\},\{z_1,w_2\}\}$. The first term in (\ref{Eq:PropertyXPattern}) survives and provides the rhs. of (\ref{Eq:2x1y1z1wVertexTerm2}). The new vertex for this term is 
\bea
\begin{minipage}{1.5cm}  \includegraphics[width=1.5cm]{2x1y1z1wVertex} \end{minipage}= \left(-{1\over 2}\right)^2\eta^{\mu\nu}\eta^{\W\nu\W\tau}~~~~~~~( y_1\prec w_1, w_1\prec z_1 ), \Label{Eq:2x1y1z1wVertex2} 
\eea
which is the dual of the vertex (\ref{Eq:2x1y1z1wVertex1}).

\subsection{New four-point vertices}
Comparing with the case $|\mathsf{W}|=1$, the $w_2$ introduces new four-point vertices which involve two $\mathsf{W}$ elements. Particularly, we consider the following terms with partitions $\{w_1\text{-}w_2\}$ and $\{w_1,w_2\}$
\bea
&&\left(-{1\over 2}\right)\begin{minipage}{1.4cm}  \includegraphics[width=1.4cm]{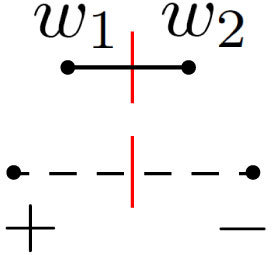} \end{minipage}\times\begin{minipage}{1cm}  \includegraphics[width=1cm]{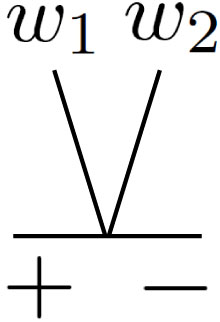} \end{minipage}\times\left[\,\begin{minipage}{1.4cm} \includegraphics[width=1.4cm]{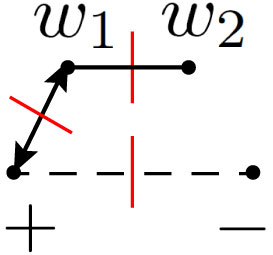} \end{minipage}+\begin{minipage}{1.4cm} \includegraphics[width=1.4cm]{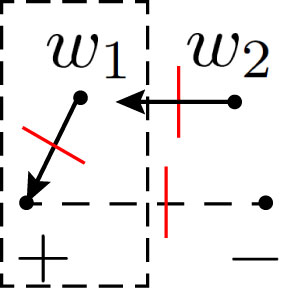} \end{minipage}\,\right]\,+\,\begin{minipage}{1.4cm}  \includegraphics[width=1.4cm]{EQW2EG4R2} \end{minipage}\times\begin{minipage}{1.25cm}  \includegraphics[width=1.25cm]{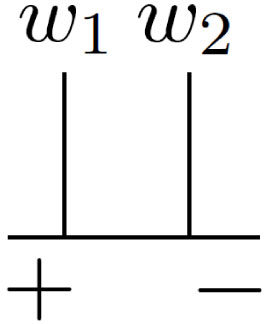} \end{minipage}\times\begin{minipage}{1.4cm} \includegraphics[width=1.4cm]{EQW2EG4R1} \end{minipage}\nn
&=&{1\over l^2}\left(-{1\over 2}\right)\left[(-1) \left(k_{w_1}\cdot l\right) \text{Tr}[T(w_1)\cdot T(w_2)]+ \W l\cdot T(w_1)\cdot T(w_2)\cdot {\W l}_{w_1}+l\cdot T(w_1)\cdot T(w_2)\cdot l_{w_1}\right],
\eea
where the first line has been directly localized into the second line. This local expression provides a $2x$-$2w$ vertex
\bea
\begin{minipage}{1.6cm}  \includegraphics[width=1.6cm]{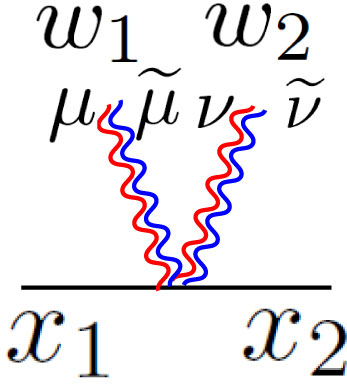} \end{minipage}=\left(-{1\over 2}\right)\left[(-1) \left(k_{w_1}\cdot k_{x_1}\right) \eta^{\mu\nu}\eta^{\W\nu\W\mu}+ k_{x_1}^{\W\mu}\eta^{\mu\nu}\left(k_{x_1}^{\W\nu}+k_{w_1}^{\W\nu}\right)+ k_{x_1}^{\mu}\eta^{\W\mu\W\nu}\left(k_{x_1}^{\nu}+k_{w_1}^{\nu}\right)\right],\Label{Eq:2wVertex}
\eea
in which both left and right reference orders are chosen as $w_1\prec w_2$.

 Another four-point vertex structure is given by the following localization 
\bea
\left(-{1\over 2}\right)\begin{minipage}{2.1cm}  \includegraphics[width=2.1cm]{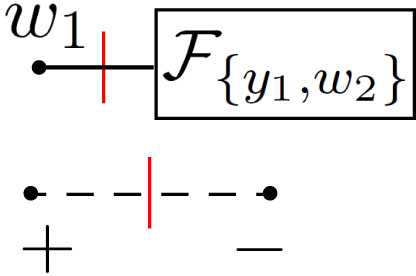} \end{minipage}\times\begin{minipage}{1.25cm}  \includegraphics[width=1.25cm]{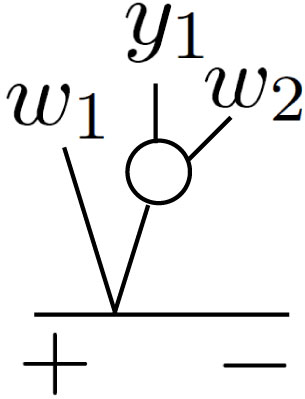} \end{minipage}\times\begin{minipage}{1.4cm} \includegraphics[width=1.4cm]{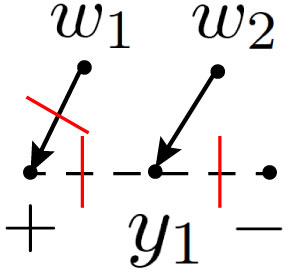} \end{minipage}={1\over l^2}\left(-{1\over 2}\right)T(y_1,w_2)\cdot T(w_1)\cdot \W l,
\eea
which induces a new $2x$-$1y$-$1w$ vertex
\bea
\begin{minipage}{1.6cm}  \includegraphics[width=1.6cm]{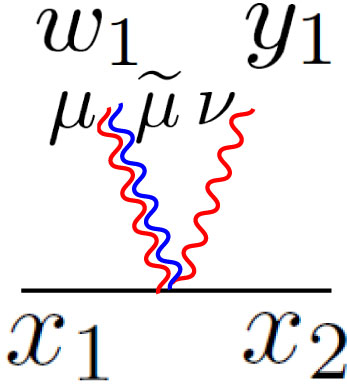} \end{minipage}=\left(-{1\over 2}\right)\eta^{\nu\mu}k_{x_1}^{\W\mu}.\Label{Eq:New2x1y1wVertex}
\eea
This vertex is different from the $2x$-$1y$-$1w$ vertex (\ref{Eq:2x1y1wVertex}), since in the $|\mathsf{W}|=1$ case we always have the left reference order  $y\prec w_1$ element. When the other $\mathsf{W}$ element $w_2$ ($w_1\prec w_2$) is added, the subcurrent $T^{\mu}(y_1,w_2)$ plays as a $\mathsf{Y}$ element, but its weight is higher than that of $w_1$. Thus, this vertex only appears if there exists such a subcurrent.

The dual of the vertex (\ref{Eq:New2x1y1wVertex}) is created by the following localization
\bea
\begin{minipage}{1.4cm}  \includegraphics[width=1.4cm]{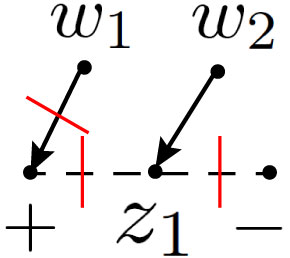} \end{minipage}\times\begin{minipage}{1.4cm}  \includegraphics[width=1.4cm]{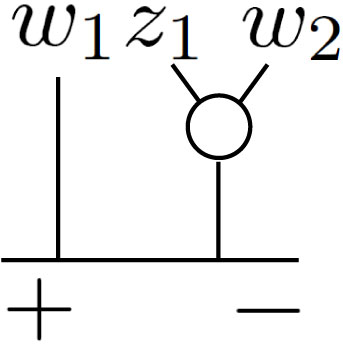} \end{minipage}\times\begin{minipage}{2.1cm} \includegraphics[width=2.1cm]{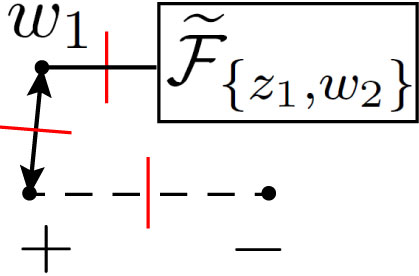} \end{minipage}={1\over l^2}\left(-{1\over 2}\right)T(z_1,w_2)\cdot T(w_1)\cdot l.
\eea
Apparently, this term provides a new $2x$-$1z$-$1w$ vertex with the reference order $w_1\prec z_1$
\bea
\begin{minipage}{1.6cm}  \includegraphics[width=1.6cm]{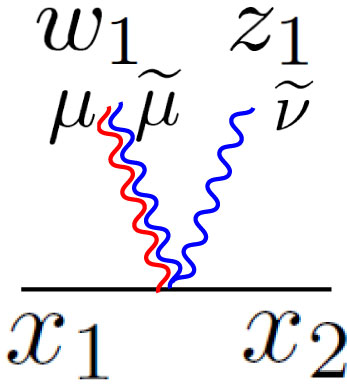} \end{minipage}=\left(-{1\over 2}\right)\eta^{\W\nu\W\mu}k_{x_1}^{\mu}.\Label{Eq:New2x1z1wVertex}
\eea
As in (\ref{Eq:New2x1y1wVertex}), the $z_1$ line in (\ref{Eq:New2x1z1wVertex}) must be contracted with a subcurrent containing both $\mathsf{Z}$
and  $\mathsf{W}$ elements. The weight of this subcurrent should be higher than that of $w_1$.

\subsection{General local formula for partial and full integrands with $|\mathsf{W}|=2$}\label{sec:7.4}
Now we extend the $|\mathsf{W}|=2$ examples to general local formula of partial and full integrands. The local partial integrand $\mathcal{A}^{\,\text{dYMS},|\mathsf{W}|=2}$ is written as 
\bea
\mathcal{A}^{\,\text{dYMS},|\mathsf{W}|=2}=\,\Sl_{\{A_1A_2...A_I\}}{1\over l^2}\,J[A_1]\, {1\over s_{A_1,l}}\,J[A_2]\cdots\,{1\over s_{A_1...A_{I-1},l}}\,J[A_I], \Label{Eq:doubleYMSW2Finalab}
\eea
where $J[A_i]$ is given by
\bea%
J[A_i]&=&J_{(3)}[A_i]+J_{(4)}[A_i]+J_{(5)}[A_i]+J_{(6)}[A_i].
\eea
The $J_{(a)}[A_i]$ ($a=3,4,5,6$) are $a$-point vertex structures. All  $J_{(3)}$ are those already constructed in the cases of $|\mathsf{W}|\leq 1$. If $A_i$ contains less than two $\mathsf{W}$ elements, the $J_{(4)}[\mathsf{Y}_i]$ ,  $J_{(4)}[\mathsf{Z}_i]$, $J_{(4)}[\mathsf{Y}_i\cup\{w_j\}]$ ,  $J_{(4)}[\mathsf{Z}_i\cup\{w_j\}]$,  $J_{(5)}[\mathsf{Y}_i\cup\mathsf{Z}_i\cup\{w_j\}]$ ($j=1,2$) are also defined in the $|\mathsf{W}|\leq 1$ case. For a subset $A_i$ involving two   $\mathsf{W}$ elements, the following new terms are included.

{\bf Six-point vertex structures} are given by
\bea
J_{\text{(6)}}[\mathsf{Y}_i\cup \mathsf{W}]&=&\Sl_{\{\mathsf{Y}_{i_1}\mathsf{Y}_{i_2}\}}\left(-{1\over2}\right)^3 T(\mathsf{Y}_{i_1})\cdot T(w_2)\cdot T(w_1)\cdot T(\mathsf{Y}_{i_2}),\\
J_{\text{(6)}}[\mathsf{Z}_i\cup \mathsf{W}]&=&\Sl_{\{\mathsf{Z}_{i_1}\mathsf{Z}_{i_2}\}}\left(-{1\over2}\right)^3 T(\mathsf{Z}_{i_1})\cdot T(w_2)\cdot T(w_1)\cdot T(\mathsf{Z}_{i_2}),
\eea
where we have summed over all partitions $\mathsf{Y}_i\to \{\mathsf{Y}_{i_1}\mathsf{Y}_{i_2}\}$ ($\mathsf{Z}_i\to \{\mathsf{Z}_{i_1}\mathsf{Z}_{i_2}\}$) such that the relative ordering of elements in $\mathsf{Y}_1$ and $\mathsf{Y}_2$ ($\mathsf{Z}_1$ and $\mathsf{Z}_2$) agrees with the right (left) permutations, respectively.

{\bf Five-point vertex structures} are given by
\bea
J_{\text{(5)}}[\mathsf{Y}_i\cup \mathsf{W}]&=&\left(-{1\over2}\right)^2 \Big[T(\mathsf{Y}_i)\cdot T(w_1)\cdot T(w_2)\cdot \left(X_{\mathsf{Y}_i\cup \mathsf{W}}+k_{w_1}\right)\nn
&&+T(\mathsf{Y}_i)\cdot T(w_2)\cdot T(w_1)\cdot \left(X_{\mathsf{Y}_i\cup \mathsf{W}}+k_{\,\mathsf{Y}_i}\right)-T(\mathsf{Y}_i)\cdot k_{w_1}\text{Tr}[T(w_1)\cdot T(w_2)]\Big],\\
J_{\text{(5)}}[\mathsf{Z}_i\cup \mathsf{W}]&=&\left(-{1\over2}\right)^2 \Big[T(\mathsf{Z}_i)\cdot T(w_1)\cdot T(w_2)\cdot \left(X_{\mathsf{Z}_i\cup \mathsf{W}}+k_{w_1}\right)\nn
&&+T(\mathsf{Z}_i)\cdot T(w_2)\cdot T(w_1)\cdot \left(X_{\mathsf{Z}_i\cup \mathsf{W}}+k_{\,\mathsf{Z}_i}\right)-T(\mathsf{Z}_i)\cdot k_{w_1}\text{Tr}[T(w_1)\cdot T(w_2)]\Big],\\
J_{\text{(5)}}[\mathsf{Y}_i\cup \mathsf{Z}_i\cup \mathsf{W}]&=&\left(-{1\over2}\right)^2\Big[ T(\mathsf{Y}_i\cup \{w_1\})\cdot T(w_2)\cdot T(\mathsf{Z}_i)+ T(\mathsf{Y}_i)\cdot T(w_2)\cdot T(\mathsf{Z}_i\cup \{w_1\})\\
&&+ T(\mathsf{Y}_i)\cdot T(\mathsf{W})\cdot T(\mathsf{Z}_i)+ T(\mathsf{Y}_i\cup \{w_2\})\cdot T(w_1)\cdot T(\mathsf{Z}_i)+ T(\mathsf{Y}_i)\cdot T(w_1)\cdot T(\mathsf{Z}_i\cup \{w_2\})\Big].\nonumber
\eea
In the expression of $J_{\text{(5)}}[\mathsf{Y}_i\cup \mathsf{Z}_i\cup \mathsf{W}]$, the first three terms are the structures in which subcurrents are attached to the vertex (\ref{Eq:2x1y1z1wVertex}), while the last two terms correspond to the vertices (\ref{Eq:2x1y1z1wVertex1}) and  (\ref{Eq:2x1y1z1wVertex2}).

{\bf Four-point vertex structures} are displayed as follows
\bea
J_{\text{(4)}}[\mathsf{Y}_i\cup \mathsf{W}]&=&\left(-{1\over2}\right)\bigg[\Sl_{\{\mathsf{Y}_{i_1}\mathsf{Y}_{i_2}\}}T(\mathsf{Y}_{i_1}\cup \{w_1\})\cdot T(\mathsf{Y}_{i_2}\cup \{w_2\})+T(\mathsf{Y}_{i})\cdot T(\mathsf{W})\cdot (X_{\mathsf{Y}_i\cup \mathsf{W}}+k_{\mathsf{Y}_i})\Label{Eq:W2J41}\\
&&~~~~~~~\,+T(\mathsf{Y}_{i}\cup \{w_1\})\cdot T(w_2)\cdot \left(X_{\mathsf{Y}_i\cup \mathsf{W}}+k_{\mathsf{Y}_i\cup \{w_1\}}\right)
+T(\mathsf{Y}_{i}\cup \{w_2\})\cdot T(w_1)\cdot X_{\mathsf{Y}_i\cup \mathsf{W}}
\bigg],\nn
J_{\text{(4)}}[\mathsf{Z}_i\cup \mathsf{W}]&=&\left(-{1\over2}\right)\bigg[\Sl_{\{\mathsf{Z}_{i_1}\mathsf{Z}_{i_2}\}}T(\mathsf{Z}_{i_1}\cup \{w_1\})\cdot T(\mathsf{Z}_{i_2}\cup \{w_2\})+T(\mathsf{Z}_{i})\cdot T(\mathsf{W})\cdot (X_{\mathsf{Z}_i\cup \mathsf{W}}+k_{\mathsf{Z}_i})\Label{Eq:W2J42}\\
&&~~~~~~~\,+T(\mathsf{Z}_{i}\cup \{w_1\})\cdot T(w_2)\cdot \left(X_{\mathsf{Z}_i\cup \mathsf{W}}+k_{\mathsf{Z}_i\cup \{w_1\}}\right)+T(\mathsf{Z}_{i}\cup \{w_2\})\cdot T(w_1)\cdot X_{\mathsf{Z}_i\cup \mathsf{W}}\bigg],\nn
J_{\text{(4)}}[\mathsf{W}]&=&\left(-{1\over 2}\right)\Big[X_{\,\W{\mathsf{W}}}\cdot T(w_1)\cdot T(w_2)\cdot \left(X_{\,\W{\mathsf{W}}}+k_{\,\W{w}_{1}}\right)+X_{\,\mathsf{W}}\cdot T(w_1)\cdot T(w_2)\cdot \left(X_{\mathsf{W}}+k_{\,{w_1}}\right)\nn
&&~~~~~~~\,+\left(k_{w_1}\cdot X_{\,\mathsf{W}}\right)\text{tr}\left[\,T(w_1)\cdot T(w_2)\right]\Big].
\eea
The  first three terms of (\ref{Eq:W2J41}) and (\ref{Eq:W2J42}), are vertex structures based on (\ref{Eq:2x1y1wVertex}) and (\ref{Eq:2x1z1wVertex}), respectively. The last terms of (\ref{Eq:W2J41}) and (\ref{Eq:W2J42}) correspond to the vertices (\ref{Eq:New2x1y1wVertex}) and (\ref{Eq:New2x1z1wVertex}).

As already pointed out in the $|\mathsf{W}|=0$ and  $|\mathsf{W}|=1$ cases, once the localization of partial integrands has been accomplished, one can always find the local partial integrands which are related to  (\ref{Eq:doubleYMSW2Finalab}) by cyclic permutations. Thus, according to (\ref{Eq:partial}), these partial integrands finally result in the local full integrand with quadratic propagators
\bea
\mathcal{I}^{\,\text{dYMS},|\mathsf{W}|=2}=\,\Sl_{\{A_1A_2...A_I\}}{1\over l^2}\,J[A_1]\, {1\over l^2_{A_1}}\,J[A_2]\cdots\,{1\over l^2_{A_1...A_{I-1}}}\,J[A_I]. \Label{Eq:doubleYMSW2Final}
\eea
%

%
\section{Double YMS with $|\mathsf{W}|=3$}\label{sec:W3}
In this section, we study the dYMS partial integrand involving three $\mathsf{W}$ particles $w_1$, $w_2$ and $w_3$. Comparing with $|\mathsf{W}|\leq 2$, when a new $\mathsf{W}$ element is added, new vertex structures will be introduced. In principle, we could obtain all local vertices by canceling the nonlocalities from the right graphs of (\ref{Eq:doubleYMS4}), following a similar approach to that in the previous sections. Nevertheless, the complexity of the localization increases with the number of $\mathsf{W}$ elements. In this section, we provide an alternative algorithm to create new vertices for the integrands with $|\mathsf{W}|=3$. Those vertices produced in the previous sections can also be verified by this algorithm.

As an example, we demonstrate the extraction of $2x$-$3w$ vertex from the double-YMS partial integrand  $\mathcal{A}^{\,\text{dYMS}}\big(\big.+,-||\{w_1,w_2,w_3\}\,\big|\,+,-||\{w_1,w_2,w_3\}\,\big)$. On one hand, the double-YMS partial integrand can be directly evaluated by (\ref{Eq:doubleYMS20}), which is nonlocal and is expressed by linear propagators. On another hand, when we try to construct the local partial integrand $\mathcal{A}^{\,\text{dYMS}}$, those vertices which have been induced for the cases with less than three $\mathsf{W}$ elements must be involved. Hence a new vertex structure with three $\mathsf{W}$ elements can be obtained by subtracting the terms with known vertices from the full partial integrand
\bea
\mathcal{A}^{\,\text{dYMS}}_{\,\text{forward}}\big(\big.+,-||\{w_1,w_2,w_3\}\,\big|\,+,-||\{w_1,w_2,w_3\}\,\big)-\mathcal{A}^{\,\text{dYMS}}_{\text{know}}\big(\big.+,-||\{w_1,w_2,w_3\}\,\big|\,+,-||\{w_1,w_2,w_3\}\,\big),\Label{Eq:difference}
\eea
where $\mathcal{A}^{\,\text{dYMS}}_{\,\text{forward}}$ is the one expressed by (\ref{Eq:doubleYMS20}), $\mathcal{A}^{\,\text{dYMS}}_{\text{know}}$ is the local expression constructed by all vertices which have been introduced via $|\mathsf{W}|\leq 2$. Explicitly, this $\mathcal{A}^{\,\text{dYMS}}_{\text{know}}$ is given by 
\bea
\mathcal{A}^{\,\text{dYMS}}_{\text{know}}\big(\big.+,-||\{w_1,w_2,w_3\}\,\big|\,+,-||\{w_1,w_2,w_3\}\,\big)=\Sl_{\{\mathsf{W}_1...\mathsf{W}_I\}}{1\over l^2}J'[\mathsf{W}_1] {1\over s_{\mathsf{W}_1,l}}\cdots{1\over s_{\mathsf{W}_1...\mathsf{W}_I,l}}J'[\mathsf{W}_I],
\eea
where we have summed over all partitions of $\{w_1,w_2,w_3\}$, noting that $I\leq 3$. The expression $J'[\mathsf{W}_i]$ is the sum of possible vertex structures which have already been introduced in the cases of $|\mathsf{W}|\leq 2$
\bea
J'[\mathsf{W}_i]=J_{(3)}[\mathsf{W}_i]+J_{(4)}[\mathsf{W}_i].
\eea
Since the localization result of the full partial integrand $\mathcal{A}^{\,\text{dYMS}}_{\,\text{forward}}$ is expected to be expressed by all possible vertices, and $\mathcal{A}^{\,\text{dYMS}}_{\text{know}}$ includes the contributions of known vertices, the difference (\ref{Eq:difference}) must be a contact term (a term which does not contain loop propagator, except for $1\over l^2$) which is a new vertex structure, diagramatically expressed as 
\bea
\begin{minipage}{1.6cm}  \includegraphics[width=1.6cm]{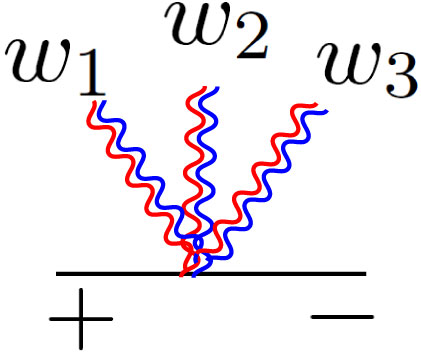} \end{minipage}=\mathcal{A}^{\,\text{dYMS}}_{\,\text{forw}}\big(\big.+,-||\{w_1,w_2,w_3\}\,\big|\,+,-||\{w_1,w_2,w_3\}\,\big)-\left[\begin{minipage}{1.6cm}  \includegraphics[width=1.6cm]{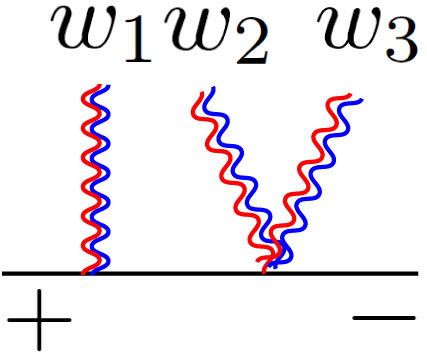} \end{minipage}+\begin{minipage}{1.6cm}  \includegraphics[width=1.6cm]{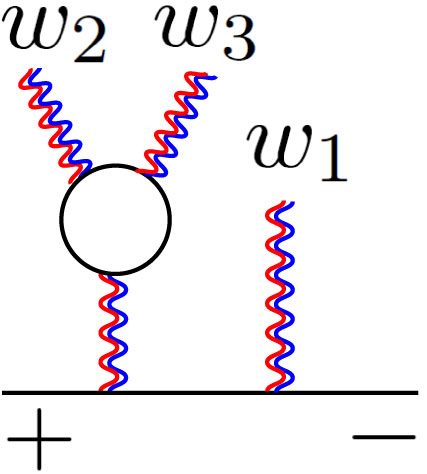} \end{minipage}+...\right].
\eea
The explicit result, after simplification,  provides the $J_{(5)}[\mathsf{W}]$ as
\bea
J_{\text{(5)}}[\,\mathsf{W}\,]&=&\left(-{1\over2}\right)^2 \Big[X_{\mathsf{W}}\cdot T(w_1)\cdot T(w_3)\cdot T(w_2)\cdot \left(X_{\W{\mathsf{W}}}+k_{\W w_1}\right)
+ X_{\W{\mathsf{W}}}\cdot T(w_1)\cdot T(w_3)\cdot T(w_2)\cdot \left(X_{\mathsf{W}}+k_{w_1}\right)\nn
&&+X_{\mathsf{W}}\cdot T(w_2)\cdot T(w_3)\cdot T(w_1)\cdot k_{\W w_2}
+X_{\W{\mathsf{W}}}\cdot T(w_2)\cdot T(w_3)\cdot T(w_1)\cdot k_{w_2}\nn
&&+X_{\mathsf{W}}\cdot T(w_2)\cdot T(w_1)\cdot T(w_3)\cdot k_{\W w_1}
+X_{\W{\mathsf{W}}}\cdot T(w_2)\cdot T(w_1)\cdot T(w_3)\cdot k_{w_1}\nn
&&+X_{\mathsf{W}}\cdot T(w_1)\cdot T(w_2)\cdot T(w_3)\cdot \left(X_{\W{\mathsf{W}}}+k_{\W w_2}\right)
+X_{\W{\mathsf{W}}}\cdot T(w_1)\cdot T(w_2)\cdot T(w_3)\cdot \left(X_{\mathsf{W}}+k_{w_2}\right)\nn
&&-\text{Tr}[\,T(w_1)\cdot T(w_3)\,] X_{\mathsf{W}}\cdot T(w_2)\cdot k_{\W w_1}
-\text{Tr}[\,T(w_1)\cdot T(w_3)\,] X_{\W{\mathsf{W}}}\cdot T(w_2)\cdot k_{w_1}\nn
&&-\text{Tr}[\,T(w_2)\cdot T(w_3)\,] X_{\mathsf{W}}\cdot T(w_1)\cdot k_{\W w_2}
-\text{Tr}[\,T(w_2)\cdot T(w_3)\,] X_{\W{\mathsf{W}}}\cdot T(w_1)\cdot k_{w_2}\Big].
\eea
Thus, the $2x$-$3w$ vertex induced by this contact term is 
\bea
\begin{minipage}{1.8cm}  \includegraphics[width=1.8cm]{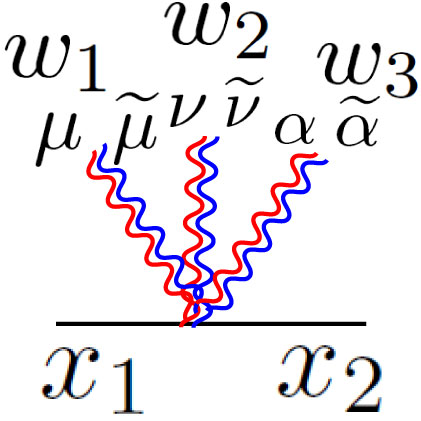} \end{minipage}&=&\left(-{1\over2}\right)^2 \left[ k_{x_1}^{\mu}\eta^{\W\mu\W\alpha}\eta^{\alpha\nu} \left(k_{x_1}^{\W\nu}+k^{\W\nu}_{w_1}\right)+k_{x_1}^{\nu}\eta^{\W\nu\W\alpha}\eta^{\alpha\mu} k^{\W\mu}_{w_2}+ k_{x_1}^{\mu}\eta^{\W\mu\W\nu}\eta^{\nu\alpha} \left(k_{x_1}^{\W\alpha}+k^{\W\alpha}_{w_2}\right)\right.\nn
&&\left.+k_{x_1}^{\nu}\eta^{\W\nu\W\mu}\eta^{\mu\alpha} k^{\W\alpha}_{w_1}-\eta^{\mu\alpha}\eta^{\W\mu\W\alpha}k_{x_1}^{\nu}k_{w_1}^{\W \nu}-\eta^{\nu\alpha}\eta^{\W\nu\W\alpha}k_{x_1}^{\mu}k_{w_2}^{\W \mu}\right]+(\mu\leftrightarrow\W\mu,\nu\leftrightarrow\W\nu,\alpha\leftrightarrow\W\alpha).
\eea
By analogous subtractions, other new vertices induced by $|\mathsf{W}|=3$ are displayed as follows.
\bea
&&\text{\bf $2x$-$1y$-$1z$-$3w$ vertex}\nn
&&\begin{minipage}{2.7cm}  \includegraphics[width=2.7cm]{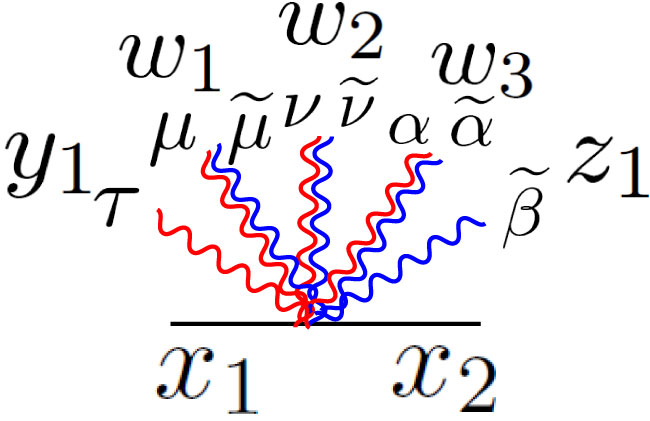} \end{minipage}=\left(-{1\over 2}\right)^4\left[\eta^{\tau\mu}\eta^{\W\mu\W\nu}\eta^{\nu\alpha}\eta^{\W\alpha\W\beta}+\eta^{\tau\mu}\eta^{\W\mu\W\alpha}\eta^{\alpha\nu}\eta^{\W\nu\W\beta}+\eta^{\tau\nu}\eta^{\W\nu\W\alpha}\eta^{\alpha\mu}\eta^{\W\mu\W\beta}+\eta^{\tau\alpha}\eta^{\W\alpha\W\nu}\eta^{\nu\mu}\eta^{\W\mu\W\beta}\right].\Label{Eq:New2x1y1z3wVertex}\\
&&\nn
&&\text{\bf $2x$-$1y$-$3w$ vertex}\nn
&&\begin{minipage}{2.2cm}  \includegraphics[width=2.2cm]{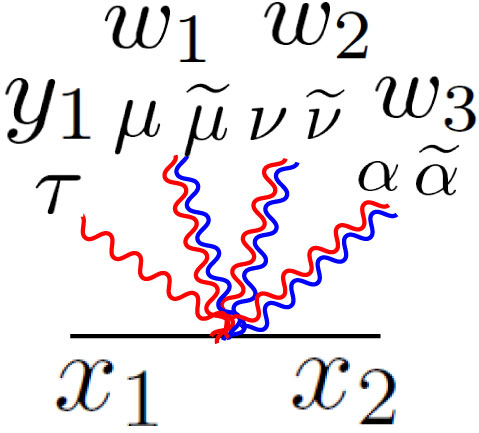} \end{minipage}=\left(-{1\over 2}\right)^3\left[\eta^{\tau\mu}\eta^{\W\mu\W\nu}\eta^{\nu\alpha}\left(k_{x_1}^{\W\alpha}+k_{y_1}^{\W\alpha}+k_{w_2}^{\W\alpha}\right)+\eta^{\tau\mu}\eta^{\W\mu\W\alpha}\eta^{\alpha\nu}\left(k_{x_1}^{\W\nu}+k_{w_1}^{\W\nu}\right)+\eta^{\tau\nu}\eta^{\W\nu\W\mu}\eta^{\mu\alpha}k_{w_1}^{\W\alpha}\right.\Label{Eq:New2x1y3wVertex}\\
&&\left.~~~~~~~~~~~~~~~~~~~+\eta^{\tau\nu}\eta^{\W\nu\W\alpha}\eta^{\alpha\mu}\left(k_{x_1}^{\W\mu}+k_{w_2}^{\W\mu}\right)+\eta^{\tau\alpha}\eta^{\W\alpha\W\nu}\eta^{\nu\mu}\left(k_{x_1}^{\W\mu}+k_{y_1}^{\W\mu}\right)-\eta^{\mu\alpha}\eta^{\W\mu\W\alpha}\eta^{\tau\nu}k^{\W\nu}_{w_1}-\eta^{\nu\alpha}\eta^{\W\nu\W\alpha}\eta^{\tau\mu}k^{\W\mu}_{w_2}
\right].\nn
&&\nn
&&\text{\bf $2x$-$1z$-$3w$ vertex}\nn
&&\begin{minipage}{2.2cm}  \includegraphics[width=2.2cm]{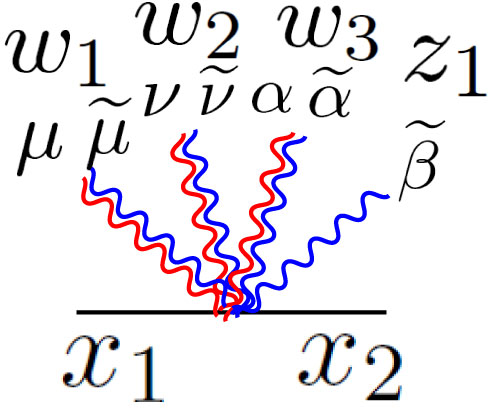} \end{minipage}=\text{(\ref{Eq:New2x1y3wVertex})}\left(\mu\leftrightarrow\W\mu,\nu\leftrightarrow\W\nu,\alpha\leftrightarrow\W\alpha,\tau\to\W\beta,k_{y_1}\to k_{z_1}\right).\\
&&\nn
&&\text{\bf New $2x$-$2y$-$2w$ vertex}\nn
&&\begin{minipage}{2.3cm}  \includegraphics[width=2.3cm]{NewFW2EG1D2} \end{minipage}= \left(-{1\over 2}\right)^3\eta^{\alpha\mu}\eta^{\W\mu\W\nu}\eta^{\nu\beta}~~~~(y_1\prec w_2\prec w_1\prec y_2,{w}_2\prec {w}_1).\Label{Eq:New2x2y2wVertex}\\
&&\nn
&&\text{\bf New $2x$-$2z$-$2w$ vertex}\nn
&&\begin{minipage}{2.3cm}  \includegraphics[width=2.3cm]{NewFW2EG2D1} \end{minipage}=\left(-{1\over 2}\right)^3\eta^{\W\alpha\W\mu}\eta^{\mu\nu}\eta^{\W\nu\W\beta}~~~~( w_2\prec w_1,z_1\prec w_2\prec w_1\prec z_2). \Label{Eq:New2x2z2wVertex}\\
&&\nn
&&\text{\bf New $2x$-$1y$-$2w$ vertex}\nn
&&\begin{minipage}{2.3cm}  \includegraphics[width=2.3cm]{NewFW2EG3D1} \end{minipage}\!\!=\!\!\left\{\begin{matrix}
\left(-{1\over 2}\right)^2\left[\eta^{\alpha\nu}\eta^{\W\nu\W\mu}(k^{\mu}_{x_1}+k^{\mu}_{y_1})+\eta^{\alpha\mu}\eta^{\W\mu\W\nu}k^{\nu}_{w_1}-\eta^{\mu\nu}\eta^{\W\mu\W\nu}k^{\alpha}_{w_1}\right]&(w_1\prec y_1\prec w_2, w_1\prec  w_2)\\
\left(-{1\over 2}\right)^2\eta^{\alpha\nu}\eta^{\W\nu\W\mu}k^{\mu}_{x_1}& (w_1\prec w_2\prec y_1,  w_1\prec w_2)\end{matrix}\right..\Label{Eq:New2x1y2wVertex}\\
&&\nn
&&\text{\bf New $2x$-$1z$-$2w$ vertex}\nn
&&\begin{minipage}{2.3cm}  \includegraphics[width=2.3cm]{NewFW2EG3D2} \end{minipage}\!\!=\!\!\left\{\begin{matrix}
\left(-{1\over 2}\right)^2\left[\eta^{\W\alpha\W\nu}\eta^{\nu\mu}(k^{\W\mu}_{x_1}+k^{\W\mu}_{z_1})+\eta^{\W\alpha\W\mu}\eta^{\mu\nu}k^{\W\nu}_{w_1}-\eta^{\W\mu\W\nu}\eta^{\mu\nu}k^{\W\alpha}_{w_1}\right]&( w_1\prec w_2,w_1\prec z_1\prec w_2)\\
\left(-{1\over 2}\right)^2\eta^{\W\alpha\W\nu}\eta^{\nu\mu}k^{\W\mu}_{x_1}& (w_1\prec w_2, w_1\prec w_2\prec z_1 )\end{matrix}\right..
\eea

All vertex structures $J_{(a)}[A_i]$ ($a=3,..,7$) are written  as contractions of effective currents with vertices. Thus the final local expression of partial integrand with $|\mathsf{W}|=3$ reads
\bea
\mathcal{A}^{\,\text{dYMS},|\mathsf{W}|=3}=\,\Sl_{\{A_1A_2...A_I\}}{1\over l^2}\,J[A_1]\, {1\over s_{A_1,l}}\,J[A_2]\cdots\,{1\over s_{A_1...A_{I-1},l}}\,J[A_I], \Label{Eq:PartialDoubleYMSW3Final}
\eea
where $J[A_i]$ is given by
\bea%
J[A_i]&=&J_{(3)}[A_i]+J_{(4)}[A_i]+J_{(5)}[A_i]+J_{(6)}[A_i]+J_{(7)}[A_i].
\eea
The full integrand with quadratic propagators is 
\bea
\mathcal{I}^{\,\text{dYMS},|\mathsf{W}|=3}=\,\Sl_{\{A_1A_2...A_I\}}{1\over l^2}\,J[A_1]\, {1\over l^2_{A_1}}\,J[A_2]\cdots\,{1\over l^2_{A_1...A_{I-1}}}\,J[A_I]. \Label{Eq:FullDoubleYMSW3Final}
\eea
The vertices of the partial and full integrands with $|\mathsf{W}|\geq 4$ can either be derived by the method provided in the previous sections or extracted by the algorithm in this section. We leave a full approach to the general pattern of vertices in a future work.

\section{From loop to tree}\label{sec:MoreDiscussions}

All the discussions on the localization of partial integrands at one-loop level also apply to tree amplitudes. At tree level, EYM and GR amplitudes can be expanded in terms of tree-level dYMS amplitudes. Particularly, a tree-level EYM amplitude is expressed by 
\bea
A_{\text{tree}}^{\,\text{EYM}}(g_1,\pmb{\rho},g_r||\mathsf{H})&=&
(\epsilon_{g_1}\cdot\epsilon_{g_r})A_{\text{tree}}^{\,\text{dYMS}}\big(g_1,g_r||(\mathsf{G}\cup\mathsf{H})\setminus\{g_1,g_r\}\,\big|\,g_1,\pmb{\rho},g_r||\mathsf{H}\big)\nn
&&~~+\Sl_{l=2}^{n}(-1)^l\Sl_{{\small{(j_1...j_l)}\in\text{S}_l}}\left(\epsilon_{g_1}\cdot{F}_{j_1}\cdot...\cdot {F}_{j_l}\cdot\epsilon_{g_r}\right)\nn
&&~~~~~~~~~~\times A_{\text{tree}}^{\,\text{dYMS}}\big(g_1,j_1,...,j_l,g_r||(\mathsf{G}\cup\mathsf{H})\setminus\{g_1,j_1,...,j_l,g_r\}\,\big|\,g_1,\pmb{\rho},g_r||\mathsf{H}\big),\Label{Eq:EYMDoubleYMS1tree}
\eea
where $g_1$ and $g_r$ are two gluons fixed as the two ends, $\mathsf{G}$ and $\mathsf{H}$ respectively denote the gluon and graviton sets. Tree-level GR amplitude $A_{\text{tree}}^{\,\text{GR}}(\mathsf{H})$ is expanded as 
\bea
A_{\text{tree}}^{\,\text{GR}}(\mathsf{H})&=&\left(\epsilon_{h_1}\cdot\epsilon_{h_s}\right)\left(\W\epsilon_{h_1}\cdot\W\epsilon_{h_s}\right)\,A_{\text{tree}}^{\,\text{dYMS}}\big(h_1,h_s||\mathsf{H}\setminus\{h_1,h_s\}\,\big|\,h_1,h_s||\mathsf{H}\setminus\{h_1,h_s\}\big)\Label{Eq:GRDoubleYMS1tree}\\
&&+\Sl_{l=2}^{n}(-1)^l\Sl_{{\small{(i_1...i_l)}\in\text{S}_l}}\left(\epsilon_{h_1}\cdot\epsilon_{h_s}\right)\big(\W\epsilon_{h_1}\cdot\W{F}_{i_1}\cdot...\cdot \W{F}_{i_l}\cdot\W\epsilon_{h_s}\big)\,\nn
&&~~~\times A_{\text{tree}}^{\,\text{dYMS}}\big(h_1,h_s||\mathsf{H}\setminus\{h_1,h_s\}\,\big|\,h_1,i_1,...,i_l,h_s||\mathsf{H}\setminus\{h_1,i_1,...,i_l,h_s\}\big)\nn
&&+\Sl_{l=2}^{n}(-1)^l\Sl_{{\small{(j_1...j_l)}\in\text{S}_l}}\big(\epsilon_{h_1}\cdot {F}_{j_1}\cdot...\cdot {F}_{j_l}\cdot\epsilon_{h_s}\big)\left(\W\epsilon_{h_1}\cdot\W\epsilon_{h_s}\right)\nn
&&~~~\times A_{\text{tree}}^{\,\text{dYMS}}\big(h_1,j_1,...,j_l,h_s||\mathsf{H}\setminus\{h_1,j_1,...,j_l,h_s\}\,\big|\,h_1,h_s||\mathsf{H}\setminus\{h_1,h_s\}\big)\nn
&&+\Sl_{l,m=2}^{n}(-1)^{l+m}\Sl_{\substack{\small{(j_1...j_l)}\in\text{S}_l\\\small{(i_1...i_m)}\,\in\,\text{S}_m}}\big(\epsilon_{h_1}\cdot {F}_{j_1}\cdot...\cdot {F}_{j_l}\cdot\epsilon_{h_s}\big)\,\big(\W\epsilon_{h_1}\cdot \W{F}_{i_1}\cdot...\cdot \W{F}_{i_m}\cdot\W\epsilon_{h_s}\big)\nn
&&~~~\times A_{\text{tree}}^{\,\text{dYMS}}\big(h_1,j_1,...,j_l,h_s||\mathsf{H}\setminus\{h_1,j_1,...,j_l,h_s\}\,\big|\,h_1,i_1,...,i_m,h_s||\mathsf{H}\setminus\{h_1,i_1,...,i_m,h_s\}\big),\nonumber
\eea
where two gravitons $h_1$ and $h_s$ are fixed as the two ends.

The tree-level dYMS amplitudes $A_{\text{tree}}^{\,\text{dYMS}}$ in  (\ref{Eq:EYMDoubleYMS1tree}) and (\ref{Eq:GRDoubleYMS1tree}), with at least two $\mathsf{X}$ elements, are defined by
\bea
A_{\text{tree}}^{\,\text{dYMS}}\big(\big.x_1,\pmb{\sigma},x_r||\mathsf{G}\,\big|\,x_1,\pmb{\rho},x_r||\W{\mathsf{G}}\,\big)\equiv\int \text{d}\mu\,I^{\,\text{dYMS}}_L\,I^{\,\text{dYMS}}_R. \Label{Eq:doubleYMStree}
\eea
In the above, both half integrands $I_L$ and $I_R$ are defined by {\it the tree-level left half integrands of YMS}
\bea
I^{\,\text{dYMS}}_L=\text{PT}(x_1,\pmb{\sigma},x_r)\,\text{Pf}\,[\Psi_{\mathsf{G}}],~~~~I^{\,\text{dYMS}}_R=\text{PT}(x_1,\pmb{\rho},x_r)\,\text{Pf}\,[\Psi_{\W{\mathsf{G}}}]. \Label{Eq:ILRtree}
\eea
In each PT factor for a half integrand, $x_1$ and $x_r$ are fixed as the two ends. When applying the expansion formula and integrating over scattering variables, we write the tree-level dYMS amplitude as
\bea
{A}_{\text{tree}}^{\,\text{dYMS}}\big(\big.x_1,\pmb{\sigma},x_r||\mathsf{G}\,\big|\,x_1,\pmb{\rho},x_r||\W{\mathsf{G}}\,\big)&=&\Sl_{{\mathcal{F}},\,\W{\mathcal{F}}}\,\Sl_{\pmb{\beta}[{\mathcal{F}}],\pmb{\gamma}[\W{\mathcal{F}}]}\,\mathcal{C}[{\mathcal{F}}]\,{A}_{\text{tree}}^{\text{BS}}\big(x_1,\pmb{\beta}[\mathcal{F}],x_r\big|x_1,\pmb{\gamma}[\W{\mathcal{F}}],x_r\big)\,\mathcal{C}\big[\W{\mathcal{F}}\big],\Label{Eq:doubleYMS1tree}
\eea
where ${A}_{\text{tree}}^{\text{BS}}$ are tree-level BS amplitudes. When ${A}_{\text{tree}}^{\text{BS}}$ are expressed by Feynman diagrams, the above expression is further arranged as 
\bea
{A}_{\text{tree}}^{\,\text{dYMS}}\big(\big.x_1,\pmb{\sigma},x_r||\mathsf{G}\,\big|\,x_1,\pmb{\rho},x_r||\W{\mathsf{G}}\,\big)&=&\Sl_{{\mathcal{F}},\,\W{\mathcal{F}}}\,\Sl_{\pmb{\beta}[{\mathcal{F}}],\pmb{\gamma}[\W{\mathcal{F}}]}\,\Sl_{\small\substack{(A_1A_2...A_I)={\pmb{\beta}}\\\small (\W A_1\W A_2...\W A_I)={\pmb{\gamma}}}}\mathcal{C}[{\mathcal{F}}]\,{\phi_{A_1|\W A_1}\phi_{A_2|\W A_2}\cdots \phi_{A_I|\W A_I}\over s_{x_1A_1}s_{x_1A_1A_2}\cdots s_{x_1A_1A_2\cdots A_{I-1}}}\,\mathcal{C}\big[\W{\mathcal{F}}\big]\nn
&=&\Sl_{{\mathcal{F}},\,\W{\mathcal{F}}}\,\Sl_{\pmb{\beta}[{\mathcal{F}}],\pmb{\gamma}[\W{\mathcal{F}}]}\,\Sl_{\small\substack{(A_1A_2...A_I)={\pmb{\beta}}\\\small (\W A_1\W A_2...\W A_I)={\pmb{\gamma}}}}\mathcal{C}[{\mathcal{F}}]\,\begin{minipage}{2.5cm}  \includegraphics[width=2.5cm]{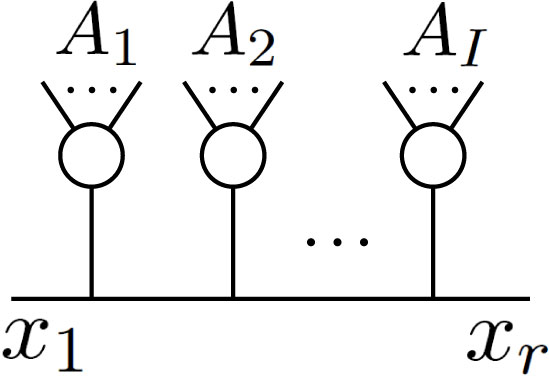} \end{minipage}\,\mathcal{C}\big[\W{\mathcal{F}}\big].\Label{Eq:doubleYMS20tree}
\eea
Here, 
the propagators between $x_1$ and $x_r$ are BS propagators in tree-level Feynman diagrams, which do not contain loop momentum.  The definition of $s_{x_1A_1...A_j}$ is $s_{x_1A_1...A_j}\equiv(k_{x_1}+k_{A_1}+...+k_{A_j})^2$. The expression (\ref{Eq:doubleYMS20tree}) also contains nonlocal terms in which Lorentz contractions of subcurrents are separated by propagators between  $x_1$ and $x_r$. Since the expression (\ref{Eq:doubleYMS20tree}) is essentially the same as the one (\ref{Eq:doubleYMS20}) when $x_1$, $x_r$ are respectively replaced by $+$, $-$, the X-patterns and BCJ-patterns are also included in (\ref{Eq:doubleYMS20tree}). A crucial observation is that the property (\ref{Eq:XCoefficient}) of the X-pattern is also satisfied in this situation
\bea
k_A\cdot X_A={1\over2}\left[(k_A+ X_A)^2-X_A^2-k_A^2\right],\Label{Eq:XCoefficientTree}
\eea
where $X_A^{\mu}$ and $(k_A+ X_A)^{\mu}$ are the momenta of the propagators to the left and the right of the subcurrent $A$, $k_A^{\mu}$ is the total momentum of the subcurrent  $A$. Hence (\ref{Eq:XCoefficientTree}) can also be used to delete propagators as in (\ref{Eq:PropertyXPattern}) (where the $+$ and $-$ are now replaced by $x_1$ and $x_r$). Consequently, the cancellations between X- and BCJ-patterns totally follow the same way as the discussions at one-loop level. When the localization-1 and -2 are applied, we get a local expression of the tree-level dYMS amplitude
\bea
{A}_{\text{tree}}^{\,\text{dYMS}}\big(\big.x_1,\pmb{\sigma},x_r||\mathsf{G}\,\big|\,x_1,\pmb{\rho},x_r||\W{\mathsf{G}}\,\big)=\,\Sl_{\{A_1A_2...A_I\}}\,J[A_1]\, {1\over s_{x_1A_1}}\,J[A_2]\cdots\,{1\over s_{x_1A_1...A_{I-1}}}\,J[A_I], \Label{Eq:LocalDoubleYMSTree}
\eea
where we have summed over all possible partitions of external particles except for $x_1$, $x_r$. Local structure $J[A_j]$ has the same definition within the one-loop case, i.e., attaching tree-level effective currents to the propagator line between $x_1$, $x_r$, via the local vertices. These effective currents are defined by (\ref{Eq:J1})-(\ref{Eq:J4}) and have double-copy structure. For $|\mathsf{W}|\leq 3$, the vertices are those explicitly induced in \secref{sec:W0} and \secref{sec:W1}-\ref{sec:W3}.

\section{Conclusions and further discussions}\label{sec:Conclusions}


In this work, we investigated the one-loop integrands for theories with gravitons, which are constructed by the forward limit approach. A specific method for canceling the nonlocalities in double Yang-Mills-scalar (dYMS) partial integrands has been provided. The explicit expressions of vertices for integrands with $|\mathsf{W}|\leq 3$ (where $|\mathsf{W}|$ denotes the number of elements which have two copies of polarization vectors) were presented. 
More complicated vertices which contain more $\mathsf{W}$ elements in principle can be constructed in a similar way. Once the local vertices have been obtained by localizations, the partial integrand with linear propagators is expressed by planting effective currents towards the loop propagator line, through these local vertices. The graphic rule for the forward limit form of one-loop integrand permits the local partial integrands related by cyclic permutations around the loop. Consequently, the partial fraction identity finally transforms these local expressions of dYMS partial integrands into full integrands with quadratic propagators. The quadratic propagator expressed one-loop integrands for Einstein-Yang-Mills (EYM) theory and Gravity (GR) are provided via expressing them by dYMS integrands. We showed that the localizations also applied at tree-level and provided the formulas for tree amplitude that are expressed through local dYMS vertex structures. An interesting feature of the effective subcurrents in the final expressions of either one-loop integrands or tree amplitudes is that they have an explicit double-copy structure. Although all results of this paper are derived with a specific choice of reference order, we emphasize that the choice of reference order does not affect the final result of one-loop amplitude. In fact, different reference orders just reflect different choices of gauge. The results in this paper provide a systematic approach to the local vertices and it seems that there exist some relations between different vertex structures.

It is worthwhile to comment on how to generalize the discussions in this work to supersymmetric Einstein-Yang-Mills (super EYM) theories. This issue encompasses two aspects: (i) How to extract the quadratic propagators for super EYM amplitudes? (ii) How to extend the localization procedures in this paper to obtain local vertices in super EYM integrands? From the perspective of the current paper, once we have the answer to (ii), we naturally get the answer to (i). {\it Nevertheless, extracting the quadratic propagators may not require the full localization.} In the following, we just analyze the explicit four-point results which were first provided in \cite{Porkert:2022efy} to demonstrate this point. 

The {\it maximally supersymmetric EYM theory} is the tensor theory of maximally supersymmetric Yang-Mills (super-YM) and YMS theory  \cite{Porkert:2022efy}. For the maximally supersymmetric EYM theory, the four-point integrands with external gluons and/or gravitons contains a left coefficient which have the form $t_8=\text{Tr}[F_1\cdot F_2\cdot F_3\cdot F_4]-{1\over 4}\text{Tr}[F_1\cdot F_2]\text{Tr}[F_3\cdot F_4]+\text{cyc}(2,3,4)$ originates from the super-YM part, and a right coefficient derived from the YMS part\footnote{ See equation (5.3) of \cite{Porkert:2022efy}.}. Because the left coefficient has an exact form with cyclic symmetry and is independent of loop momentum, it serves as a trace factor in (\ref{Eq:EYMDoubleYMS1}) (More precisely speaking, the terms with $l=n$ where the dYMS integrands become YMS integrands). Therefore, when the trace is stripped out, the remaining part is a YMS partial integrand, which is the simplest case since it is a double integrand of the form S$\otimes$YMS. The localization and the treatment of the quadratic propagators in this case have already been completed in \cite{Xie:2024pro}. One naturally arrives at a quadratic propagator form of this example.  {\it Example of half-maximally supersymmetric EYM} is more subtle, as the left coefficients involve loop momentum, see \cite{Porkert:2022efy}. Nevertheless, these coefficients have been arranged into a symmetric form that satisfies the consistency condition (see recent papers \cite{Cao:2025ygu,Du:2025yxz,Du:2025rty} as well as the paper on tensorial PT factor \cite{Feng:2022wee}). This feature allows us to treat the left coefficient as a ‘trace’ and then apply the localization procedure to the right coefficients\footnote{More detail is presented in \cite{Du:2025rty} by the authors of the current work.}. The resulting integrand again takes the form of YMS integrands with quadratic propagators accompanied by the left kinematic coefficients. 

The above two examples provide a hint for analyzing the quadratic propagators from aspect (i): Once an amplitude (or susy amplitude) is written as two copies, and the coefficient from one copy can be written as a trace or a kinematic coefficient involving the loop momentum which satisfy the consistency condition \cite{Cao:2025ygu,Du:2025yxz},  it is naturally to arrange the full amplitude into a quadratic propagator form by performing localizations with respect to the other copy of coefficients. To further proceed this approach for amplitudes with an arbitrary number of external particles, the key step is to construct the trace or consistency kinematic factors for one copy.

From aspect (ii), when we want to perform a full localization of supersymmetric amplitudes (with either external gluons, gravitons or fermions) totally following the approach proposed in the current paper, we need to establish a similar graphic expansion rule to generate the kinematic coefficients on both sides. Once this has been accomplished, we can find the ‘upper’ and ‘lower’ blocks inside a subgraph \cite{Xie:2024pro}, then the cancellations between BCJ- and X- patterns totally follow from those in the current paper. Using the localization techniques in this paper, one can get the susy version of vertices in a systematic way. It is hopeful to generate the graphic rule by the known result of supersymmetric theories (the off-shell approaches \cite{Mafra:2015vca,Lee:2015upy,Bridges:2019siz} may provide more hints along this line). We leave a systematic treatment of amplitudes with supersymmetries for future work.

To end this work, we present several topics that are related to this paper. {\it First}, it seems that the tree-level effective subcurrents can be considered as an off-shell extension of the double-copy formula of on-shell tree amplitudes. Therefore, a tree amplitude or one-loop integrand can be constructed by planting these subcurrents towards the propagator line between two fixed particles or around the loop, through conventional Feynman diagram vertices. Nevertheless, it is not easy to obtain (calculate) the multi-point vertex contributions because of the complexity. This work in fact provided a general approach to constructing local vertices in a recursive way: applying the localizations to diagrams with fewer-point vertices and then extracting a higher-point vertex. In future work, we will study the general recursive relations between vertices in a general way. {\it Second}, although we have achieved the localizations of dYMS, which essentially determined a quadratic propagator form of EYM and GR,  a straightforward localization of EYM and GR still deserves further investigation since the traces of field strength tensors also bring nonlocality into the amplitudes. One interesting observation is that the final result of this step localization may be related to an explicit solution of the recursive approach  \cite{Gomez:2022dzk,Gomez:2024xec} to one-loop integrands. A systematic study of the localization of EYM and GR is expected.
{\it Last}, this work may provide a new angle to a future study of the BCJ double-copy \cite{Bern:2008qj,Bern:2010ue} with respect to quadratic propagators, by regarding the dYMS integrands as a bridge. Especially,  it is worth finding a systematic way to decompose these multi-point vertices into cubic vertex structures which satisfy Jacobi identities.

\section*{Acknowledgements}

We would like to thank (in alphabetical order) Qu Cao, Gang Chen, Bo Feng, Chih-Hao Fu, Song He, Chang Hu, Yihong Wang, Konglong Wu, Fan Zhu for helpful comments and discussions. We appreciate Qu Cao, Song He, Yong Zhang and Fan Zhu for informing us about their new work to appear, on the double copy relations from single-cuts. The $|\mathsf{W}|=1$ example was first presented by YD in``5th seminar on field theory and string theory
in China'' (2024, HeFei, China). YD would like to thank the organizers of ``QUIST-VIII'' for kind invitation. The authors thank the organizers of ``3rd Frontier Workshop on Quantum Field Theory''(2025, Xi'an, China). This work is supported by NSFC under Grant No. 11875206.

\appendix

\section{Examples for cancellation in the case  $|\mathsf{W}|=2$}

The cancellations associated with partitions $\{y_1\text{-}w_1,y_2\text{-}w_2\}$ and $\{z_1,z_2,w_1\text{-}w_2\}$ are described below.

{\bf Cancellation induced by partition $\{y_1\text{-}w_1,y_2\text{-}w_2\}$}~~For partition $\{y_1\text{-}w_1,y_2\text{-}w_2\}$, there exists term
\bea
\mathrm{T}_1&=&\left(-{1\over 2}\right)^2\begin{minipage}{2.1cm}  \includegraphics[width=2.1cm]{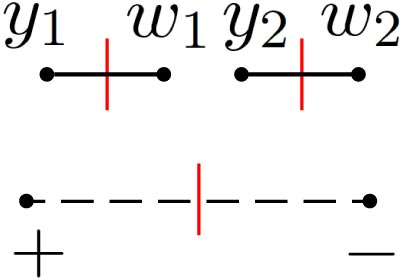} \end{minipage}\times\begin{minipage}{2.2cm}  \includegraphics[width=2.2cm]{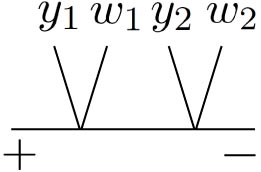} \end{minipage}\times\begin{minipage}{2.4cm} \includegraphics[width=2.4cm]{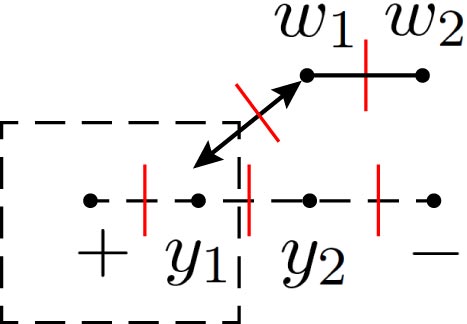} \end{minipage}\nn
&=&\left(-{1\over 2}\right)^3\begin{minipage}{2.1cm}  \includegraphics[width=2.1cm]{EQW2EG1L1} \end{minipage}\times\left[\begin{minipage}{2.2cm}  \includegraphics[width=2.2cm]{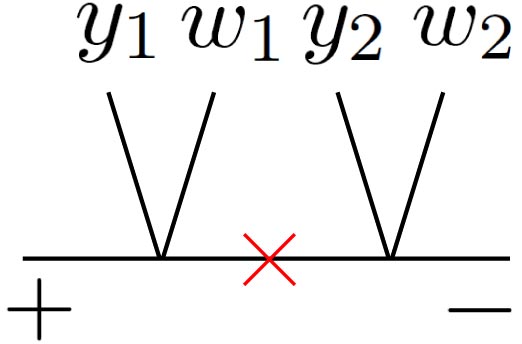} \end{minipage}-s_{y_1,l}\begin{minipage}{2.1cm}  \includegraphics[width=2.1cm]{EQW2EG1D1} \end{minipage}\right]\times\begin{minipage}{2.1cm}\includegraphics[width=2.1cm]{EQW2EG1R2} \end{minipage}\nn
&=&\mathrm{T}_{1\text{A}}-\mathrm{T}_{1\text{B}},
%
\eea
where we have applied the property (\ref{Eq:PropertyXPattern}) of X-pattern to get the second line. The $\mathrm{T}_{1\text{A}}$ term is already local with the left coefficient $(\epsilon_{y_1}\cdot\epsilon_{w_1})(\epsilon_{y_2}\cdot\epsilon_{w_2})$ and the right coefficient $\W{\epsilon}_{w_1}\cdot\W{\epsilon}_{w_2}$. However, further calculations show that both of the terms in $\mathrm{T}_1$ cancel out with the other terms and the $\mathrm{T}_{1\text{A}}$ is not a six-point vertex structure preserved in the final result. Concretely, we consider the following term with the partition $\{\{y_1,w_1\},y_2\text{-}w_2\}$
\bea
\mathrm{T}_2&=&\left(-{1\over 2}\right)\begin{minipage}{2.1cm}  \includegraphics[width=2.1cm]{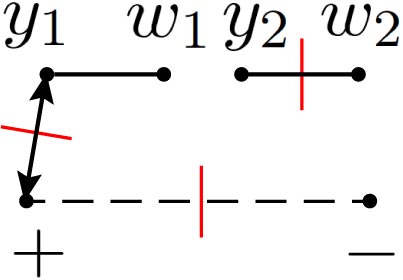} \end{minipage}\times\begin{minipage}{2.2cm}  \includegraphics[width=2.2cm]{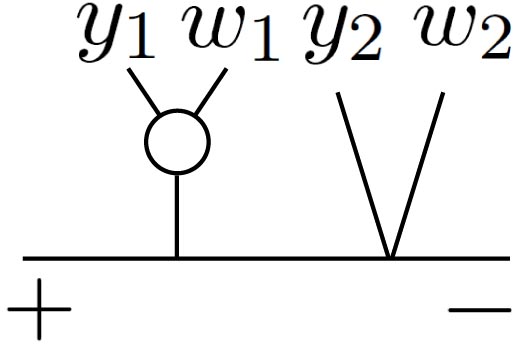} \end{minipage}\times\begin{minipage}{2.1cm} \includegraphics[width=2.1cm]{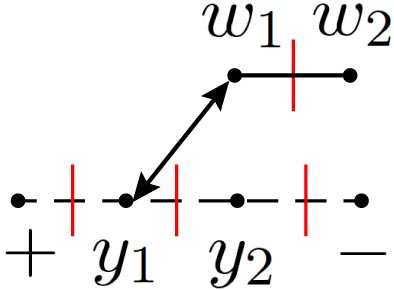} \end{minipage}\nn
&=&\left(-{1\over 2}\right)^2\begin{minipage}{2.1cm}  \includegraphics[width=2.1cm]{EQW2EG1L2} \end{minipage}\times\begin{minipage}{2.2cm}  \includegraphics[width=2.2cm]{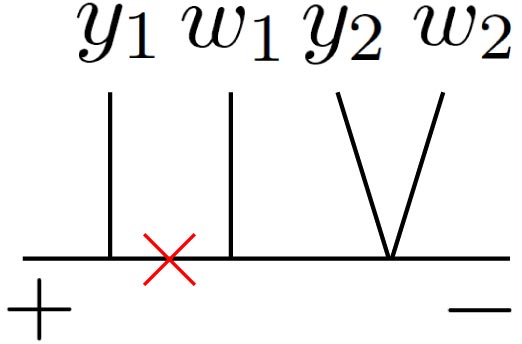} \end{minipage}\times\begin{minipage}{2cm} \includegraphics[width=2cm]{EQW2EG1R2} \end{minipage}\nn
&=&\left(-{1\over 2}\right)^3\begin{minipage}{2.1cm}  \includegraphics[width=2.1cm]{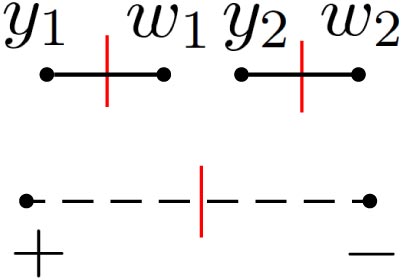} \end{minipage}\times s_{y_1,l}\begin{minipage}{2.2cm}  \includegraphics[width=2.2cm]{EQW2EG1D4} \end{minipage}\times\begin{minipage}{2cm} \includegraphics[width=2cm]{EQW2EG1R2} \end{minipage}.
\eea
The first line in the above expression becomes the second line when the off-shell BCJ relation 
\bea
(-k_{y_1}\cdot k_{w_1})\phi_{y_1w_1|y_1w_1}=\left(-{1\over 2}\right)(\phi_{y_1|y_1}\phi_{w_1|w_1}-\phi_{y_1|w_1}\phi_{w_1|y_1})
\eea
is applied.
After this step, the left coefficient which did not provide an X-pattern before, now contains an X-pattern associated with $y_1$. When the property of X-pattern is further applied, we get the third line which is the same as $\mathrm{T}_{1\text{B}}$. Thus, term $\mathrm{T}_{2}$ cancels out with $-\mathrm{T}_{1\text{B}}$. 

A third term $\mathrm{T}_3$ with the partition $\{\{y_1,y_2,w_1\}\text{-}w_2\}$, which should also be taken into account is 
\bea
\mathrm{T}_3=\left(-{1\over 2}\right)\begin{minipage}{2.1cm}  \includegraphics[width=2.1cm]{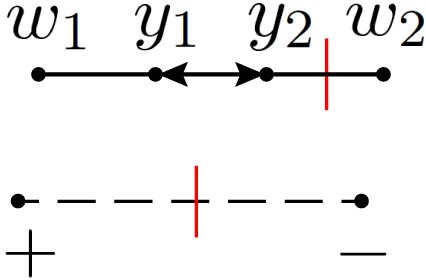} \end{minipage}\times\begin{minipage}{1.6cm}  \includegraphics[width=1.6cm]{EQW2EG1D6} \end{minipage}\times\begin{minipage}{2cm} \includegraphics[width=2cm]{EQW2EG1R3} \end{minipage}.
\eea
The left graph implies that the left permutation in the subcurrent containing $y_1$, $y_2$, $w_1$ is $(y_2 y_1 w_1)$, while the right graph implies an off-shell BCJ relation for this subcurrent. This observation further reduces $\mathrm{T}_3$ as follows
\bea
\mathrm{T}_3=\left(-{1\over 2}\right)^2\begin{minipage}{2.1cm}  \includegraphics[width=2.1cm]{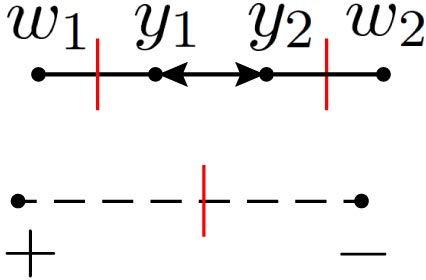} \end{minipage}\times\begin{minipage}{2.2cm}  \includegraphics[width=1.8cm]{EQW2EG1D7} \end{minipage}\times\begin{minipage}{2cm} \includegraphics[width=2cm]{EQW2EG1R2} \end{minipage}.
\eea
The type-3 line between $y_1$ and $y_2$ further permits one to apply the off-shell BCJ relation on the subcurrent with $y_1$ and $y_2$
\bea
(-k_{y_1}\cdot k_{y_2})\phi_{y_2y_1|y_1y_2}=\left(-{1\over 2}\right)(\phi_{y_2|y_1}\phi_{y_1|y_2}-\phi_{y_1|y_1}\phi_{y_2|y_2}).
\eea
As a result, $\mathrm{T}_3$ is finally expressed as 
\bea
\mathrm{T}_3=\left(-{1\over 2}\right)^3\begin{minipage}{2.1cm}  \includegraphics[width=2.1cm]{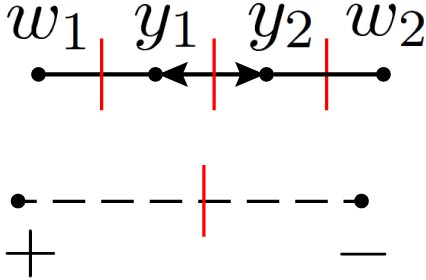} \end{minipage}\times(-1)\begin{minipage}{2cm}  \includegraphics[width=2cm]{EQW2EG1D8} \end{minipage}\times\begin{minipage}{2cm} \includegraphics[width=2cm]{EQW2EG1R2} \end{minipage},
\eea
where the extra minus is a result of the off-shell BCJ relation. Apparently, $\mathrm{T}_3$ cancels out with $\mathrm{T}_{1\text{A}}$. The sum of $\mathrm{T}_1$, $\mathrm{T}_2$, $\mathrm{T}_3$ is zero. These terms do not contribute to vertex structures.

{\bf Cancellation induced by partition $\{z_1,z_2,w_1\text{-}w_2\}$}~~We first consider the following term with partition $\{z_1,z_2,w_1\text{-}w_2\}$
\bea
\left(-{1\over 2}\right)\begin{minipage}{2cm}  \includegraphics[width=2cm]{EQW2EG2L1} \end{minipage}\times\begin{minipage}{2cm}  \includegraphics[width=2cm]{EQW2EG2D1} \end{minipage}\times\begin{minipage}{1.4cm} \includegraphics[width=1.4cm]{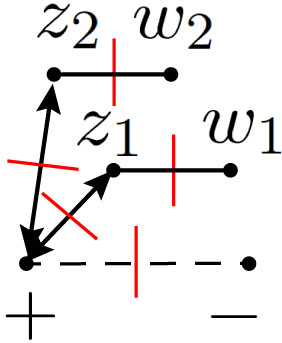} \end{minipage}=\left(-{1\over 2}\right)^3\begin{minipage}{2cm}  \includegraphics[width=2cm]{EQW2EG2L1} \end{minipage}\times\begin{minipage}{2cm}  \includegraphics[width=2cm]{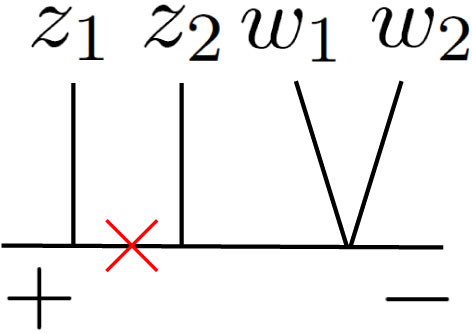} \end{minipage}\times\begin{minipage}{2.1cm} \includegraphics[width=2.1cm]{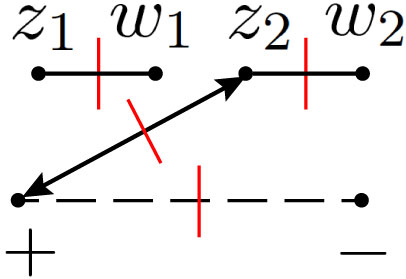} \end{minipage},\Label{Eq:2x2z2wCancellation1}\nn
\eea
where the property of X-patterns associated with $z_1$ has been used. 
%
%
%
%
Another term under consideration is the one with partition $\{\{z_1,z_2\},w_1\text{-}w_2\}$
\bea
\left(-{1\over 2}\right)\begin{minipage}{2cm}  \includegraphics[width=2cm]{EQW2EG2L2} \end{minipage}\times\begin{minipage}{1.8cm}  \includegraphics[width=1.8cm]{EQW2EG2D4} \end{minipage}\times\begin{minipage}{1.5cm} \includegraphics[width=1.5cm]{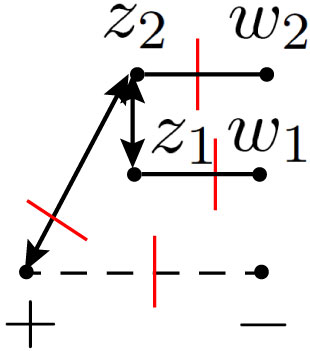} \end{minipage}
\rightarrow(-1)\left(-{1\over 2}\right)^3\begin{minipage}{2cm}  \includegraphics[width=2cm]{EQW2EG2L1} \end{minipage}\times\begin{minipage}{2cm}  \includegraphics[width=2cm]{EQW2EG2D2New} \end{minipage}\times\begin{minipage}{2cm} \includegraphics[width=2cm]{EQW2EG2R2New} \end{minipage}.\Label{Eq:2x2z2wCancellation2}\nn
\eea
In the above expression, we have applied the off-shell BCJ relation to the subcurrent containing $z_1$ and $z_2$. Apparently, (\ref{Eq:2x2z2wCancellation1}) cancels out with (\ref{Eq:2x2z2wCancellation2}).

\bibliographystyle{JHEP}
\bibliography{reference.bib}

\end{document}